\documentclass[11pt,reqno]{amsart}

\usepackage{amsmath,amsfonts,amsthm,amssymb,amsxtra}
\usepackage[margin=1.1in]{geometry}
\usepackage{color}
\usepackage{stmaryrd}
\usepackage{nicematrix}
\usepackage{float}
\usepackage{bbold}
\usepackage{graphicx,subfigure}
\usepackage[font=small]{caption}
\usepackage[colorlinks,citecolor=red,urlcolor=blue]{hyperref}

\newtheorem{theorem}{Theorem}%[section]
\newtheorem{conjecture}{Conjecture}
\newtheorem{proposition}{Proposition}
\newtheorem{lemma}{Lemma}[section]
\newtheorem{corollary}{Corollary}
\newtheorem{claim}{Claim}[section]

\theoremstyle{definition}

\newtheorem{definition}{Definition}
\newtheorem{example}{Example}
\newtheorem{assumption}{Assumption}

\theoremstyle{remark}

\newtheorem{remark}{Remark}[section]

\numberwithin{equation}{section}

\renewcommand{\epsilon}{\varepsilon}

\newcommand{\N}{\mathbb{N}}
\newcommand{\V}{\mathbb{V}}

\renewcommand{\phi}{\varphi}
\newcommand{\R}{\mathbb{R}}

\newcommand{\Z}{\mathbb{Z}}

\newcommand{\E}{\mathbb{E}}

\newcommand{\one}{\mathbb{1}}

\newcommand{\wt}{\widetilde}

\newcommand{\intbr}[2]{\llbracket  #1 ,   #2 \rrbracket }

\newcommand{\ipc}[2]{\left \langle #1 , #2 \right \rangle }
\newcommand{\braket}[3]{\left \langle #1  | #2 | #3 \right \rangle }
\newcommand{\Ev}[1]{\mathbb{E} \left( #1 \right)}  %% produces \E( # )

\renewcommand{\P}{\mathbb{P}}
\renewcommand{\Pr}[1]{ {\P}\left\{\, #1 \, \right\}}

\newcommand{\eps}{\varepsilon}

\renewcommand{\P}{\mathbb{P}}

\newcommand{\bigzero}{\mbox{\normalfont\Large\bfseries 0}}
\newcommand{\rvline}{\hspace*{-\arraycolsep}\vline\hspace*{-\arraycolsep}}

\newcommand{\vertiii}[1]{{\left\vert\kern-0.25ex\left\vert\kern-0.25ex\left\vert #1
    \right\vert\kern-0.25ex\right\vert\kern-0.25ex\right\vert}}

\newcommand\numberthis{\stepcounter{equation}\tag{\theequation}}

\newtheorem{defn}[definition]{Definition}

\newtheorem{ex}[example]{Example}

\newcommand{\pdelta}{\Delta^{(p)}}

\usepackage{tikz}
\usepackage[shortlabels]{enumitem}
\usepackage{pgfplots}

\newcommand{\wall}{\mathcal{W}}
\newcommand{\isl}{\mathcal{I}}
\newcommand{\isll}{\mathcal{I}_L}
\newcommand{\islh}{\mathcal{I}_H}

\pgfplotsset{compat=1.18}

\begin{document}

\title[Landscape approximation of the ground state eigenvalue]{{Landscape approximation of the ground state eigenvalue for graphs and random hopping models}}
\author[L. Shou, W. Wang, S. Zhang]{Laura Shou, Wei Wang, Shiwen Zhang}
\date{}
\maketitle

\begin{abstract}
  We consider the localization landscape function $u$ and ground state eigenvalue $\lambda$ for operators on graphs. We first show that the maximum of the landscape function is comparable to the reciprocal of the ground state eigenvalue if the operator satisfies certain semigroup kernel upper bounds. This implies general upper and lower bounds on the landscape product $\lambda\|u\|_\infty$ for several models, including the Anderson model and random hopping (bond-disordered) models, on graphs that are roughly isometric to $\Z^d$, as well as on some fractal-like graphs such as the Sierpinski gasket graph. Next, we specialize to a random hopping model on $\Z$, and show that as the size of the chain grows, the landscape product $\lambda\|u\|_\infty$ approaches $\pi^2/8$ for Bernoulli off-diagonal disorder, and has the same upper bound of $\pi^2/8$ for $\operatorname{Uniform}([0,1])$ off-diagonal disorder. 
  We also numerically study  the random hopping model when the band width (hopping distance) is greater than one, and provide strong numerical evidence that a similar approximation holds for low-lying energies in the spectrum. 
\end{abstract}

{\hypersetup{linkcolor=black}
\tableofcontents
}
\section{Introduction}\label{sec:intro}

The landscape function, first introduced in \cite{filoche2012universal},  is a solution $u$ to the equation $Hu(x)=1$ for an elliptic operator $H$. 
In  the series of works \cite{arnold2016effective,arnold2019computing,arnold2019localization},  the localization landscape theory was developed to study the spectrum and eigenvectors of a large class of operators via the landscape function $u$,
without explicitly solving the eigenvalue problem. The central class of interest in these works is the Schr\"odinger operator $-\Delta+V$, which includes the notable (continuous) Anderson model,  with a nonegative potential $V$.
In \cite{arnold2019computing}, Arnold et al. studied  continuous Anderson models on some bounded domain $\Omega\subset\R^d$, and observed  that 
${\lambda^{(n)}}\cdot{\max^{(n)} {u}}  \approx  1+ \frac{d}{4}$,
where $\lambda^{(n)}$ is the $n$-th eigenvalue in ascending order and $\max^{(n)} {u}$ is the $n$th local maximum of the landscape function $u$.  
This observation was studied further in \cite{CWZ} for the 1D continuous Anderson Bernoulli model $-\frac{d^2}{dx^2}+V$, where the authors showed that the landscape product
$\lambda_N\|u_N\|_{L^\infty([0,N])}\to \pi^2/8$ almost surely as $N\to\infty$, where $\lambda_N$ is the ground state eigenvalue and $u_N$ is the landscape function of the Dirichlet Anderson Bernoulli model on $[0,N]$.  
Numerical experiments supporting similar asymptotic behavior for the excited state energies were also performed. 
More recently, the landscape product was studied for the discrete Anderson model on $\Z^d$ in \cite{sanchez2023principal} with more general distributions for the potential $V$.   
The above approximations of eigenvalues via the landscape function have good numerical accuracy \cite{arnold2019computing,CWZ}, and are computationally cheap to apply. 
The landscape function was further used in \cite{david2021landscape,arnold2022landscape} to develop a landscape law to approximate the integrated density of states.

In the case of zero potential $V$, one has just the Dirichlet Laplacian  $-\Delta$ on $\Omega\subset\R^d$, and the associated landscape function $u$ that solves $-\Delta u(x)=1$ is known as the torsion function. The torsion function can be realized as the expected exit time of a Brownian motion started at $x$ and has its own interest in a variety of contexts such as in elasticity theory, heat conduction, and minimal submanifolds, see e.g. \cite{banuelos2002torsional,markvorsen2006torsional,PolyaSzego}. In \cite{vdc}, van den Berg and Carroll showed that the supremum of the torsion function $\|u\|_{L^\infty(\Omega)}$ is comparable to the reciprocal of the ground state eigenvalue $\lambda$ of $-\Delta$ on $\Omega\subset\R^d$, in the sense that $e^{-1}\le  \lambda\|u\|_{L^\infty}\le 4+3d\log 2$. They also used this to relate $L^p$ norms of the torsion function to the distance to the boundary of $\Omega$ under a Hardy type inequality assumption. In \cite{vogt}, Vogt generalized this work from the Dirichlet Laplacian to more general operators $H$ on $\Omega\subseteq\R^d$ that satisfy certain Gaussian upper bounds on the integral kernel of $e^{-tH}$, and proved that $1\le  \lambda\|u\|_{L^\infty}\le d/8+c\sqrt d+1$ with an absolute constant $c\approx0.61$, for the associated ground state $\lambda$ and torsion function $u$ of $H$. The leading order $d/8$ in the upper bound is optimal as $d\to\infty$.

In this paper we investigate the {landscape product}  
$\lambda\|u\|_{L^\infty}$ for discrete operators on graphs, which include the Anderson model and random hopping (bond-disordered) models. 
Such random hopping models are relevant in semiconductor physics to describe systems with varying hopping matrix elements but nearly constant on-site energies \cite{ITA}, and are related to the random conductivity models used to study disordered media \cite{Kirkpatrick}.
The off-diagonal disorder is known in the physics literature to cause Anderson localization, though with a
possible singularity at the center of the band energy that is not present with just on-site diagonal disorder \cite{Dyson,ITA,TheodorouCohen}. 
Here we will focus on the low-lying energies and states in the spectrum, which appear to behave similarly to that of the Anderson model, for example by presenting with Lifshitz tails in the integrated density of states \cite{KloppNakamura}. Also in this low energy regime, Agmon-type estimates for eigenvector localization were obtained via the landscape function in \cite{filoche2021effective} for certain $M$-matrices, which include for example random
band hopping models. 
We mention also that localization and the integrated density of states in related models with off-diagonal disorder have been of interest in the mathematics literature, for example \cite{FigotinKlein,AizenmanMolchanov,MuellerStollman}, among others.

Our first result concerns general graphs and operators, where we obtain similar bounds on the landscape product $\lambda\|u\|_{L^\infty}$ as obtained for continuous operators on $\R^d$ in \cite{vdc,vogt}, provided the operator satisfies certain semigroup kernel bounds. We show that these general bounds can be applied to several popular models including  
the Laplacian on graphs, the Anderson model, and the random band hopping model, on graphs that are \emph{roughly isometric} to $\Z^d$ (to be defined in Section \ref{sec:general}), as well as on some fractal-like graphs  
such as the \emph{Sierpinski gasket graph}. In the second part, we focus on 1D chains and consider the associated random band {hopping} models, in which particles may have nonzero hopping coefficients between their $2W$ nearest neighbors.   
We show first that in the nearest neighbor hopping case when the band width is 1, that if the off-diagonal disorder is drawn from a  Bernoulli-type  distribution, then 
$\lambda_N\|u_N\|_{\ell^\infty}\xrightarrow[N\to \infty]{\text{a.s.}} \pi^2/8$, where $N$ is the number of sites. We also prove the same upper bound for uniform off-diagonal disorder. We then study the case when the band width is greater than $1$ numerically, and provide strong numerical evidence that a similar approximation using the $j$th local maximum of the landscape function holds for  a large number of low-lying energies in the spectrum.

%%%%%%%%%%%%%%%%%%%%%%%%%%%%%%%%%%%%%%%%%%%%%%%%%%%%%%%%%%%%%%%%main 
\subsection{Main results on graphs}\label{sec:mainresult}
We now introduce some definitions in order to state our main results.   
We consider a (non-weighted) graph $\Gamma=(\mathbb{V},E)$, where the vertex set $\mathbb{V}$ is countably infinite.    
The natural graph metric, denoted $d(\cdot,\cdot)$, gives the length $n$ of the shortest path between two points. We write $x\sim y$ to mean $\{x,y\}\in E$, and in this case say that $y$ is a neighbor of $x$. We denote by $\deg (x)=\#\{y: x  \sim y\}$
the degree of a vertex $x\in \V$. We will assume that $\Gamma$ is connected and that each vertex has finite degree uniformly bounded above, i.e., $M_\Gamma:=\sup_{x\in\V} \deg(x)<\infty$.  We consider the  Jacobi  operator $H$ acting on 
$C(\V):=\R^{\V}=\{f: \V\to \R\}$, defined as
\begin{align}\label{eq:H-graph}
    Hf(x)=\frac{1}{\deg(x)}\sum_{y:y\sim x}\Big( f(x)-a(x,y)f(y)\Big)+ V(x) f(x),
\end{align}
where  $V\in 
\{g: \V\to {\R_{\ge0}}\}$ is a non-negative on-site potential, and $a\in [0,1]^{\V \times \V}$ is an off-diagonal hopping satisfying $a(x,y)=a(y,x)\in [0,1]$ if $x\sim y$ and $a(x,y)=0$ otherwise. 
\begin{remark}
    If $a(x,y)=1$ for $x\sim y$ and $V(x)\equiv0$, then $H$ is the standard negative (probabilistic) Laplacian $-\pdelta$. The operator in the general form \eqref{eq:H-graph} includes models such as Schr\"odinger  operators with an onsite potential $V$, and bond-disordered models with hopping strengths $a(x,y)$. We state our main result in terms of \eqref{eq:H-graph} to avoid technicalities in this introduction.  We  will see in Section \ref{sec:general} that the general result  only relies on positivity and ellipticity of the operator and does not need the specific form of \eqref{eq:H-graph}.  
\end{remark}

Let $A\subset \V$ be a subset of vertices and $\Gamma_A\subset \Gamma$ the subgraph induced by $A$. We denote by $H_A$ the restriction of $H$ to $A$ with Dirichlet boundary conditions. Under these mild assumptions, one can easily check that if $A$ is finite, then $H_A$ is a bounded self-adjoint operator on $\ell^2(A)$, with a strictly positive ground state eigenvalue $\lambda_A>0$ and nonegative Green's function, $G_A(x,y)=H_A^{-1}(x,y)\ge0,\, \forall x,y\in A$. A direct consequence is that there is a unique vector $u_A\in C_+(A)=\R_+^{A}=\{f: A\to {\R_{>0}}\}$, called the landscape function of $H_A$, solving $H_Au_A(x)=1,\, \forall x\in A$. The following generic lower bound on the landscape product will then follow quickly from basic properties discussed in Section~\ref{sec:pre}; see Lemma~\ref{lem:u-positive}.
\begin{proposition}\label{prop:lam-u-gen}
 For any finite subset $A\subset \V$, let $\lambda_A,u_A$ be given as above. Then  
\begin{align}
    \lambda_A \|u_A\|_{\ell^\infty(A)}\ge 1.
\end{align}   
\end{proposition}
 
In order to obtain an upper bound for $ \lambda_A \|u_A\|_{\ell^\infty(A)}$, we need additional assumptions on the graph $\Gamma$ and the subgraph operator $H_A$. 
\begin{assumption}[semigroup kernel upper bounds]
We will primarily be interested in $H_A$ with the following semigroup kernel upper bounds:
\begin{align}\label{eqn:gaussian-kernel-def}
\langle x|e^{-tH_A}|y\rangle=(e^{-tH_A}\one_y)(x) &\le \begin{cases}
c_1t^{-z/2}\exp(-c_2d(x,y)^2/t),& d(x,y)\ge1\text{ and }t\ge d(x,y)\\
c_3\exp(-c_4 d(x,y)),&d(x,y)\ge1\text{ and }t< d(x,y) \\
{c_5t^{-z/2},}&x=y
\end{cases},
\end{align}
for some constants $z,c_i>0$, and for all $t>0$ and $x,y\in A$.
If $z,c_i$ only depend on $\Gamma$, and are independent of $A$, then we called it a uniform kernel upper bound. 
\end{assumption}
For $H_A$ the probabilistic Laplacian $-\pdelta_A$, the semigroup $e^{-tH_A}=e^{t\pdelta_A}$ is the heat semigroup, and the quantity $\langle x|e^{t\pdelta_A}|y\rangle$ is simply the probability that a continuous time simple random walk $X_t$ started at site $x$ and killed upon exiting $A$, is at site $y$ at time $t$.

Our first result, which will be proved in Section \ref{sec:general}, contains the key upper bound for $ \lambda_A \|u_A\|_{\infty}$. 
\begin{theorem}[landscape product for graphs]\label{thm:gen-bound-intro}
Retain the definitions in Proposition \ref{prop:lam-u-gen}. Suppose that in addition, the graph $\Gamma$ satisfies the volume growth $\#\{y:d(x,y)\le r\}\le C_Ur^{z}$ for $r\in\N$ and some constants $C_U$ and $z$, and the $H_A$  satisfy a  uniform upper kernel bound as in \eqref{eqn:gaussian-kernel-def}, with the same constant $z$ and constants $c_i>0$ only depending on $\Gamma$.   Then there is a constant $C<\infty$ only depending on $C_U,z,{M_\Gamma}$, and the $c_i$ such that
\begin{align}\label{eq:gen-bound-intro}
 1\le  \lambda_A \|u_A\|_{\ell^\infty(A)}  \le C. 
\end{align}
\end{theorem}
\begin{remark}
\makeatletter
\hyper@anchor{\@currentHref}%
\makeatother
\label{rmk:hardy}
\begin{enumerate}[(i),leftmargin=*]    
\item We will state and prove a more general version
of the upper bound   in Section \ref{sec:general}. The proof will follow from using heat kernel bounds to show an upper bound on the desired landscape product, following the method for the $\R^d$ case in \cite{vdc}, see also \cite{GiorgiSmits,vogt}. 
Note that unlike the continuous case on $\R^d$, Gaussian heat kernel bounds (the first equation of \eqref{eqn:gaussian-kernel-def}) do not hold for all $t$ and $d(x,y)$ on graphs \cite{Davies93}; one has Poisson tail decay 
for $t< d(x,y)$ (a slightly stronger statement than the second equation of \eqref{eqn:gaussian-kernel-def}); however this does not end up causing any major obstructions.
We will also discuss examples where the semigroup kernel conditions are met (with uniform constants), including on $\Z^d$ or the hexagonal lattice.
We will see in Section \ref{sec:general} that the form of the operator in \eqref{eq:H-graph} and the Gaussian heat kernel bounds in \eqref{eqn:gaussian-kernel-def} can be slightly relaxed, the latter of which will allow us to obtain an upper bound similar to that of \eqref{eq:gen-bound-intro} for regular fractal type graphs, such as the Sierpinski gasket graph (Figure~\ref{fig:sierpinski}).
\def\points{{0,0},{1,0},{.5,.866025},{1.5,.866025},{1,0},{2,0},{1,1.73205},{0,0}}
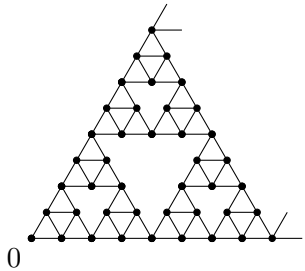
\begin{figure}[!ht]
\begin{tikzpicture}[scale=.4]
\foreach \shift in {(0,0),(2,0),(4,0),(6,0),(1,1.73205),(5,1.73205),(2,3.4641),(4,3.4641),(3,5.19615)}{
	%%%% draw little triangle
	\foreach \c[remember=\c as \clast (initially {0,0})] in \points
	{
		\draw[shift=\shift cm] (\clast) -- (\c);
		\filldraw[shift=\shift cm, black] (\c) circle (3pt);
	}
	%%%% end draw little triangle
}
\node[below left] at (0,0) {$0$};
\draw (4,6.9282) -- ++(60:1cm);
\draw (4,6.9282) -- ++(0:1cm);
\draw (8,0)--(9,0);
\draw (8,0) -- ++(60:1cm);
\end{tikzpicture}
\caption{Sierpinski gasket graph. This is a graph embedded in $\R^2$ whose edges are essentially the edges of triangles in a Sierpinski triangle. The bottom left corner is $0$, each edge is length $1$, and the graph grows unboundedly in the positive $x$ and $y$ directions.}\label{fig:sierpinski}
\end{figure}
\item  For the Dirichlet Laplacian on $[1,N]\cap \Z$, one can compute the associated landscape function $u_N$ explicitly, as it is the exit time of a simple random walk, resulting in $\|u_N\|_{\infty} \propto N^2$, and also the ground state eigenvalue $\lambda_N \propto N^{-2}$. For the probabilistic Laplacian on graphs, as a consequence of Theorem \ref{thm:gen-bound-intro}, we can show similarly that if a Hardy type inequality is satisfied, then $\|u_A\|_{\infty} $ is comparable to $ \rho_A^2$ and $\lambda_A $ is comparable to $ \rho_A^{-2}$, where $\rho_A$ is the inradius of a subset $A\subset \V$.  This will be Corollary \ref{cor:hardy} in Section \ref{sec:general}.
\end{enumerate}
\end{remark}

\subsection{Main results on band matrices and 1D hopping models}
Our general bound in Theorem~\ref{thm:gen-bound-intro} can be applied to graphs such as $\Z^d$ or those which are roughly isometric (to be defined in Section \ref{sec:general}) to $\Z^d$.  In particular, we are interested in the following band model with band width $W\in \N$ on $\Z^d$. Let the matrix $H_{N,W}^{d}=(h_{ij})_{N^d\times N^d}$ 
be defined as
\begin{align}
    h_{ii}&=2dW,  \notag \\
    h_{ij}&=-a_{ij}\in [-1,0],\quad \text{if }1\le \|i-j\|_2\le W,\label{eq:HdWN}\\
    h_{ij}&=0,\quad{\rm otherwise}, \notag
\end{align}
where {$ \|i-j\|_{2}$
    is the standard Euclidean norm applied to $i,j\in \Z^d$.} This corresponds to the lattice model with sites at the points of $\Z^d$ and edges between points $i,j$ that are within $\ell^2$ distance $W$ of each other, when $\Z^d$ is viewed as embedded in $\R^2$. In the case that $W=1$, this just becomes a nearest neighbor model on $\Z^d$.
As an example, if $d=1, W=3$, and $N=7$, then the matrix $H^1_{7,3}$ is of the following form.
\begin{align}\label{eq:H73}
    H_{7,3}^{1}=&\begin{pmatrix}
	6 & -a_{12} & -a_{13}  & -a_{14} & 0 & 0 &  0 \\
	-a_{21} &   6 &-a_{23} &\ddots &\ddots&\ddots  &  \ddots \\
-a_{31}	& -a_{32}  & 6 &  \ddots & \ddots  & \ddots & 0\\
-a_{41} & \ddots& \ddots& 6  & \ddots& \ddots& -a_{47} \\
0  &\ddots & \ddots & \ddots& 6 & \ddots   & \ddots  \\
0  &\ddots & \ddots & \ddots& -a_{65}  & 6  & \ddots  \\
0  &\ddots & 0 & -a_{74}&  \ddots &  -a_{76}  & 6
\end{pmatrix}.
\end{align}

We denote  by $\lambda_{N,W}^{d}$ 
the ground state energy of $H_{N,W}^{d}$, and  by $u_{N,W}^{d}$ the landscape function of $H_{N,W}^{d}$. The positivity of $\lambda_{N,W}^{d}$ and $u_{N,W}^{d}(n)$ are essentially immediate, see Lemma \ref{lem:maxP}. 
As a direct consequence of Theorem \ref{thm:gen-bound-intro}, one has 
\begin{corollary}[landscape product for band matrix model]\label{cor:HNW-gen-intro}
     There is a constant $C=C(d,W)$ depending only on $d$ and $W$ such that for any $N,W$ and $a_{ij}\in [0,1]$, 
    \begin{align}\label{eq:HNW-general-bound-intro}
      1 \le  \,      \lambda_{N,W}^{d} \|u_{N,W}^{d}\|_{\ell^\infty(\Lambda)} \, \le \, C. 
    \end{align}
\end{corollary}
\begin{remark}\label{rmk:Zd-graph-band}
The operator $H_{N,W}^{d}$ is defined on the standard $\Z^d$ graph  with longer range local interactions than the usual nearest neighbor model. We can denote the induced graph by $\Gamma'=(\Z^d,E_d)$, where the set of edges is $E_d=\{(x,y): 1\le\|x-y\|_{\Z^d}\le W\}$.
Then after normalizing by the degree $2W$ of each vertex, the operator $H_{N,W}^{d}$ can be realized on $\Gamma'$ as a nearest-neighbor interaction model in the form of \eqref{eq:H-graph}, to which Theorem~\ref{thm:gen-bound-intro} applies. 
\end{remark}

\begin{figure}[!ht]
	\centering
	\includegraphics[width=0.48\linewidth]{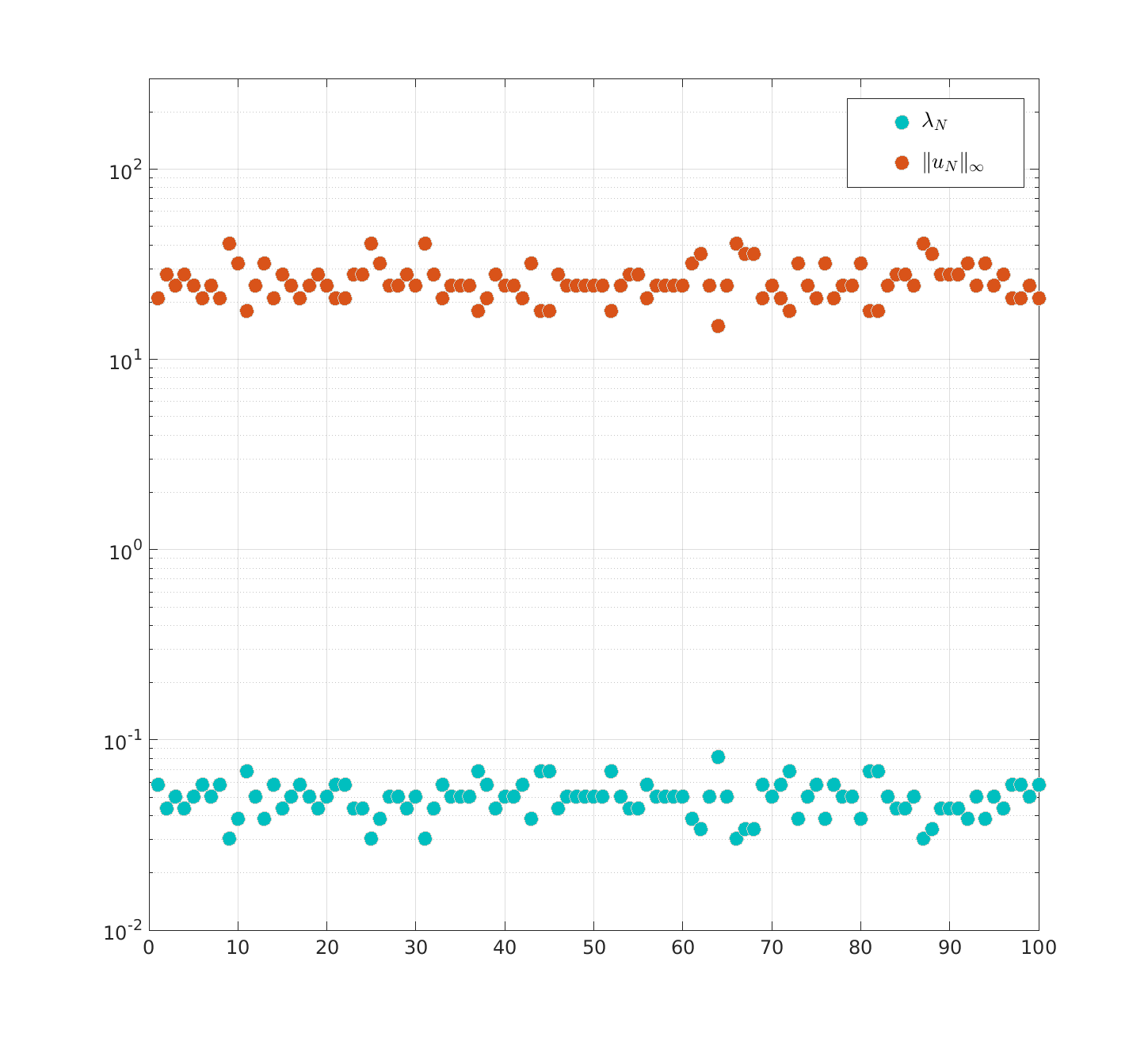}
        \includegraphics[width=0.48\linewidth]{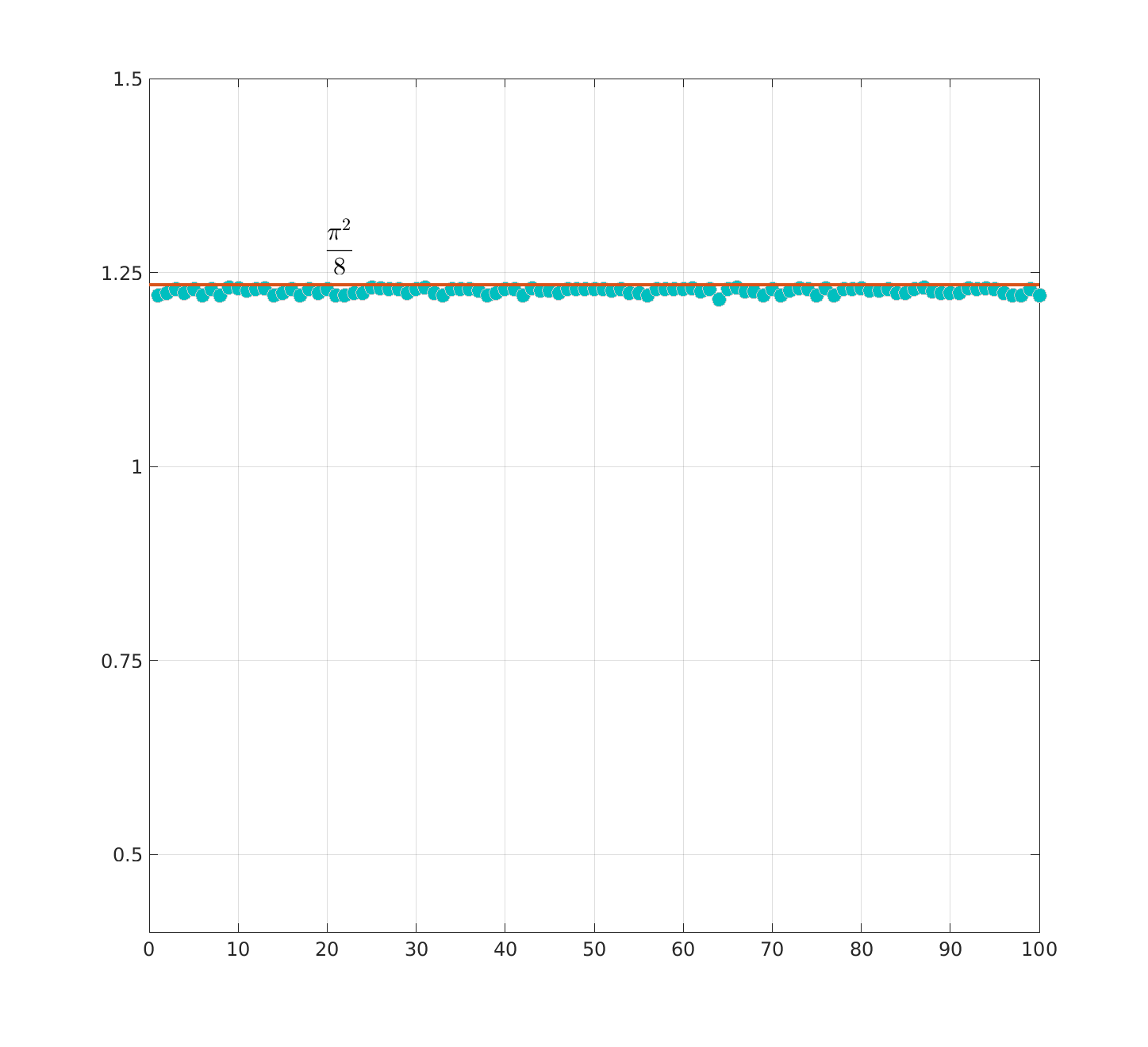}
	\caption{ We fix $N=10^4$, and run 100 random realizations with Bernoulli disorder in the 1D nearest neighbor hopping model \eqref{eq:HW1}, in which each $a_j,(j=2,3,\cdots,N)$ is chosen as 1 with probability $1/2$ and 0 with probability $1/2$.  Left: $\lambda_N$ and $\|u_N\|_\infty$ of $H_N$; Right: the associated $\lambda_N \|u_N\|_\infty$. }\label{fig:ber1-intro}	 
\end{figure}

\begin{figure}[!ht]
	\centering
	\includegraphics[width=0.5\linewidth]{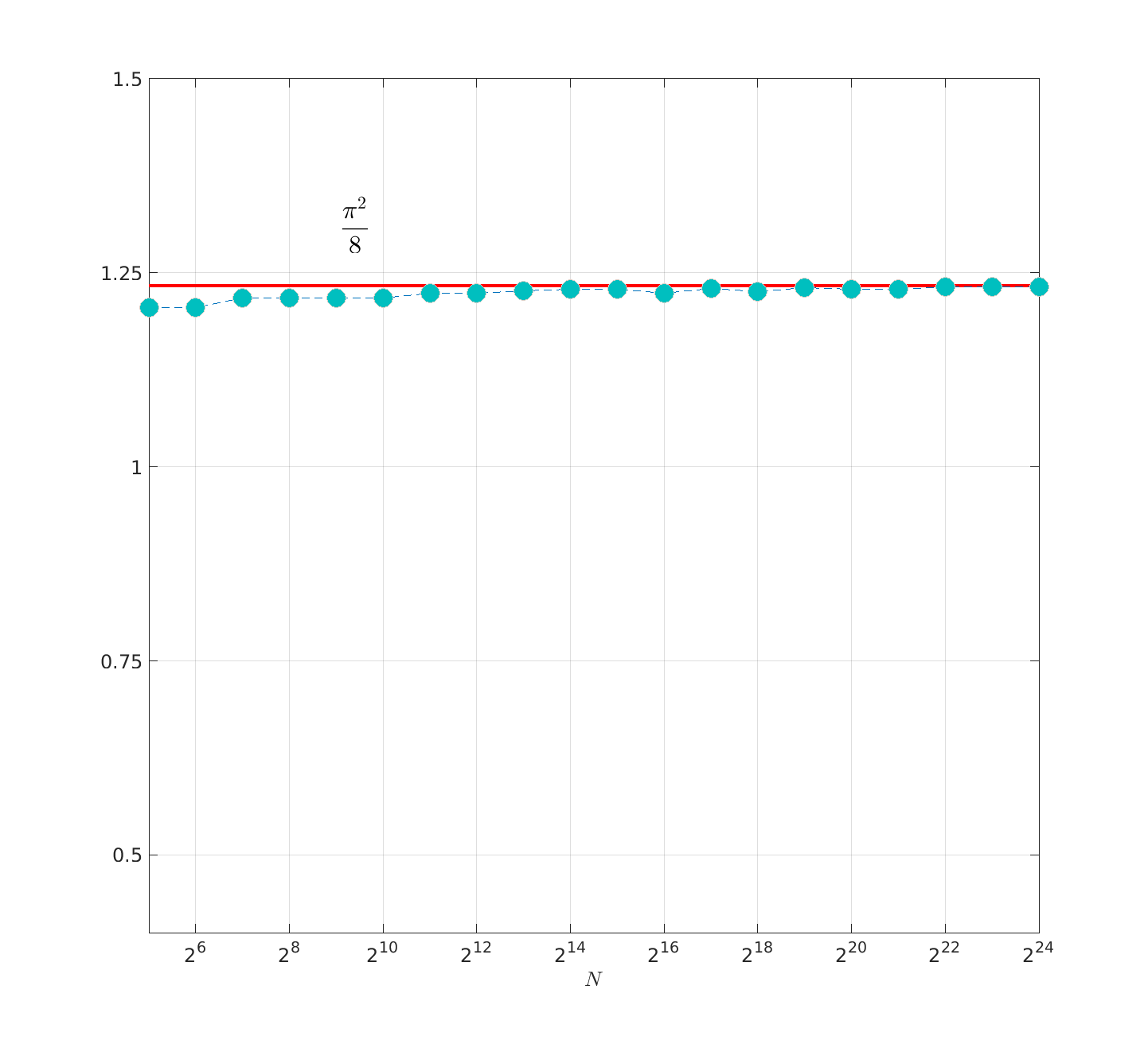}
	\caption{ The dependence of the product $\lambda_N \|u_N\|_\infty$ on $N$, for the 1D nearest neighbor hopping model \eqref{eq:HW1} with Bernoulli disorder ($p=1/2$). To demonstrate the convergence to $\dfrac{\pi^2}{8}$ as $N\rightarrow\infty$, we plot the behavior of $\lambda_N \|u_N\|_\infty$ over a series of $N$. Specifically, we vary $N$ from 32 to $2^{24}=16\,777\,216$, and plot the landscape product for just a single random realization for each $N$.}
	 \label{fig:ber2-intro}
\end{figure}

%%%%%%%%%%%%%%%%%%%%%%%%%%%%%%%%%%%%%%%%%%%%%%%%%%%%%%%%%%%%%%%%%%%%%%%%%%%%%%%%%%%%%%%%%%%%%%%%%%% d=1 
Next, we focus on band model \eqref{eq:HdWN} in the case $d=1$, for which we obtain more accurate bounds on the landscape product $\lambda_N\|u_N\|_\infty$. We start with the case  $W=1$, which corresponds to adding random hopping elements to the usual nearest neighbor Laplacian on a chain. To simplify the notation, instead of writing $H_{N,1}^{1}$ as in \eqref{eq:HdWN}, we denote by  $H_N=-\Delta+B$ the $N\times N$ matrix of interest, where $\Delta$ is the combinatorial Dirichlet
Laplacian on $\intbr{1}{N}$, where $\intbr{a}{b}:=[a,b]\cap \Z$ for any $a<b$, and $B$ holds the off-diagonal hopping elements, in the form:
\begin{align}\label{eq:HW1}
    H_N=&\begin{pmatrix}
	2 & -a_2 &0  & \cdots &  0 \\
	-a_2 &   2 &\ddots   &  \vdots \\
	0 & \ddots& \ddots&\ddots  & 0 \\
	\vdots  &\ddots & \ddots&  2   &-a_N \\
	0  &\cdots & 0&-a_N & 2
\end{pmatrix}\\ 
=&\begin{pmatrix}
	2 & -1 &0  & \cdots &  0 \\
	-1 &   2 &\ddots   &  \vdots \\
	0 & \ddots& \ddots&\ddots  & 0 \\
	\vdots  &\ddots & \ddots&  2   &-1 \\
	0  &\cdots & 0&-1 & 2
\end{pmatrix}+ \begin{pmatrix}
	  0 & b_2 &0  & \cdots &  0 \\
	b_2 &   0 & 0 &\ddots   &  \vdots \\
	0 & \ddots& \ddots&\ddots  & 0 \\
	\vdots  &\ddots & \ddots&  0   & b_N \\
	0  &\cdots & 0& b_N &   0 
\end{pmatrix},
\end{align}
where $a_j\in[0,1]$ and $b_j=1-a_j$ for $j=2,\cdots,N$. Denote by $ \lambda_N$ and $u_N=\{u_N(n)\}\in \R^N$, the ground state energy and the landscape function of $H_N$ respectively.  
We also denote by $\|u_N\|_\infty= \max _{n \in \intbr{1}{N}} u_N(n)$ the $\ell^\infty$-norm of $u$. We are most interested in the case where $\{a_j\}_{j\in\Z}$ are i.i.d. random variables with a certain common distribution. 
\begin{theorem}[landscape product for 1D random hopping model] \label{thm:HNW1-intro}
Let $H_N$ be given as in \eqref{eq:HW1}, where the $\{a_j\}$ are i.i.d. random variables supported on $[a,1]$ for some $0\le a <1$. 
\begin{enumerate}[(i)]
    \item If $\{a_j\}$ has the $\{a,1\}$-Bernoulli distribution with some $p=\P(a_j=1)\in (0,1)$, then \begin{align} \label{eq:ratio-ber}
    \lim_{N\to \infty} \lambda_N \|u_N\|_{\infty}=\frac{\pi^2}{8}, \ \ a.s.
\end{align}
\item 
If $\{a_j\}$ has the $[a,1]$-uniform distribution, then \begin{align}
1 \le \liminf_{N\to \infty} \lambda_N \|u_N\|_{\infty}\le  \limsup_{N\to \infty} \lambda_N \|u_N\|_{\infty}\le \frac{\pi^2}{8}, \ \ a.s.
\end{align}
\end{enumerate}\end{theorem}
Note by $\{a,1\}$-Bernoulli distribution we simply mean the distribution with $\P(X=1)=p$ and $\P(X=a)=1-p$, so that $(X-a)/(1-a)$ has the usual Bernoulli $\operatorname{Ber}(p)$ distribution.
Figures~\ref{fig:ber1-intro} and \ref{fig:ber2-intro} below demonstrate a numerical illustration of Theorem~\ref{thm:HNW1-intro}. This theorem can be viewed as a strengthening of the general bound \eqref{eq:gen-bound-intro} for the specific random hopping model \eqref{eq:HW1}. It also extends results for the Anderson model (with only diagonal disorder) from \cite{CWZ,sanchez2023principal} to the off-diagonal disorder model \eqref{eq:HW1}. 

\begin{figure}[!ht]\label{fig:uniform1-intro}
	\centering
	\includegraphics[width=0.5\linewidth]{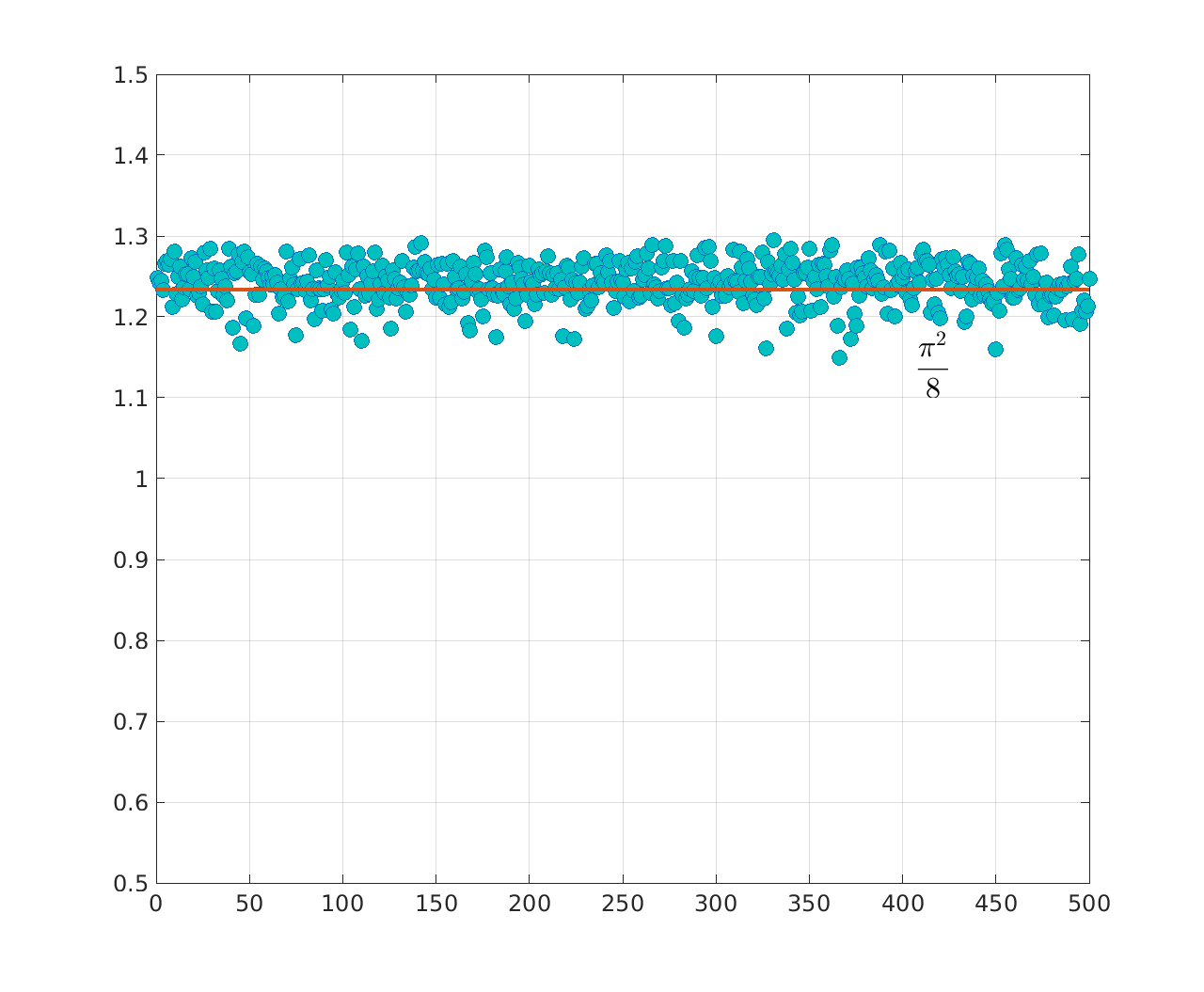}
	\caption{\scriptsize  The behavior of $\lambda_N  \|u_N\|_\infty$ by 500 random realizations, in which $N=10^4$, $a_j,(j=2,3,\cdots,N)$ are uniformly random numbers from [0,1]. }
\label{uniform}	 
\end{figure}

The key observation  leading to the limit $\pi^2/8$ in 1D as  
developed in \cite{CWZ} is as follows: first, for the continuous 1D Laplacian $-\Delta=-\frac{d^2}{dx^2}$ on the unit interval $[0,1]$ with the zero boundary condition, the first (smallest) eigenvalue is $\lambda=\pi^2$, and the landscape function is $u(x)=x(1-x)/2$ with a maximum value $\| u\|_\infty=1/8$. 
The free landscape product is thus $\lambda\,\| u\|_\infty=\pi^2/8$.
For the Anderson model, intuitively, a localized eigenfunction can be approximated by the fundamental mode of the largest potential well with zero or low potential. By estimating the size of the largest potential well on certain scales, the preceding observations were combined to obtain the a.s. limit $\lambda_N\,\| u_N\|_\infty \to  \pi^2/8$   for the Anderson Bernoulli model. 
This heuristic of finding the largest low potential well was also later used in \cite{sanchez2023principal} to prove convergence of the landscape product for the Anderson model on $\Z$ with more general potential distributions, and to obtain lower bounds on the product for the higher dimensional lattices $\Z^d$, $d\ge 2$. The proof of Theorem \ref{thm:HNW1-intro} uses ideas and techniques from both \cite{CWZ} and \cite{sanchez2023principal}, resulting in the desired limit $\pi^2/8$ for the Bernoulli case and the same upper bound of $\pi^2/8$ for the uniform case.
We expect \eqref{eq:ratio-ber} to hold in the uniform case as well (see Figure \ref{uniform}), although this would require much more precise information on the ground state eigenvalue, akin to the Lifshitz tails result from \cite{BiskupKoenig} used in the landscape product lower bound for the Anderson model on $\Z^d$.

\begin{conjecture}\label{conj:unif}
Retain the hypotheses in part (2) of Theorem \ref{thm:HNW1-intro}, i.e., if $\{a_j\}$ has the uniform distribution on $[a,1]$, then  
\begin{align}
    \lim_{N\to \infty} \lambda_N \|u_N\|_{\infty}=\frac{\pi^2}{8}, \ \ a.s.
\end{align}
\end{conjecture}

\begin{figure}
    \centering
     \begin{tikzpicture}
 \foreach \x in {0,1,2,3,4,5,6,7,8}
   \draw (\x cm,1pt) -- (\x cm,-1pt) node[anchor=north] {$\x$};
\draw (1,0) -- (2,0)--(3,0) -- (4,0)--(5,0) -- (6,0)--(7,0);
   \draw (1,0) .. controls (2,1)  .. (3,0);
 \draw (2,0) .. controls (3,-1)  .. (4,0);
   \draw (3,0) .. controls (4,1)  .. (5,0);
   \draw (4,0) .. controls (5,-1)  .. (6,0);
  
    \draw (5,0) .. controls (6,1)  .. (7,0);
    \draw[dashed] (0,0) -- (1,0);
    \draw[dashed] (7,0) -- (8,0);
       \draw[dashed] (-1,0) .. controls (0,1)  .. (1,0);
       \draw[dashed] (0,0) .. controls (1,-1)  .. (2,0);
        \draw[dashed] (7,0) .. controls (8,1)  .. (9,0);
         \draw[dashed] (6,0) .. controls (7,-1)  .. (8,0);
\end{tikzpicture}
    \caption{A graph $\Gamma=(\Z,E)$ consists of vertices $\Z$ and edges $E=\{(i,j)\in \Z^2: 0<|i-j|\le 2\}$. We consider the subset $\{1,\cdots,7\}$ and the Dirichlet (sub-graph) Laplacian $-\Delta_{7,2}$ as in \eqref{eq:Lap72}. }
    \label{fig:Lap72}
\end{figure}
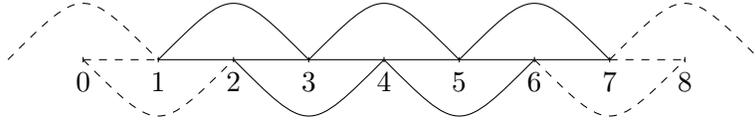

Now we turn to the case $W>1$ (with $d=1$ still) and consider $H_{N,W}=(h_{ij})_{N\times N}$ as
\begin{align}\label{eq:HWN}
    h_{ii}&=2 W,  \notag \\
    h_{ij}&=-a_{ij}\in [-1,0],\quad \text{if }1\le |i-j|\le W.\\
     h_{ij}&=0,\quad{\rm otherwise}, \notag
 \end{align}
Notice  that if $a_{ij}=1$ for all $0<|i-j|\le W$, then $H_{N,W}$ becomes a standard negative sub-graph Dirichlet Laplacian $-\Delta_{N,W}$. 
 An example of the graph for $W=2$, $N=7$ is shown in Figure \ref{fig:Lap72}, where the matrix representation of the negative Dirichlet sub-graph Laplacian $-\Delta_{7,2}$ is 
 \begin{align}\label{eq:Lap72} 
     -\Delta_{7,2}=&\begin{pmatrix}
	4 & -1 & -1  & 0 & 0 & 0 &  0 \\
	-1 &   4 & -1 & -1 &0&0  &  0 \\
-1	& -1  & 4 &  -1 & -1  & 0 & 0\\
0 & -1& -1& 4  & -1 & -1 & 0 \\
0  &0 & -1 & -1& 4 & -1   & -1  \\
0  &0 & 0 & -1& -1  & 4  & -1  \\
0  &0 & 0 & 0&  -1 & -1  & 4
\end{pmatrix}.
\end{align}  

We have the following deterministic result, proved in Section \ref{sec:num}, for the (sub-graph) Laplacian $-\Delta_{N,W}$.  
\begin{theorem}[landscape product for free band Laplacian]\label{thm:LapNW-intro}
Let $\lambda _{N,W} $ be the ground state energy and $u _{N,W} $ be the landscape function of $-\Delta_{N,W}$. For any positive integer $W$, 
   \begin{align}\label{eq:LapNW-result-intro}
       \lim_{N\to \infty} \, \lambda_{N,W} \, \|u_{N,W}\|_{\infty} = \frac{\pi^2}{8}. 
   \end{align} 
\end{theorem}
Recall when $W=1$, \eqref{eq:LapNW-result-intro} holds trivially due to the explicit expression of the  ground state energy and  the landscape function, see Lemma \ref{lem:free}. {When $W>2$, we do not solve for the ground state energy or  the landscape function (of $-\Delta_{N,W}$) explicitly.}
Instead, we obtain the correct asymptotic behavior of $\lambda_{N,W} $ and $ \|u_{N,W}\|_{\infty}$ in terms of $N$ and $W$, which leads to \eqref{eq:LapNW-result-intro}. We will prove these asymptotic behaviors in Section \ref{sec:num} and then discuss their connection to the (discrete) one-dimensional Hardy type inequality, cf. Remark \ref{rmk:hardy}(ii).

We have seen this limit $\pi^2/8$ in various different models with certain ``one-dimensional structure''. Inspired by Theorem \ref{thm:HNW1-intro} and Theorem \ref{thm:LapNW-intro}, it is natural to ask whether \eqref{eq:LapNW-result-intro} still holds if the $a_{ij}$ are chosen randomly in \eqref{eq:HWN}. This question is also motivated by the study of localized sub-regions for eigenvectors of random band matrices, e.g. \cite{filoche2021effective}.  
Numerical evidence in Section \ref{sec:num} suggests we might expect the following to hold:
\begin{conjecture}\label{conj:HW-intro}
    Let $H_{N,W}$  be given as in \eqref{eq:HWN}. Suppose $a_{ij}$ are i.i.d. random variables with either a $\{0,1\}$-Bernoulli or $[0,1]$ uniform distribution. Then a.s.,
    \begin{align}\label{eq:ratio-HNW-intro}
       \lim_{N\to \infty} \, \lambda_{N,W} \, \|u_{N,W}\|_{\infty}  = \frac{\pi^2}{8}. 
   \end{align}
\end{conjecture}
 
Additionally, our numerics suggest that similar results should hold for low-lying excited state energies.

\subsection{Outline} The rest of the paper is organized as follows. In Section \ref{sec:pre}, we introduce additional background and some preliminary lemmas concerning the landscape function and ground state eigenvalue on graphs. In Section~\ref{sec:general}, we state and prove a more general version of Theorem~\ref{thm:gen-bound-intro} for the landscape product on graphs with certain semigroup kernel bounds. Section~\ref{sec:pre1d} introduces some of the ideas and preliminary lemmas for {1D} hopping models. In Section~\ref{sec:HNW1-proof}, we prove Theorem~\ref{thm:HNW1-intro}, for the landscape product for the random hopping model on a {1D} chain. Finally, in Section~\ref{sec:num}, we investigate the random band matrices with band width $W>1$, proving the limiting product for the free non-disordered case, and numerically investigating the product for various band widths with random off-diagonal hopping terms.

\subsubsection*{Acknowledgments} 
  The authors would like to thank  Svitlana Mayboroda for useful suggestions. 
Shou is supported by Simons Foundation grant 563916, SM.  Wang is supported by Simons Foundation grant 601937, DNA. 
Zhang is supported in part by the NSF grants DMS1344235, DMS-1839077, and Simons Foundation grant 563916, SM.

\section{Background and preliminaries for operators on graphs}\label{sec:pre}

\subsection{Graphs background}
We collect some basic definitions here for the  readers' convenience. We refer to \cite{barlow,chung2000discrete} for further details. 
For a  graph $\Gamma=(\mathbb{V},E)$, we assume the vertex set $\V$ is countably infinite. The edge set $E$ is a subset of $\V \times \V$, and we write $x\sim y$ if $\{x,y\} \in E$, in which case $x$ and $y$ are said to be neighbors. The vertex degree of $x$, $\deg(x)=\#\{y: x  \sim y\}$ is the number of neighbors of $x$. Loops and multiple edges between points are not allowed.   A path $\gamma$ in $\Gamma$ is a sequence $x_0,\cdots,x_n$ with $x_{i-1} \sim x_i$ for all $i\in \intbr{1}{n}$, and the length of this path is $n$. The graph is equipped with the natural metric $d(x,y)=$ length of the shortest path from $x$ to $y$.  

Throughout we will always assume 
\begin{align*}
&\Gamma\text{ is connected, i.e. } d(x,y)<\infty\text{ for all } x,y\in\V.\\
&\Gamma\text{ has uniformly bounded vertex degree, i.e. }M_{\Gamma}:=\sup_{x\in \V}\deg(x)<\infty.
\end{align*}
The uniformly bounded vertex degree condition will also be phrased as $\Gamma$ has bounded geometry.

Continuing with definitions, a ball in $\Gamma$,  centered at $x$, with radius $r$, is defined as $B(x,r)=\{y:d(x,y)\le r\}$. For $A\subset \V$, the (exterior) boundary of $A$ is defined as $\partial A=\{y\in A^c: \exists x\in A\ {\rm with} \ x\sim y\} $, and the interior boundary is defined as $\partial^\circ A=\partial (A^c)$. In general, one can also consider a \emph{weighted} graph with some weights $\mu_{xy}:\V\times \V \to \R_{\ge0}$, though for simplicity, in this paper, we will only consider $\Gamma$ with the natural weights $\mu_{xy}=1$ if $x\sim y$ and $\mu_{xy}=0$ otherwise.

We will be interested in functions on the vertices of the graph, which will be denoted by the function space $C(\V):=\R^{\V}=\{f: \V\to \R\}$.  
We denote by  $\one \in C(\V)$ the function which is identically $1$, and by $\one_A$ the indicator function (or characteristic function) of $A\subset \V$. The convenient graph inner product is the \emph{weighted} inner product, 
\begin{align}\label{eqn:weighted}
    {({f},{g})_\mu:=\sum_{x\in \V} {f(x)}g(x)\deg(x).}
\end{align}
The $\ell^2(\V)$ space is defined via the $\ell^2$ norm induced by this weighted inner product, and the $\ell^{\infty}(\V)$ space is defined via the usual $\ell^{\infty}$ norm $\|f\|_{\infty}=\sup_{x\in \V}|f(x)|$. The subspaces $\ell^2(A)$ and $\ell^{\infty}(A)$ are defined accordingly for any finite subset $A\subset \V$.

We retain also the usual subscript-less inner product $\langle f|g\rangle=\sum_{x\in\V}f(x)g(x)$, which we distinguish from the weighted inner product $(\cdot,\cdot)_\mu$, through the use of angled brackets and absence of a subscript $\mu$. 
Using this unweighted inner product, we will frequently use the bra-ket notation to write
$\braket{x}{L}{y}=\langle\one_x|L|\one_y\rangle=(L\one_y)(x)$ for the  matrix elements of an operator $L$. 
The matrix entries are related to the weighted inner product via $\langle x|L|y\rangle=\frac{1}{\deg(x)}(\one_x,L\one_y)_\mu$.

A central operator we use is the (probabilistic) Laplacian $\pdelta$, defined on $C(\V)$ by
\begin{align}\label{eqn:plaplacian}
     \pdelta f(x)=\frac{1}{\deg(x)}\sum_{y:y\sim x}\left(f(y)- f(x)\right) =:(P-I)f(x),
\end{align}
where $I$ is the identity operator and $P$ is the transition matrix (with respect to the natural weights on $\Gamma$), which has matrix elements $\langle x|P|y\rangle=\frac{1}{\deg(x)}$ if $x\sim y$, and zero otherwise. This describes a uniform transition probability to each of $x$'s neighbors. The operators $P$ and $\pdelta$ are self-adjoint on $\ell^2(\V)$ with respect to the weighted inner product $(\cdot,\cdot)_\mu$. 

The combinatorial Laplacian $\Delta^{(c)}$ is defined by 
\[
\Delta^{(c)}f(x)=\deg(x)\pdelta f(x) = \sum_{y:y\sim x}\left(f(y)- f(x)\right)=(\mathcal{A}-{D})f(x),
\]
where $\mathcal{A}$ is the adjacency matrix of $\Gamma$ and ${D}=\operatorname{diag}(\deg(x))_x$,
and it is self-adjoint under the unweighted inner product $\langle\cdot|\cdot\rangle$. For our purposes here on graphs, we will work primarily with the probabilistic Laplacian, which corresponds to simple random walk on $\Gamma$ (discussed below). When we later look at the 1D hopping model on $\Z$ in Sections~\ref{sec:pre1d} and \ref{sec:HNW1-proof},
the probabilistic and combinatorial Laplacian will differ only by an overall constant factor as $\deg(x)=2dW$ is constant. In that case, we will use the normalization from the combinatorial Laplacian, and will omit the superscript to write $\Delta=\Delta^{(c)}$, which will match with the typical definition of the discrete Laplacian on $\Z^d$.

A discrete time simple random walk (SRW) $(X_n)_{n\in\N_0}$ starting at $x\in\V$ is the discrete time Markov chain with $X_0=x$ and transition probabilities given by $\P^{x}(X_{n+1}=y|X_n=z)=\langle z|P|y\rangle$, i.e. $P$ is the transition matrix of $(X_n)$. A continuous time SRW $(Y_t)_{t\ge0}$ on $\Gamma$ is formed from a discrete backbone $(X_n)$ by assigning independent exponential $\operatorname{Exp}(1)$ holding times for each vertex visit between transitions, i.e. $Y_t\sim X_{N_t}$ where $(N_t)_t$ is an independent Poisson process with intensity 1. 
The law of $(Y_t)$ with initial position $Y_0=x$ is also denoted by $\P^x$ when clear from context. 
The transition probabilities of $Y_t$, denoted as $p_t(x,y)$, are
\begin{align*}
   {p_t(x,y)={\P^x(Y_t=y)}=\langle x|e^{t\pdelta}|y\rangle.}
\end{align*}
The transition density is $q_t(x,y)=\frac{p_t(x,y)}{\deg(y)}$ and is also called the continuous time heat kernel on $\Gamma$, as it satisfies
\begin{align}
    \frac{\partial}{\partial t}q_t(x_0,y)=\pdelta_y q_t(x_0,y).
\end{align}
Since we assume $\Gamma$ to be connected and have uniformly bounded vertex degree, we refer to bounds on either $p_t(x,y)$ or $q_t(x,y)$ as heat kernel bounds.

\subsection{Restriction to subsets}
For any operator $H$ on $C(\V)$, its restriction $H_A$ to a subset $A\subset\V$ will be obtained by imposing Dirichlet boundary conditions, i.e. if $\Pi_A$ is the projection $\psi\mapsto \psi\one_A$, then $H_A=\Pi_AH\Pi_A$. For example, the probabilistic Laplacian restricted to $A$ is given by $\pdelta_A=P_A-I_A$, where $I_A$ is the identity restricted to $A$, and $P_A$ is the (sub-Markovian) transition matrix of the SRW that is killed upon exiting $A$.
Written explicitly, the probabilistic Laplacian and combinatorial Laplacian restricted to $A$ are given by the formulas, for $x\in A$,
\begin{align*}
\pdelta_Af(x)&=\bigg(\frac{1}{\deg(x)}\sum_{y\in A:y\sim x}f(y)\bigg)-f(x),\\
\Delta_A^{(c)}f(x)&=\bigg(\sum_{y\in A:y\sim x}f(y)\bigg)-\deg(x)f(x),
\end{align*}
where in both of these $\deg(x)$ is always the degree of the vertex $x$ in the \emph{original} graph $\Gamma$. Writing $\deg_A(x)$ for the degree of $x$ when restricted to vertices in $A$, then
\begin{align*}\label{eqn:eform}
\numberthis
(f,-\pdelta_Af)_\mu=\langle f|-\!\Delta_A^{(c)}|f\rangle &= \frac{1}{2}\sum_{x\in A}\sum_{y\in A:y\sim x}|f(y)-f(x)|^2+\sum_{x\in A}\kappa_x|f(x)|^2,
\end{align*}
where $\kappa_x=\deg(x)-\deg_A(x)\ge0$.
Additionally, using the exit time (for the continuous time SRW) $\tau_A:=\inf\{t\ge0:Y_t\not\in A\}$,
\begin{align}\label{eqn:Akernel}
\langle x|e^{t\pdelta_A}|y\rangle &= \P^x(Y_t=y,\tau_A>t). 
\end{align}
The Jacobi operator defined in \eqref{eq:H-graph} has the restriction, for $x\in A$,
\begin{align}\label{eqn:H-res}
H_Af(x) &= f(x)-\bigg(\frac{1}{\deg(x)}\sum_{y\in A:y\sim x}a(x,y)f(y)\bigg)+V(x)f(x).
\end{align}

For any $A\subset \V$, if $H_A:\ell^2(A)\to\ell^2(A)$ is a (strictly) positive self-adjoint operator, then the ground state energy, or smallest eigenvalue $\lambda_A$ of $H_A$, can be computed by the min-max (or variational) principle,
\[
\lambda_A=\inf_{0\ne f\in\ell^2(A)}\frac{(f,H_Af)_\mu}{(f,f)_\mu}>0.
\]
Since $H_A$ is invertible, the landscape function $u_A=H^{-1}_A\one_A$ is then well-defined. For example,  the above is the case for the negative Laplacian or Jacobi operator \eqref{eq:H-graph} on a finite domain $A$ (cf. Lemma~\ref{lem:maxP}). 
In the negative Laplacian case, the landscape function is the same as the torsion function and is then given by the expected value of the SRW exit time (either the continuous time SRW exit time $\tau_A$, or discrete time SRW exit time $T_A=\min\{m\in\N_0:X_m\not\in A\}$),  
\begin{align}\label{eqn:exit}
u_A(x) &= \langle x|(I_A-P_A)^{-1}|\one_A\rangle = \sum_{m=0}^\infty \sum_{y\in A}\langle x|P_A^m|y\rangle
=\sum_{m=0}^\infty \P^x[X_m\in A]=\E^x[T_A],\\
u_A(x)&=\int_0^\infty\langle x|e^{t\pdelta_A}|\one_A\rangle\,dt=\int_0^\infty\P^x[\tau_A>t]\,dt=\E^x[\tau_A].
\end{align}

\subsection{Maximum principle}
The following maximum principle will be useful in obtaining bounds on the landscape function.
\begin{definition}\label{def:maxP}
 We say $H_A$ satisfies the (weak) maximum principle if $\inf _{x\in A}(H_Af)(x)\ge 0$ implies $\inf _{x\in A} f(x)\ge 0$ . 
\end{definition}
\begin{lemma}[lower bound]\label{lem:u-positive}
Let $A\subset\V$ be a finite subset.
  If $H_A$ {is strictly positive and}  satisfies the (weak) maximum principle, then $G_A(x,y):=\braket{x}{H_A^{-1}}{y}\ge 0$ for all $x,y\in A$. As a consequence, then $u_A(x)>0$ for all $x\in A$, and
    \begin{align}\label{eq:general-lower}
        \|u_A\|_\infty = \|H_A ^{-1}\|_{\ell^{\infty}\to \ell^{\infty}} \ge \lambda_A^{-1}. 
    \end{align}
\end{lemma}
\begin{proof}
For any $y\in A$ and column vector $G_A(\cdot,y)=H_A^{-1}|y\rangle\in \ell^2(A)$  of $G_A$, first note that $H_AG_A(\cdot,y)(x)=\delta_{xy}\ge0$. Then  the maximum principle implies that $G_A(x,y)\ge 0$ for all $x,y\in A $. 
Clearly, there is no row (or column) of $G_A$ identically zero since $G_A$ is invertible. Therefore,  $ u_A(x)=\sum_{y\in A }G_A(x,y)>0$   
for any $x\in A $. 

Then by the definition of the matrix $\infty$-norm, 
\begin{align*}
    \|H_A ^{-1}\|_{\ell^{\infty}\to \ell^{\infty}}=\max_{x\in A } \sum_{y\in A }\left|G(x,y) \right|=\max_{x\in A } \sum_{y\in A }G(x,y) =\max_{x\in A } u_A(x)=\|u_A\| _{\infty}.
\end{align*}
On the other hand, let $\phi_0$ be the ground state of $H_A $ associated with $\lambda_A$, then 
\begin{align*}
     \|H_A ^{-1}\|_{\ell^{\infty}\to \ell^{\infty}}=\sup_{{0\neq\phi}}\frac{\|(H_A )^{-1}\phi \|_\infty}{\|\phi\|_{\infty}} \ge \frac{\|H_A ^{-1}\phi _0\|_\infty}{\|\phi_0\|_{\infty}}=\frac{\|\lambda_A^{-1}\phi _0\|_\infty}{\|\phi_0\|_{\infty}}=\lambda_A^{-1},
\end{align*}
which completes the proof of \eqref{eq:general-lower}. 
\end{proof}

The (weak) maximum principle holds widely, including  for $H$ (and its restriction $H_A$) given as in \eqref{eq:H-graph}. The   maximum principle or comparison test for $ H_A$ then guarantees the lower bound $\|u_A\|_\infty\lambda_A\ge1$ for the landscape product. 
\begin{lemma}[maximum principle for Jacobi]\label{lem:maxP}
Let $H$ be as in \eqref{eq:H-graph}, and let $H_A$ be its restriction \eqref{eqn:H-res} on $\ell^2(A)$ for some finite subset $A\subset \V$. Assume that $0\le a(x,y)=a(y,x)\le 1$ and $V(x)\ge 0$ for all coefficients in $H_A$. Then $H_A$ is self-adjoint (with respect to the weighted inner product) and has positive spectrum, and satisfies the (weak) maximum principle as in Definition \ref{def:maxP}. As a consequence, the lower bound \eqref{eq:general-lower} holds for $H_A$.  
\end{lemma}
\begin{proof}
    For the maximum principle, the idea is similar to the (weak) maximum principle for a subharmonic function on $\R^d$. 
    The proof for  discrete Schr\"odinger operators on $\Z^d$ with on-site disorder can be found in e.g. Lemma 2.12 of \cite{WZ}. The more general case here can be proved in a similar way.  
    Let $f \in \ell^2(A)$ satisfy $(H_A f)(x)\ge 0$. Suppose $f$ attains a strictly negative global minimum at $x_0\in A $, i.e. $f(x_0)=\min_{x\in A} f(x)<0$.  Combining 
    \[ 
(H_Af)(x_0)={f(x_0)-\bigg(\frac{1}{\deg(x_0)}\sum_{y\in A:y\sim x_0}a(x_0,y)f(y)\bigg)}+ V(x_0) f(x_0)\ge 0, \]
    with $a(x,y)\in [0,1]$ and $V(x)\ge 0$, one can conclude that $V(x_0)=0$, $a(x_0,y)= 1$  and $f(y)= f(x_0)$ for all $y\sim x_0$. Inductively, one can obtain $V(x)=0$, $a(x,y)= 1$  and $f(y)= f(x_0)$ for all $x,y\in A  $ (away from the interior boundary), which will contradict  $(H_A f)(x)\ge 0$ for those $x$ at the  interior  boundary of   $A $, as the sum over $\{y\in A:y\sim x\}$ for those $x$ will have fewer than $\deg(x)$ terms.
    
Self-adjointness follows by checking $(g,H_Af)_\mu=(H_Ag,f)_\mu$ using $a(x,y)=a(y,x)$. 
Moreover, as in \eqref{eqn:eform}, 
\begin{align}
\begin{aligned}  
(f,H_Af)_\mu&=\sum_{x\in 
 A}\deg(x)(1+V(x)) f(x)^2- \sum_{x,y\in A:y\sim x}a(x,y)f(x)f(y)\\
 &=(f,-\pdelta_Af)_\mu+\sum_{x\in A}\deg(x)V(x)f(x)^2+ \sum_{x,y\in A:y\sim x}\big(1-a(x,y)\big)f(x)f(y).
 \end{aligned}
\end{align}

Positivity of the ground state eigenvalue $\lambda_A$ follows from the variational characterization of the lowest eigenvalue, using non-negativity of the ground state\footnote{One can take the ground state non-negative since taking absolute values only decreases the quadratic form.} and a comparison to the lowest eigenvalue of the negative Laplacian $-\pdelta_A$.
\end{proof}
\begin{remark}\label{rmk:maxP}
The maximum principle will later be used to obtain \emph{upper bounds} on the landscape function, in Corollary~\ref{cor:hardy} as well as in Sections~\ref{sec:HNW1-proof}  and \ref{sec:num}.  
If $H_A$ satisfies the (weak) maximum principle, then any $v$ such that $(H_Av)(x)\ge1$ for all $x\in A$, is an upper bound for the landscape function $u_A$ on $A$, as $(H_A(v-u_A))(x)\ge0$.
\end{remark}

%%===============================
\section{Proofs for the landscape product on graphs}\label{sec:general}
%%===============================

In this section we prove Theorem \ref{thm:gen-bound-intro}. The lower bound in Theorem \ref{thm:gen-bound-intro} is a direct consequence of the non-negativity of the Green's function, Lemma~\ref{lem:maxP}.  The main work is then to show the general upper bound on the product, which will rely on semigroup kernel estimates. 
While the resulting constants in the bound are not optimal, they apply to fairly general lattices and operators, as long as they allow semigroup kernel bounds of a similar gaussian form as for the free Laplacian on $\Z^d$ (or more generally, a subgaussian form). This includes the random hopping model, random band hopping model, and Anderson model with positive on-site potential, on graphs that are \emph{roughly isometric} to $\Z^d$ (to be defined below), as well as some fractal-like graphs. Obtaining these semigroup kernel bound is discussed in Section~\ref{sec:kernel}.
 
In what follows, we consider a graph $\Gamma=(\mathbb{V},E)$ as described in Section~\ref{sec:pre}.  In particular, we will always assume $\Gamma$ is connected with bounded geometry.
\begin{theorem}[general landscape product bounds]\label{thm:gen-bound-graph}
Let $\Gamma$ be a graph with natural graph metric $d(\cdot,\cdot)\in\N_0$ and polynomial volume growth: $V(x,r):=\#\{y:d(x,y)\le r\}\le C_Ur^{z}$ for all $x\in\V$ and $r\in\N$, and some constants $z\ge1$ and $C_U>0$ ($C_U$ may depend on $z$). 
Let $\mathcal{A}$ be a collection of finite subsets $A\subset \V$, and let $\{H_A\}_{A\in \mathcal{A}}$ be a collection of 
positive bounded self-adjoint operators $H_A$ on $\ell^2(A)$ for each $A\in \mathcal{A}$. 
If $\braket{x}{H_A^{-1}}{y}\ge 0$ for all $x,y\in A$, then 
\begin{align}
    \lambda_A \|u_A\|_\infty\ge 1,
\end{align}
where  $\lambda_A >0$ is the ground state eigenvalue of $H_A$, and $u_A(x)=\langle x|H_A^{-1}|\one_A\rangle$  is the  landscape function of $H_A$.

If, in addition, there are the semigroup kernel upper bound,  
\begin{align}\label{eqn:gaussian-poisson}
\sup_{A\in\mathcal{A}}|\langle x|e^{-tH_A}|y\rangle| &\le \begin{cases}
c_1t^{-z/2}\exp(-c_2d(x,y)^2/t),&d(x,y)\ge1\text{ and }t\ge d(x,y)\\
c_3\exp(-c_4 d(x,y)),&d(x,y)\ge1\text{ and }t< d(x,y) \\
c_5t^{-z/2},&x=y
\end{cases},
\end{align}
for some constants $c_i>0$, and for all $t>0$ and $x,y\in A$, 
then  {there is a constant $C$ depending only on $C_U,z$ and the $c_i$, such that} for any $A\in\mathcal{A}$,
\begin{equation}\label{eqn:gen-upper-bound}
\lambda_A  \|u_A\|_\infty \le C <\infty.
\end{equation}
\end{theorem}
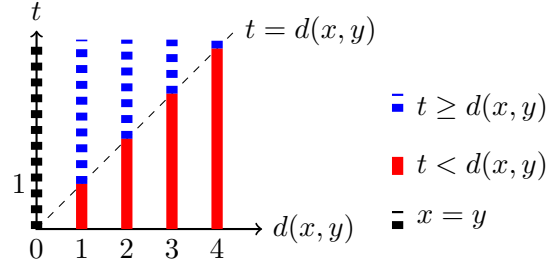
\begin{figure}[!ht]
\begin{center}
    \begin{tikzpicture}[scale=1.2]
    % Draw axes 
    \centering 
    \draw [<->,thick] (0,2.2) node (yaxis) [above] {$t$}
        |- (2.5,0) node (xaxis) [right] {$d(x,y)$};
        \draw[dashed] (0,0)--(0.5,0.5)--(2.2,2.2) node[right]{$t=d(x,y)$};
        \draw (0,0.5) node [left]{1}; 
        \draw (0,0) node [below]{0}; 
         \draw (0.5,0) node [below]{1};
          \draw (1,0) node [below]{2};
           \draw (1.5,0) node [below]{3};
            \draw (2,0) node [below]{4};
 \draw [dashed, line width=1.5mm, blue ] (4,1.3)--(4,1.5) ;
\draw (4.1,1.35) node[right]{$t\ge d(x,y)$};

\draw [line width=1.5mm, red ]  (4,0.6)--(4,0.8) ;
    \draw (4.1,0.7) node[right]{ $t< d(x,y)$};

    \draw [dashed,line width=1.5mm]  (4,0)--(4,0.2) ;
    \draw (4.1,0.1) node[right]{ $x=y$};
    
         \draw [dashed, line width=1.5mm ] (0,0)--(0,2.1);
          \draw [dashed, line width=1.5mm, blue ] (0.5,0.5)--(0.5,2.1);
           \draw [dashed, line width=1.5mm, blue ] (1,1)--(1,2.1);
            \draw [dashed, line width=1.5mm, blue ] (1.5,1.5)--(1.5,2.1);
            \draw [dashed, line width=1.5mm, blue ] (2,2)--(2,2.1);
          \draw [line width=1.5mm, red ] (0.5,0)--(0.5,0.5);
  \draw [line width=1.5mm, red ] (1,0)--(1,1);
  \draw [line width=1.5mm, red ] (1.5,0)--(1.5,1.5);
  \draw [line width=1.5mm, red ] (2,0)--(2,2);
    
\end{tikzpicture}
\end{center}
\caption{An illustration showing the three subdivisions in \eqref{eqn:gaussian-poisson}.} 
\end{figure}
Theorem \ref{thm:gen-bound-intro} is then a special case of Theorem \ref{thm:gen-bound-graph} since $H_A$ defined in \eqref{eq:H-graph} satisfies  $\langle x|H_A^{-1}|y\rangle\ge 0$    by Lemmas~\ref{lem:u-positive} and \ref{lem:maxP}. The lower bound $ \lambda_A \|u_A\|_\infty\ge 1$ is then a direct consequence of the non-negativity of the Green's function as in Lemma \ref{lem:u-positive}, and so it is enough to show the upper bound on $ \lambda_A \|u_A\|_\infty$ under \eqref{eqn:gaussian-poisson}.

\begin{remark}
The same proof of \eqref{eqn:gen-upper-bound} will show that one can replace the Gaussian bound in \eqref{eqn:gaussian-poisson} by the \emph{subgaussian} bound (modifying also the on-diagonal bound as below),
\begin{align}\label{eqn:subgaussian}
    \sup_{A\in\mathcal{A}}|\langle x|e^{-tH_A}|y\rangle| &\le \begin{cases}c_1t^{-z/\beta}\exp\left(-c_2\left(\frac{d(x,y)^\beta}{t}\right)^{\frac{1}{\beta-1}}\right),& {d(x,y)\ge1\text{ and }t\ge d(x,y)}\\
c_5t^{-z/\beta},&x=y
\end{cases},
\end{align}
for some $\beta\ge2$.
This will allow for {landscape product}  bounds for fractal-like graphs like the Sierpinski gasket graph (Remark~\ref{rmk:othergraphs}), which satisfies subgaussian heat kernel bounds with $z=\log 3/\log 2$ and $\beta=\log 5/\log 2$, and the volume growth bound $V(x,r)=\#\{y:d(x,y)\le r\}\le Cr^z$, see \cite{barlow}.
\end{remark}
\begin{corollary}[landscape product bound with subgaussian kernel]\label{cor:subgaussian}
The landscape product bound equation~\eqref{eqn:gen-upper-bound} in Theorem~\ref{thm:gen-bound-graph} holds (with a possibly different constant $C$ that may also depend on $\beta$) if the kernel bounds in \eqref{eqn:gaussian-poisson} are replaced with the subgaussian and on-diagonal bound \eqref{eqn:subgaussian} for some $\beta\ge2$, and the volume growth condition is replaced with $V(x,r)\le Cr^\beta$ for $r\ge1$.
\end{corollary}

Before presenting the proof of Theorem~\ref{thm:gen-bound-graph}, we discuss a consequence on the landscape function for the probabilistic Dirichlet Laplacian $\Delta_A^{(p)}$ on $A\subset \V$; in this case, the landscape function is just the torsion function $u(x)=\langle x|(-\pdelta_A)^{-1}|\one_A\rangle$. For $x\in A$, let $\delta(x)=\min_{y\in \partial A} d(x,y)$ be the distance from $x$ to the boundary $\partial A$, and let $\rho_A=\max_{x\in A}\delta(x)$ be the \emph{inradius} of $A$. The Dirichlet Laplacian $\Delta_A^{(p)}$ is said to satisfy a Hardy type inequality, with constant $C_A>0$ if 
  \begin{align}\label{eq:hardy}
   {C_A \sum_{x\in A}\sum_{y\in A:y\sim x}|f(y)-f(x)|^2\ge \sum_{x \in A} \frac{\deg(x)f^2(x)}{\delta^2(x)},\ \ f\in  \ell^2(A)  .}
  \end{align}
  The following result extends Corollary 1 in \cite{vdc} from $\R^n$ to graphs $\Gamma$. 
\begin{corollary}[Hardy inequality consequence for graphs]\label{cor:hardy}
    Suppose $\Gamma$ is connected with bounded geometry and has polynomial volume growth $\#\{y:d(x,y)=r\}\le C_Ur^{z-1}$ for $r\in\N$ and some constants $C_U>0,z\ge1$. {For a finite subset $A\subset\V$,} suppose $-\Delta_A^{(p)}$ satisfies the semigroup kernel bounds \eqref{eqn:gaussian-poisson}, and a Hardy type inequality \eqref{eq:hardy} with constant $C_A>0$. 
Let $u_A$ and $\lambda_A$ be the associated landscape function and ground state energy of $-\Delta_A^{(p)}$, and let $\wt u_A$ and $\wt \lambda_A$ be the associated landscape function and ground state energy of the combinatorial Dirichlet Laplacian $-\Delta_A^{(c)}$.
Then  there are constants  $c_i$ only depending on $C_A$ and $\Gamma$ such that  
  \begin{align}\label{eq:hardy-u}
 c_1\rho_A^2   \le \|u_A\|_{\ell^\infty(A)},\|\wt u_A\|_{\ell^\infty(A)} \le c_2   \rho_A^2, 
  \end{align}    
  and 
  \begin{align}\label{eq:hardy-lam}
  c_3{\rho_A^{-2}}   \le \lambda_A,\wt \lambda_A \le   c_4{\rho_A^{-2}}.
  \end{align}   
\end{corollary}
\begin{proof} 
 First note that  $\lambda_A \le \wt \lambda_A \le M_\Gamma \lambda_A$ and $ M_\Gamma^{-1}\|u_A\|_\infty\le \|\wt u_A\|_\infty \le \| u_A\|_\infty$, where $ M_\Gamma=\sup_x \deg(x)$, since $-\pdelta$ and $-\Delta_A^{(c)}$ differ just by multiplication by the diagonal operator with the vertex degrees on the diagonal. Thus it suffices to prove the bounds \eqref{eq:hardy-u} and \eqref{eq:hardy-lam} for just the probabilistic Laplacian. 
 
For the upper bound on $\|u_A\|_\infty$, using \eqref{eqn:eform} and the Hardy inequality \eqref{eq:hardy},
\begin{align*}
\lambda_A =\inf_{(f,f)_\mu=1}(f,-\pdelta_A f)_\mu &\ge\inf_{(f,f)_\mu=1}\frac{1}{2}\sum_{x\in A}\sum_{y\in A:y\sim x}|f(y)-f(x)|^2
\ge\frac{1}{2C_A\rho_A^2}.
\end{align*}
Combining with the upper bound $\lambda_A\|u_A\|_\infty\le C$ from Theorem~\ref{thm:gen-bound-graph} yields the upper bound $\|u_A\|_\infty\le 2C\cdot C_A\rho_A^2$.

For the lower bound on $\|u_A\|_\infty$, let $T_A$ be the exit time from $A$ of a discrete time SRW, so that the landscape (torsion) function $u_A(x)$ is the expected exit time $\E^x[T_A]$.  Then
\begin{align*}
       u_A(x)=\E^x(T_A)\ge \E^x(T_{B(x,\delta(x))})\ge{c\delta(x)^2},
\end{align*} 
where the last inequality is a consequence of the heat kernel upper bounds and polynomial volume growth, e.g. \cite[Lemma 4.21]{barlow}.
Therefore, $\|u_A\|_\infty \ge c \rho_A^2$, and the upper bound $\lambda_A \| u_A\|_\infty\le C$ from Theorem~\ref{thm:gen-bound-graph} implies $\lambda_A \le c \rho_A^2$. 
\end{proof}

\subsection{Proof of the general upper bound}\label{subsec:proof-ub}
We now prove the upper bound \eqref{eqn:gen-upper-bound} in Theorem~\ref{thm:gen-bound-graph}.
We will just need to show the estimate,
\begin{equation}\label{eqn:kernelsum}
\sum_{y\in A}|\langle x|e^{-tH_A}|y\rangle| \le C'(\Gamma) e^{-c\lambda_At},
\end{equation}
for some constants $c,C'$ only depending on $\Gamma$,  since then
\begin{align*}
u_A(x)=\langle x|H_A^{-1}|\one_A\rangle &= \int_0^\infty dt\, \langle x|e^{-tH_A}|\one_A\rangle\le \frac{C'(\Gamma)}{c\lambda_A}.
\end{align*}
{In fact, if one has $\langle x|e^{-tH_A}|\one_A\rangle\le \langle x|e^{-t(-\pdelta_A)}|\one_A\rangle=\P^x(X_t\in A)\le1$, one can obtain the better estimate by optimizing\footnote{Without the bound $\langle x|e^{-tH_A}|\one_A\rangle\le 1$, one can still optimize by first bounding $\langle x|e^{-tH_A}|\one_A\rangle\le C$ similar to \eqref{eqn:split}, which will generate a factor $CT$ instead of $T$ in \eqref{eqn:isplit}.} over a split in the integral,}
\begin{align}\label{eqn:isplit}
u_A(x)
& \le T+\int_T^\infty \sum_{y\in A}|\langle x|e^{-tH_A}|y\rangle|\,dt
\le T+\frac{C'(\Gamma)}{c\lambda_A}e^{-c\lambda_A T},
\end{align}
and taking $T=\frac{\left(\log C'(\Gamma)\right)_+}{c\lambda_A }$ gives the bound
\begin{equation}\label{eqn:ulog}
|u_A(x)| \le \frac{\min\left\{C'(\Gamma),(1+\log C'(\Gamma))\right\}}{c\lambda_A }.
\end{equation}
While we do not make much attempt to optimize the constants in what follows, we do mention the above improvement, as it provides a significantly smaller factor of $\log C'(\Gamma)$ instead of $C'(\Gamma)$, which for the continuous analogue on $\R^d$ studied in \cite{vdc,vogt}, produced the correct asymptotic growth order as $d\to\infty$.

To obtain \eqref{eqn:kernelsum}, we will make use of the estimates for $t,\gamma\ge0$,
\begin{align}\label{eqn:property1}
\langle x|e^{-tH_A}|y\rangle &\le \Big(\frac{\deg y}{\deg x}\Big)^{1/2}e^{-t\lambda_A }\le M_{\Gamma}^{1/2}e^{-t\lambda_A},\\ 
\langle x|e^{-tH_A}|x\rangle&\le e^{-\gamma t\lambda_A }\langle x|e^{-(1-\gamma)tH_A}|x\rangle,\label{eqn:properties}
\end{align}
which follow from {$\langle x|e^{-tH_A}|y\rangle\deg(x)=(\one_x,e^{-tH_A}\one_y)_\mu$} and Cauchy--Schwarz or the spectral theorem.
As done in \cite{vdc}, in the Gaussian kernel regime $t\ge\max(1,d(x,y))$, equations \eqref{eqn:gaussian-poisson}, \eqref{eqn:property1}, and \eqref{eqn:properties} provide the estimate,
\begin{align*}
|\langle x|e^{-tH_A}|y\rangle| &= |\langle x|e^{-tH_A}|y\rangle|^{1/2}\cdot|\langle x|e^{-tH_A}|y\rangle|^{1/2}\\
&\le c_1^{1/2}t^{-z/4}\exp(-c_2d(x,y)^2/(2t))\cdot M_{\Gamma}^{1/4}|\langle x|e^{-tH_A}|x\rangle|^{1/4}|\langle y|e^{-tH_A}|y\rangle|^{1/4}\\ 
&\le e^{-t\lambda_A /4} \cdot c_1^{1/2}c_5^{1/2}2^{z/4}M_{\Gamma}^{1/4} \cdot   t^{-z/2} \exp(-c_2d(x,y)^2/(2t)),\numberthis\label{eqn:gs-gaussian} 
\end{align*}
where the second line follows from the Gaussian kernel bound applied to one factor and Cauchy--Schwarz for non-negative operators applied to the other, and the third line follows from applying  \eqref{eqn:properties} with $\gamma=1/2$ followed by the on-diagonal ($x=y$) bound from \eqref{eqn:gaussian-poisson} to each of the factors raised to the power $1/4$.

In the regime where $d(x,y)\ge1$ and $t< d(x,y)$, using \eqref{eqn:property1} and \eqref{eqn:gaussian-poisson} yields,
\begin{align}\label{eqn:gs-tail}
|\langle x|e^{-tH_A}|y\rangle| &= |\langle x|e^{-tH_A}|y\rangle|^{1/2}\cdot|\langle x|e^{-tH_A}|y\rangle|^{1/2}
\le M_{\Gamma}^{1/4}e^{-t\lambda_A/2}c_3^{1/2}\exp(-c_4d(x,y)/2).
\end{align}

The semigroup kernel bounds in \eqref{eqn:gs-gaussian} and \eqref{eqn:gs-tail} only depend on $d(x,y)$, so we can set $R=d(x,y)$ and split up the sum over $y\in A$ in \eqref{eqn:kernelsum} into the three regimes $R=0$, $R\in\intbr{1}{t}$ (Gaussian kernel bounds \eqref{eqn:gs-gaussian}), and $R>t$ (exponential bound \eqref{eqn:gs-tail}). 
Letting $v(x,r)=\#\{y:d(x,y)=r\}$ then yields,
\begin{multline}\label{eqn:split}
\sum_{y\in A}|\langle x|e^{-tH_A}|y\rangle| \le M_\Gamma^{1/2}e^{-t\lambda_A }
+e^{-t\lambda_A /4}c_1^{1/2}c_5^{1/2}2^{z/4}M_{\Gamma}^{1/4}\sum_{R=1}^t v(x,R) t^{-z/2}e^{-c_2R^2/(2t)} \\
+ e^{-t\lambda_A /2}c_3^{1/2}M_{\Gamma}^{1/4}\sum_{R=t+1}^\infty v(x,R)e^{-c_4R/2}.
\end{multline}
The last term is readily bounded, as $\sum_{R=t+1}^\infty v(x,R)e^{-c_4R/2}\le\sum_{R=1}^\infty v(x,R)e^{-c_4R/2}= c(\Gamma)$, using that $v(x,R)\le V(x,R)\le C_UR^z$. 
It thus remains to show the bound for the second term, that
\begin{align}\label{eqn:tsum}
\sum_{R=1}^t v(x,R)t^{-z/2}e^{-c_2R^2/(2t)}\le C(\Gamma),
\end{align}
independent of $t$ and $x$.
Using  $v(x,R)=V(x,R)-V(x,R-1)$ for $R\ge1$, a rearrangement of the sum shows
\begin{align}
\begin{aligned}\label{eqn:rearrange}
\sum_{R=1}^t v(x,R)e^{-c_2R^2/(2t)} 
&=\sum_{R=1}^{t-1} V(x,R)\left[e^{-c_2R^2/(2t)}-e^{-c_2(R+1)^2/(2t)}\right] +V(x,t)e^{-c_2t^2/(2t)}- e^{-c_2/(2t)}\\
&\le \sum_{R=1}^\infty V(x,R)\left[e^{-c_2R^2/(2t)}-e^{-c_2(R+1)^2/(2t)}\right].
\end{aligned}
\end{align}
Factoring out $e^{-c_2R^2/(2t)}$ and using $1-e^{-x} \le x$ for all $x\in\R$ yields
\begin{align}
\begin{aligned}
\left[e^{-c_2R^2/(2t)}-e^{-c_2(R+1)^2/(2t)}\right] 
&\le \frac{C'R}{t}e^{-c_2R^2/(2t)},
\end{aligned}
\end{align}
so that using $V(x,R)\le C_UR^z$, we obtain
\begin{align}\label{eqn:sint}
\sum_{R=1}^t v(x,R)e^{-c_2R^2/(2t)} \le 
\sum_{R=1}^\infty V(x,R){C'R}{t}^{-1}e^{-c_2R^2/(2t)} 
&\le C_UC't^{-1}\sum_{R=1}^\infty R^{z+1}e^{-c_2R^2/(2t)}.
\end{align}
The sum $\sum_{R=1}^\infty R^{z+1}e^{-c_2R^2/(2t)}$ is bounded by integral comparison as follows.
The function $f(r)=r^{z+1}e^{-c_2 r^2/(2t)}$  is increasing up to $r_*=t^{1/2}\left(\frac{z+1}{c_2}\right)^{1/2}$, 
where it obtains its maximum value $\left(\frac{z+1}{c_2}\right)^{\frac{z+1}{2}}e^{-\frac{z+1}{2}}t^{\frac{z+1}{2}}$, and then is decreasing afterwards. 
Thus if $r_*\le1$, then $f$ is decreasing on $[1,\infty)$ and there is the integral estimate,
\begin{align*}
\sum_{R=1}^\infty R^{z+1}\exp(-c_2R^2/(2t)) &\le \exp(-c_2/(2t))+\int_0^\infty r^{z+1}\exp(-c_2r^2/(2t))\,dr\\ 
\numberthis&\le\left(e^{-\frac{z+2}{2}}\left(\frac{z+2}{c_2}\right)^{\frac{z+2}{2}}+\frac{2^{z/2}\Gamma((z+2)/2)}{c_2^{(z+2)/2}}\right)t^{(z+2)/2}.
\end{align*}

If $r_*>1$, then $f$ increases on $[1,r_*]$ and decreases on $[r_*,\infty)$, and there is the integral estimate,
\begin{align*}
\sum_{R=1}^\infty R^{z+1}\exp(-c_2R^2/t) & \le 2f(r_*)+ \left(\int_1^{\lfloor r_*\rfloor}+\int_{\lceil r_*\rceil}^\infty\right) r^{z+1}\exp(-c_2r^2/(2t))\,dr \\ 
&\numberthis\label{eqn:gbound}\le \left(2e^{-\frac{z+1}{2}} \left(\frac{z+1}{c_2}\right)^{\frac{z+2}{2}}+\frac{2^{z/2}\Gamma((z+2)/2)}{c_2^{(z+2)/2}}\right)t^{(z+2)/2},
\end{align*}
where in the bound for $f(r_*)$ we used that $t^{-1/2}\le \left(\frac{z+1}{c_2}\right)^{1/2}$ if $r_*\ge1$.
Thus
\begin{align}
\begin{aligned}
\sum_{R=1}^t v(x,R)e^{-c_2R^2/(2t)}
&\le C_UC't^{-1}\, C(z,c_2)  t^{(z+2)/2}= C(\Gamma)t^{z/2},
\end{aligned}
\end{align}
which gives \eqref{eqn:tsum} and then \eqref{eqn:kernelsum}.
\qedhere

\subsection{Obtaining semigroup kernel bounds}\label{sec:kernel}
 
 In order to obtain the required semigroup kernel estimates to apply Theorem~\ref{thm:gen-bound-graph} to the random hopping models or Anderson model with positive on-site potential, we first note that it is enough to obtain appropriate heat kernel estimates for the free Laplacian. Such heat kernel estimates for the Laplacian on graphs have been well-studied, see for example \cite{Carne,Varopoulos, Davies, HS, CKS, Woess,barlow}, among others.   { Recall that the first equation in \eqref{eqn:gaussian-poisson} is referred to as a Gaussian  kernel bound, and the second equation in \eqref{eqn:gaussian-poisson} is referred to as a Poisson or exponential tail estimate. Poisson tail estimates in the regime $d(x,y)\ge t$ hold widely for the  Laplacian on graphs with mild conditions (\cite{Davies93}, see also \cite[Theorem 5.17]{barlow}), and so we will focus on obtaining the Gaussian kernel bounds.}

The random hopping Hamiltonians  we consider are perturbations of the probabilistic Laplacian $-\pdelta=I-P$ (introduced in equation \eqref{eqn:plaplacian}) obtained by replacing the entries $-\frac{1}{\deg(x)}$ in $P$ by $-\frac{a(x,y)}{\deg(x)}$, for some off-diagonal hopping terms $a(x,y)=a(y,x)\in[0,1]$. Letting $U$ be 
the symmetric edge hopping matrix with entries $\langle x|U|y\rangle=a(x,y)$, then the Hamiltonian can be written as
\[
H^{(\mathrm{hop})}=I-P\odot U,
\]
where $P\odot U$ is defined to be the entry-wise product, $\langle x|P\odot U|y\rangle=\langle x|P|y\rangle\langle x|U|y\rangle$. 
Similarly, we consider the Anderson model with nonnegative potential $V\ge0$,
\[
H^{(\mathrm{And})}=I-P+V,
\]
where $V$ is an on-site multiplication operator. Any distribution of the random (or even non-random) hopping terms $a(x,y)$ or potential terms $V(x)$ is allowed as long as they meet the above requirements.

For the graph $\Gamma=(\mathbb{V},E)$, consider now a finite subset $A\subset \mathbb{V}$, and the corresponding restrictions $H_A^{(\mathrm{hop})}$ and $H_A^{(\mathrm{And})}$ obtained by imposing Dirichlet boundary conditions. In both of these cases the restricted matrix $P_A$ is now the (sub-Markovian) transition matrix describing simple random walk on $\Gamma$ that is killed upon exiting $A$. 
Then there are the semigroup kernel bounds, for $\star\in\{\mathrm{hop},\mathrm{And}\}$, 
\begin{align}\label{eq:kernel-bound-FK}
    \langle x|e^{-tH^{(\star)}_A}|y\rangle &\le \langle x|e^{-t(I_A-P_A)}|y\rangle\equiv\langle x|e^{-t(-\pdelta_A)}|y\rangle,
\end{align}
   by either direct evaluation or via the Feynman--Kac formula.  
For example, for $x,y\in A$, direct evaluation yields,
\begin{align*}
\langle x|e^{-tH_A^\mathrm{(hop)}}|y\rangle=
\langle x|e^{-t(I_A-P_A\odot U_A)}|y\rangle 
&=e^{-t}\sum_{m=0}^\infty \frac{t^m}{m!}\langle x|(P_A\odot U_A)^m|y\rangle \\
&\le e^{-t}\sum_{m=0}^\infty\frac{t^m}{m!}\langle x|P_A^m|y\rangle=\langle x|e^{-t(-\pdelta_A)}|y\rangle,
\end{align*}
using that all entries of $P_A$ are non-negative, and that all entries of $U_A$ are non-negative and bounded above by $1$. Similarly, for a potential $V$, the discrete Feynman--Kac formula yields,
\begin{align}\label{eqn:fk}
\langle x|e^{-tH_A^\mathrm{(And)}}|y\rangle
=\langle x|e^{-t(-\pdelta_A+V)}|y\rangle &=\mathbb{E}^x\left[\exp\bigg(-\int_0^tV(X_s)\,ds\bigg)\one_{X_t=y}\,\one_{\tau_A>t}\right],
\end{align}
which achieves its maximum over non-negative $V$ by taking $V\equiv0$ which corresponds to just the negative Laplacian.

Additionally, if $A\subseteq B$, then by considering SRW probabilities as in \eqref{eqn:Akernel},
\begin{align}
    \langle x|e^{-t(-\pdelta_A)}|y\rangle \le \langle x|e^{-t(-\pdelta_B)}|y\rangle,\quad x,y\in\Gamma,
\end{align}
where we consider $e^{-t(-\pdelta_A)}$ or $e^{-t(-\pdelta_B)}$ as extended by zero to all of $\Gamma$.
Thus appropriate heat kernel bounds on the full graph $\Gamma$ will be enough to apply Theorem~\ref{thm:gen-bound-graph} in the random hopping or Anderson models on growing sets. 

\subsubsection{Heat kernel bounds}
As mentioned before, we will gain heat kernel bounds for graphs that are roughly isometric to $\Z^d$, defined as follows:
\begin{defn}[roughly isometric]\label{def:isometric}\mbox{}
\begin{itemize}
\item Let $(X_1,d_1)$ and $(X_2,d_2)$ be metric spaces. A map $\varphi:X_1\to X_2$ is a \emph{rough isometry} if there exists constants $C_1,C_2$ such that
\begin{align}
C_1^{-1}\big(d_1(x,y)-C_2\big) &\le d_2(\varphi(x),\varphi(y)) \le C_1\big(d_1(x,y)+C_2\big),\\
&\bigcup_{x\in X_1}B_{d_2}(\varphi(x),C_2)=X_2.
\end{align}
If there exists a rough isometry between two spaces then they are \emph{roughly isometric}, and this is an equivalence relation.

\item Let $\Gamma_1$ and $\Gamma_2$ be connected graphs whose vertices have uniformly bounded degrees. A map $\varphi:\mathbb{V}_1\to\mathbb{V}_2$ is a \emph{rough isometry} if:
\begin{enumerate}
\item $\varphi$ is a rough isometry between the metric spaces $(\mathbb{V}_1,d_{\Gamma_1})$ and $(\mathbb{V}_2,d_{\Gamma_2})$ with constants $C_1$ and $C_2$.
\item there exists $C_3<\infty$ such that for all $x\in\mathbb{V}_1$,
\begin{equation}
C_3^{-1}\operatorname{deg}(x) \le \operatorname{deg}(\varphi(x)) \le C_3\operatorname{deg}(x).
\end{equation}
\end{enumerate}
Two graphs are \emph{roughly isometric} if there is a rough isometry between them, and this is an equivalence relation.
\end{itemize}
\end{defn}

\begin{ex}\label{ex:band-graph}
 As discussed in Remark \ref{rmk:Zd-graph-band}, let $\Gamma=(\Z^d,E_d)$ be the standard $\Z^d$ graph. We denote by $\Gamma'=(\Z^d,E'_d)$ the graph  induced by a band matrix $H_{N,W}^d$ \eqref{eq:HdWN}, with the edge set  $E'_d=\{x\sim y\ {\rm if}\ 0<\|x-y\|_{2}\le W\}$. 
The graph $\Gamma'$  is roughly isomorphic to $\Z^d$ (as the graph $\Gamma$). Let $\varphi:{\Z^d}\to{\Z^d}$ be the identity. For $\Gamma'$, the neighbors of a point $x$ are all points $y$ within Euclidean distance $W\ge1$ of $x$. Let $\#\mathcal{B}_{W,d}$ denote the number of lattice points within distance $W$ of $0$, which is a constant depending only on $d$ and $W$. Since $\deg_{\Gamma'}(x)=\#\mathcal{B}_{W,d}$, and also
\begin{align*}
\frac{1}{W\sqrt{d}}d_{\Gamma}(x,y)\le \frac{1}{W}\|x-y\|_2\le d_{\Gamma'}(x,y)&\le d_{\Gamma}(x,y)=\|x-y\|_1,
\end{align*}
then $\Gamma'$ is roughly isometric to $\Z^d$.

\end{ex}

\begin{ex}\label{ex:hex}
2D lattices such as the triangular lattice or honeycomb lattice are roughly isometric to $\Z^2$.  Additionally, finite stacks of such 2D lattices (with only local interactions/edges between layers) are also roughly isometric to $\Z^2$. 
\end{ex}

Graphs that are roughly isometric to $\Z^d$ have heat kernels that behave similarly to that of $\Z^d$. Such graphs then meet the hypothesis of Theorem~\ref{thm:gen-bound-graph} and have a bounded {landscape product}. 

\begin{theorem}[Theorems 6.28, 5.14 in \cite{barlow}]\label{thm:isometric}
Let $\Gamma$ be connected with uniformly bounded vertex degree, and be roughly isometric to $\Z^d$.
Let $p_t(x,y)=\mathbb{P}^x[X_t=y]=\langle x|e^{-t(I-P)}|y\rangle$.
There exist constants $c_i$ and $b_i$ {(which depend on the dimension $d$ and constants in the rough isometry)} such that the following estimates hold.
\begin{enumerate}[(a)]
\item If $d(x,y)\ge1$ and $t\ge d(x,y)$, then
\[
b_1t^{-d/2}\exp(-b_2d(x,y)^2/t)\le p_t(x,y)\le c_1t^{-d/2}\exp(-c_2d(x,y)^2/t).
\]

\item If $d(x,y)\ge1$ and $t< d(x,y)$, then writing $R=d(x,y)$,
\[
b_3\exp(-b_4 R(1+\log(R/t)))\le p_t(x,y)\le c_3\exp(-c_4R(1+\log(R/t))).
\]
\item On-diagonal bounds: For any $t>0$, $p_t(x,x)\le c_5t^{-d/2}$. 
\end{enumerate}
\end{theorem}

Note since we assume $\Gamma$ is connected and has uniformly bounded vertex degree, the factor of $\deg(y)$ in the difference between $p_t(x,y)$ and $q_t(x,y)=p_t(x,y)/\deg(y)$ can be absorbed into the constants $c_i$.
We immediately obtain,

\begin{corollary}\label{cor:semi-bound-lap}
      Let $\Gamma$ be connected with uniformly bounded vertex degree, and be roughly isometric to $\Z^d$. Let $\{H_A\}$ be given as in Theorem \ref{thm:gen-bound-graph}. If 
      \begin{align}
          |\langle x|e^{-tH_A}|y\rangle|\le |\langle x|e^{-t(-\pdelta)}|y\rangle|,\; \forall x,y\in A,
      \end{align} 
then $H_A$ satisfies the upper semigroup kernel bound \eqref{eqn:gaussian-poisson}, and therefore, satisfies the landscape product bound \eqref{eqn:gen-upper-bound}.  
\end{corollary}
\begin{proof}
 A graph roughly isometric to $\Z^d$ must satisfy a volume growth bound $V(x,r)\le C_U r^d$ (cf. \cite[Exercise 4.16 or Lemma 4.22]{barlow}). 
 Thus with the semigroup kernel bounds, the result follows from Theorem~\ref{thm:gen-bound-graph}.
\end{proof}

Combining Corollary \ref{cor:semi-bound-lap}, the bound \eqref{eq:kernel-bound-FK},  and the discussion in Example \ref{ex:band-graph}, proves Corollary \ref{cor:HNW-gen-intro}. Note that the constants in the semigroup kernel bound for the graph $\Gamma'$ induced by $H_{N,W}^d$ also depend on $W$, which leads to the dependence in $W$ in the upper bound of \eqref{eq:HNW-general-bound-intro}.

\begin{remark}\label{rmk:Zd}
For the graph $\Z^d$, one can compute explicit heat kernel bounds, such as those computed in \cite[Appendix]{mourrat} based on Davies' method and a Nash inequality \cite{Nash,Davies87,CKS},
\begin{align}\label{eqn:Zdkernel}
    p_t(x,y) &\le \begin{cases}
    \frac{C_2}{t^{d/2}}\exp\left(-\frac{3\|x-y\|_1^2}{32e^2\cdot dt}\right),&8e^2 dt\ge \|x-y\|_1\\
    \frac{C_2}{t^{d/2}}\exp\left(-\frac{3}{4}\|x-y\|_1\right),&8e^2dt\le\|x-y\|_1
    \end{cases}.
\end{align}
The constant $C_2$ can be taken to be no worse than $d^{d/2}e^{Cd}(2d)$ for some non-dimensional constant $C$, by keeping track of constants following the proofs in \cite{CKS} and \cite{mourrat}.
The proof of Theorem~\ref{thm:gen-bound-graph}, using \eqref{eqn:ulog}, then yields the upper bound for the {landscape product}  on $\Z^d$,
\begin{align}\label{eqn:Zdratio}
    \lambda_A\|u_{A}\|_\infty &\le Cd\log d(1+o(1)),
\end{align}
as $d\to\infty$.
This is however likely not the expected growth in the dimension $d$; based on the continuous case \cite{vdc,vogt} and lower bound from \cite{sanchez2023principal}, one might expect the ratio growth to be linear in $d$, at least for balls and certain general potential classes. 
For the continuous case in $\R^d$, the optimal leading order coefficient as $d\to\infty$ was obtained in \cite{vogt} using heat kernel methods.
For the discrete case in $\Z^d$, it was conjectured in \cite{sanchez2023principal} that the ratio for the Anderson model with certain potential classes converges to $\frac{\mu_d}{2d}$ (with the lower bound proved there), where $\mu_d$ is the principal eigenvalue for the continuous Laplacian on the unit ball with Dirichlet boundary conditions. The value $\mu_d$ is given by the square of the first zero of the Bessel function of order $\frac{d}{2}-1$ (see e.g. \cite[II.5]{Chavel}), so that for $d\to\infty$, the quantity $\frac{\mu_d}{2d}$ is $\frac{d}{8}(1+o(1))$. We see the bound in equation~\eqref{eqn:Zdratio} thus differs by a factor of order $\log d$. 
Nevertheless, the bound suggests that in high dimensions, the landscape product does not really see the potential or hopping terms too strongly in the leading order, allowing the free Laplacian heat kernel bounds to result in a landscape product bound that is relatively close to the actual leading order.

\end{remark}

\begin{remark}[subgaussian heat kernels and weighted graphs]\label{rmk:othergraphs} 
In the previous section, we considered mainly only graphs that are roughly isometric to $\Z^d$. However, there exist a variety of general methods to obtain heat kernel estimates based on the graph geometry. For example, Davies' method mentioned in Remark~\ref{rmk:Zd} applies to more general graphs (and continuous models \cite{CKS}). Several other references and methods for obtaining Gaussian upper bounds are described in \cite[\S6.4]{barlow}. One particular type of graph that is not roughly isometric to $\Z^d$, but that can satisfy subgaussian heat kernel bounds, is fractal-like graphs, including the {Sierpinski gasket graph} drawn in Figure~\ref{fig:sierpinski}.
As demonstrated in \cite{BarlowPerkins} (see also \cite[\S16]{Woess}), the Sierpinski gasket graph satisfies subgaussian heat kernel bounds, and thus by Corollary~\ref{cor:subgaussian} has a bounded {landscape product}. For similar heat kernel bounds of other fractal-like graphs, see the overviews \cite{BarlowFS,Bass,Grigoryan}.

In this paper we have only considered graphs with \emph{natural weights}, where each neighbor of a vertex $x$ has equal probability of being visited in a simple random walk on $\Gamma$. One can also consider certain \emph{weighted graphs} with varying probabilities, producing a weighted Laplacian, and consider the corresponding heat kernel bounds.
\end{remark}

%%%%%%%%%%%%%%%%%%%%%%%%%%%%%%%%%%%%%%%%%%%%%%%%%%%%%%%%%%%%%%%%%%%%%%%%%%%%%%%%%%%%%%%%%%%%%%%%%%%%%%%%%%%%%%%%%%%%%%%%%%%%%%%
\section{Overview and preliminaries for the 1D hopping model}\label{sec:pre1d}

In this section, we introduce some of the main ideas and preliminary lemmas for proving Theorem~\ref{thm:HNW1-intro}.
Let $H_N=H_N(a_2,\cdots,a_N)$ be   as in \eqref{eq:HW1}; recall this is
 \[   H_N=\begin{pmatrix}
 	2 & -a_2 &0  & \cdots &  0 \\
 	-a_2 &   2 &\ddots   &  &\vdots \\
 	0 & \ddots& \ddots&\ddots  & 0 \\
 	\vdots  &\ddots & \ddots&  2   &-a_N \\
 	0  &\cdots & 0&-a_N & 2
 \end{pmatrix}.
\]
We start with an especially simple case of Theorem~\ref{thm:HNW1-intro}(i) to give an overview of some ideas that will be used in more complicated ways in the main proof in Section~\ref{sec:HNW1-proof}.

\subsection{ {Warm-up: $\{0,1\}$-Bernoulli case when $d=1$ and $W=1$}} \label{subsec:warm}
We start with the simple Bernoulli case where $a_j\in\{0,1\}$  are i.i.d. random variables satisfying $\P(a_j=1)=p\in (0,1)$ and $\P(a_j=0)=1-p$. This corresponds to taking $a=0$ in Theorem~\ref{thm:HNW1-intro}(i).
In this case however, we see that because the $a_j$ only take the values $0$ or $1$, then $H_N$ decouples as a direct sum of negative Laplacians, splitting into a new block whenever $a_j=0$. The matrix $H_N$ is of the block form,
\[
H_N=\left(\begin{smallmatrix}
-\Delta_{j_1} \\
& -\Delta_{j_2} \\
&&   -\Delta_{j_3} \\ 
&&&\;\ddots\\
&&&& \;\;-\Delta_{j_k}
\end{smallmatrix}\right)=\bigoplus_{i=1}^k(-\Delta_{j_i}),
\]
where $-\Delta_{j_i}$ is the $j_i\times j_i$ negative Laplacian, written for concreteness in \eqref{eq:free-Lap}. (If $j_i=1$, then this is just the diagonal entry $2$.)
Since $H_N=\bigoplus_{i=1}^k(-\Delta_{j_i})$, then the ground state eigenvalue is simply the minimum eigenvalue taken from the $-\Delta_{j_i}$, and the landscape maximum is the maximum of the free landscape function on each block.

One can compute both the ground state eigenvalue and landscape maximum explicitly for the free Laplacian block $-\Delta_{\ell}$; as $\ell\to\infty$ these are $\lambda=\frac{\pi^2}{\ell^2}(1+o(1))$ and $\|u\|_\infty=\frac{\ell^2}{8}(1+o(1))$ (more precise statements in Lemma~\ref{lem:free}). Thus the larger the block size, the smaller the lowest eigenvalue and the larger the landscape.
Let $-\Delta_{\ell_{\max}}$ be a block which has the largest size among $-\Delta_{j_1},\cdots, -\Delta_{j_k}$. 
By the explicit block form, the ground state eigenvalue $\lambda_N$ of $H_N$ is the ground state eigenvalue of $-\Delta_{\ell_{\max}}$, and the maximum of the landscape function $u_N$ of $H_N$ is the maximum of the local landscape function of $-\Delta_{\ell_{\max}}$. 
So as long as $\ell_{\max}\to\infty$ as $N\to\infty$, the landscape product for $H_N$ is
\[
\lambda_N\|u_N\|_\infty =\frac{\pi^2}{\ell_{\max}}\cdot\frac{\ell_{\max}^2}{8}(1+o(1))=\frac{\pi^2}{8}(1+o(1)).
\]
Since $\ell_{\max}$ is just the longest run of $1$s in $N$ Bernoulli trials, then we have $\ell_{\max}\sim\frac{\log N}{\log(1/p)}\to\infty$ as $N\to\infty$, \cite{schilling1990longest}.

{For the $\{a,1\}$-Bernoulli case with $a>0$, one uses a similar idea of considering free Laplacian blocks separated where $a_j\ne1$, but in this case the interactions between the free Laplacian blocks of strength $a$ must be taken into account. As some examples of the difference, the lower bound for the ground state eigenvalue will have to be analyzed using the `heavy island' argument also used in \cite{BW,CWZ}, and the upper bound for $\|u_N\|_\infty$ will involve more intervals than just the longest run of $a_j=1$. The analysis for this case will be done in Section~\ref{subsec:ber}.

In the uniform disorder case (Section~\ref{subsec:unif}), one again considers diagonal blocks, but this time of operators that are ``$\varepsilon_N$-close'' to the negative free Laplacian, for an appropriate scale $\varepsilon_N$. The arguments are more delicate in this case, and also involve precise Green's function estimates like those used in \cite{sanchez2023principal}. 
}

\subsection{Preliminary lemmas} 
In this part, we state several preliminary lemmas for the hopping model \eqref{eq:HW1} when $d=1,W=1$. We recall  some notation first. Given $\{a_j\}_{j\in \Z}$, denote by $H=H(\cdots,a_{-1},a_0,a_1,\cdots)$ the operator on $\ell^2(\Z)$ given by 
\begin{align}\label{eq:HW=1-pre}
    Hf(n)=2f(n)- \Big( a_{n}f(n-1)+ a_{n+1}f(n+1) \Big),\ n\in\Z.
\end{align}
For any $\Lambda=\intbr{s}{t} \subset \Z$, denote by  $H_{\Lambda}=H_{\Lambda}(a_{s+1},\cdots,a_2,\cdots,a_N,\cdots,a_t)$ the restriction of $H$ \eqref{eq:HW=1-pre} on $\Lambda$:
\begin{align}\label{eq:H-ext}
    H_{\Lambda}=&\begin{pmatrix}
	2 & -a_{s+1} &0  & \cdots &  0 \\
	-a_{s+1} &   2 &\ddots   &  \vdots \\
	0 & \ddots& \ddots&\ddots  & 0 \\
	\vdots  &\ddots & \ddots&  2   &-a_t \\
	0  &\cdots & 0&-a_t & 2
\end{pmatrix}, 
\end{align}
and denote by $H_N=H_{\intbr{1}{N}}(a_2,\cdots,a_N)$ for simplicity as in \eqref{eq:HW1}. 
Since we no longer have differing vertex degrees as we did for graphs $\Gamma$ in Sections~\ref{sec:pre} and \ref{sec:general}, we now only use the usual inner product $\langle f,g\rangle=\sum_x\overline{f(x)}g(x)$.
The explicit form $H_N$ implies  
\begin{align}
    \ipc{\phi}{H_N\phi}=& \sum_{k=1}^N \phi^2(k) +2\sum_{k=1}^{N-1}  a_{k+1} \phi(k+1) \phi(k) \notag\\
  =&  2\phi^2(1)+2\phi^2(N)+ \sum_{k=2}^{N-1}(2-a_k-a_{k-1})\phi^2(k)+ \sum_{k=2}^N(\phi(k)-\phi(k-1))^2 .  \label{eq:q2}
\end{align}
The   quadratic form \eqref{eq:q2} gives $ \ipc{|\phi|}{H_N|\phi|}\le \ipc{\phi}{H_N\phi} $
where $|\phi|:=(|\phi(1)|,\cdots,|\phi(N)|)$. This implies that the ground state $\phi$ is always pointwise nonnegative.

We will frequently use the monotonicity of the Green's function and the landscape function with respect to the off-diagonal sequence $a_j$. 
\begin{lemma}[monotonicity of $G$ and $u$]\label{lem:G-monot}
For $\Lambda=\intbr{s}{t}\subset \Z$, let $H_{\Lambda}=H_{\Lambda}(a_{s+1},\cdots,a_t)$ and $\wt H_{\Lambda}=H_{\Lambda}(\wt a_{s+1},\cdots,\wt a_t)$ be as in \eqref{eq:H-ext}.
If $\wt  a_j\ge a_j$ for all $j=s+1,\cdots,t$, then 
\[  G(n,m):=  H_{\Lambda}^{-1}(n,m)\le \wt H_{\Lambda}^{-1}(n,m):=\wt G(n,m)\]
for all $n,m\in \Lambda$
and 
\[ u_{\Lambda}(n):=  H^{-1}\one_{\Lambda}(n)\le \wt H^{-1}\one_{\Lambda}(n):=\wt u_{\Lambda}(n) \]
for all $n\in {\Lambda}$
\end{lemma}
\begin{proof} Write $  H_{\Lambda}=\wt H_{\Lambda}+E$
where  
\[E=\begin{pmatrix}
	0 &  \ddots  &   &  0  \\
  \ddots &   \ddots   &   \wt a_j-a_j  &   \\
  &  \wt a_j-a_j &  &  \ddots  \\
 0 &  &  \ddots & 0    
\end{pmatrix}, \ \wt a_j\ge a_j.\]
Since all matrix entries of $H_{\Lambda}^{-1},\wt H_{\Lambda}^{-1},E$ and  $ \wt u_{\Lambda} $  are nonnegative and  $\wt H_{\Lambda}^{-1}=H_{\Lambda}^{-1}+H_{\Lambda}^{-1}E\wt H_{\Lambda}^{-1} $,  then (pointwise for all matrix elements)
\[  \wt H_{\Lambda}^{-1}\one_{\Lambda}=H_{\Lambda}^{-1}\one_{\Lambda}+H_{\Lambda}^{-1}E\, \big(\wt H_{\Lambda}^{-1}\one_{\Lambda}\big) \ge H_{\Lambda}^{-1}\one_{\Lambda},  \]
which is $\wt u_{\Lambda}(n)\ge u_{\Lambda}(n) $ for all $n\in\Lambda$.

\end{proof}

The next lemma is the geometric resolvent identity for the Green's function and the domain monotonicity of the landscape function. 
\begin{lemma}[resolvent identity and domain monotonicity]\label{lem:domain-mono}
  Let $H_{\Lambda}=H_{\Lambda}(a_{s+1},\cdots,a_t)$  be as in \eqref{eq:H-ext}.  Let $I=\intbr{i}{j}\subset J \subsetneq \Lambda$ be two sublattices. Let $H_I,H_{J}$ be the restriction of $H$ on $I,J$, respectively, and let $G_I=H_I^{-1}, G_{J}=H_{J}^{-1}$ be the associated Green's functions.   For any $n,m\in I$, 
\begin{align}\label{eq:GRI}
    G_J (n,m)=G_I(n,m)+a_iG_I(n,i)G_J (i-1,m)+a_{j+1}G_I(n,j)G_J (j+1,m).
\end{align}
As a consequence of the positivity of $a_i$ and $G$, one has $G_J (n,m)\ge G_I(n,m)$ for any $n,m\in I\subset J$. 
Let $u_I=H_I^{-1}  \one_I,u_J=H_J^{-1} \one_J$ be the associated landscape functions. Then 
\[  \max_{n\in J} u_J(n) \ge \max_{n\in I} u_I(n). \]
In particular, if $ J=\intbr{1}{N}$, and $u_N=H^{-1}_N \one_N$, then for any $I\subset \intbr{1}{N}$,
\[  \|u_N\|_{\ell^{\infty}(\intbr{1}{N})} \ge   \|u_I \|_{\ell^{\infty}(I)} .\]
\end{lemma}
The  geometric resolvent identity \eqref{eq:GRI} is a standard result, see for example \cite[Theorem 5.20]{kirsch2008invitation}. The consequences on the landscape function then also follow from the definition.\\

Next, we have the following properties for the ground state eigenvalue and landscape function of the free $\ell\times\ell$ Laplacian.
\begin{lemma}[free Laplacian ground state and landscape]\label{lem:free}
Let 
\begin{align}\label{eq:free-Lap}
   -\Delta_{\ell}:=\begin{pmatrix}
	2 & -1 &0  & \cdots &  0 \\
	-1 &   2 &\ddots   &  \vdots \\
	0 & \ddots& \ddots&\ddots  & 0 \\
	\vdots  &\ddots & \ddots&  2   &-1 \\
	0  &\cdots & 0&-1 & 2
\end{pmatrix}_{\ell \times \ell} 
\end{align}
be the negative free Laplacian matrix of size $\ell\times\ell$, acting on $\intbr{i}{j}$ where $\ell=j-i+1$. Then the ground state energy of $-\Delta_\ell$ is 
\begin{align}\label{eq:lam-free}
    \lambda(-\Delta_\ell)=4\sin^2\Big(\frac{\pi}{2(\ell+1)}\Big) = \frac{\pi^2}{\ell^2}(1+o(1))\ \ {\rm as}\ \ell\to \infty.
\end{align}
The landscape function of $-\Delta_\ell$, denote by $u=-\Delta_\ell^{-1}\one $, is 
\begin{align}\label{eq:u-free}
    u(n)=
    \frac{1}{2}\big(n-i+1\big)\big(j+1-n\big), \ n\in\intbr{i}{j}. 
\end{align}
As a consequence, 
\begin{align}\label{eq:u-free-max}
  \|-\Delta_{\ell}^{-1}\|_{\infty}=   \max_{n\in \intbr{i}{j}} u(n)=\frac{1}{8}\Big(\ell^2+2\ell+{(1+(-1)^{\ell+1})/2}\Big) = \frac{\ell^2}{8}(1+o(1))\ \ {\rm as}\ \ell\to \infty.
\end{align}
\end{lemma}
The explicit eigenvalue \eqref{eq:lam-free} is a standard result for the discrete Laplacian matrix, see for example \cite{chung2000discrete}. Equation \eqref{eq:u-free} can be verified by direct computation of $-\Delta_{\ell}u$ using \eqref{eq:free-Lap}, or by interpreting $u(n)$ as a SRW exit time. The first equality of \eqref{eq:u-free-max} follows from the positivity of all entries of $-\Delta_{\ell}^{-1}$ and the definition of the $\infty$-norm of square matrices.  \\ 

Finally, we consider $H_N=H_N(a_2,\cdots,a_N)$ where $\{a_j\}$ are i.i.d. random variables. The following lemma is the a.s. asymptotic size of the maximal block in $H_N$ which is ``close'' to a negative free Laplacian matrix. As noted previously, if $a_j$ takes only two values, then the question is simply the length of the longest run of consecutive successes in $N$ Bernoulli trials, which is well-studied, see for example \cite{schilling1990longest} and references therein. More general distributions for the lattice $\Z^d$ were studied in \cite{sanchez2023principal}. We summarize the results for the 1D Bernoulli and uniform distribution cases here for readers' convenience. 

\begin{lemma}[see Proposition 3, \cite{sanchez2023principal}]\label{lem:l-asym}
Let $H_N=H_N(a_2,\cdots,a_N)$ be as in \eqref{eq:HW1} where $\{a_j\}$ are i.i.d. random variables in $[a,1]$ for some $0\le a<1$. 
\begin{description}
    \item[Bernoulli case] Suppose $a_j\in\{a,1\}$  are i.i.d. random variables satisfying $p=\P(a_j=1)\in (0,1)$ and $1-p=\P(a_j=a)$. Let $ S_N=\frac{\log N}{\log p^{-1}}.$ For each realization of $a_j$, let $\ell_N=\ell_N(a_2\cdots,a_N)$ be the length of a longest interval $I\subset \intbr{1}{N}$ where $a_j=1$ for all $j\in I$. 
    Then a.s. 
    \begin{align}\label{eq:l-asym-ber}
     \lim_{N\to \infty}   \frac{\ell_N}{S_N}=1.    \end{align}
     \item[Uniform case]Suppose $a_j$  are i.i.d. random variables, each uniformly distributed in $[a,1]$. Let $\eps_N=(\log N)^{-2}$ and $ T_N=\frac{\log N}{\log \eps_N^{-1}}=\frac{\log N}{2\log \log N}.$ For each realization of $a_j$, let $\ell_N=\ell_N(a_2\cdots,a_N)$ be the length of a longest interval $I\subset \intbr{1}{N}$ where $a_j\ge 1-\eps_N$ for all $j\in I$. 
    Then a.s. 
    \begin{align}\label{eq:l-asym-uni}
     \lim_{N\to \infty}   \frac{\ell_N}{T_N}=1.    \end{align}
\end{description}
\end{lemma}
\begin{remark}
    In \cite{sanchez2023principal}, the author considered ball of radius 
$Y_N$ in $\Z^d$, while here for convenience we used the ``diameter'' $\ell_N=2Y_N$ of the 1D interval. Proposition 3 of \cite{sanchez2023principal} is stated for diagonal disorder $v_j\ge 0$, where the asymptotic behavior of $\ell_N$ is characterized by the tail probability of $v_j\sim 0$. Here, for the off diagonal disorder $a_j$, the singularity is near $a_j\sim 1$, for which the proof can be adapted from the diagonal case by considering the distribution of $1-a_j$. 
\end{remark}
\begin{remark}\label{rem:l-asym}
We define $\ell_N$ to be the length of a longest interval $I=\intbr{m}{n}\subset \intbr{1}{N}$ where $a_j\ge 1-\eps_N$ for all $j\in \intbr{n}{m}$ and $a_{n-1},a_{m+1}<1-\eps_N$. One can also exclude the boundaries $n$ or $m$ and consider for example $\wt{I}=\intbr{n+1}{m}$, or $\intbr{n+1}{m-1}$. These different choices of the ``maximal block'' will make the maximal length $\ell_N$ only differ by at most $2$, which will still lead to the same asymptotic behavior \eqref{eq:l-asym-ber} or \eqref{eq:l-asym-uni}.   
\end{remark}
\begin{remark}\label{rem:wt-eps}
    In \eqref{eq:l-asym-uni}, we chose $\eps_N=(\log N)^{-2}$ to get the sharp estimates for the ground state energy and the landscape function, see Theorem \ref{thm:unif}.  {The asymptotic behavior of $\ell_N$ holds similarly as long as $\eps_N$ satisfies a summability condition and we have a good tail probability bound on the random variables.} Another useful choice is $\wt \eps_N=(\log N)^{-\sigma}$  for some $\sigma>0$, then $\ell_N \to \frac{\log N}{\log \wt \eps_N^{-1}}=\frac{\log N}{\sigma\log \log N}.$  
\end{remark}
%%%%%%%%%%%%%%%%%%%%%%%%%%%%%%%%%%%%%%%%%%%%%%%%%%%%%%%%%%%%%%%%%%%%%%%%%%%%%%%%%%%%%%%%%%%%%%%%%%%%%%%%%%%%%%%%%%

%%%%%%%%%%%%%%%%%%%%%%%%%%%%%%%%%%%%%%%%%%%%%%%%%%%%%%%%%%%%%%%%%%%%%%%%%%%%%%%%%%%%%%%%%%%%%%%%%%%%%%%%%%%%%%%%%%%%%%%%%%%%%%%
\section{Proofs for the landscape product for the 1D hopping model}\label{sec:HNW1-proof}
In this section we prove Theorem~\ref{thm:HNW1-intro}.

\subsection{Bernoulli case}\label{subsec:ber}

As we already discussed the $\{0,1\}$-Bernoulli case in Section~\ref{subsec:warm}, it is enough to consider the case  $0<a<1$. An example of such a matrix is,
\begin{align} \label{eq:HM-ber}
    H_8=&\left(\begin{smallmatrix}
	2 & -1 &    &   &   &   & &   \\
	-1 &   2 &-1 &  & &   &  &  \\
 	& -1  & 2 & -a   &    &   & &  \\
  &  &  -a & 2  & -1 &  &  & \\
   &  &   &  -1& 2 & -1    &  &  \\
   &  &   &  & -1  & 2  &   -1 & \\
   &  &  & &    &  -1  & 2 & -a \\
   &  &  & &    &    & -a & 2
\end{smallmatrix}\right).
\end{align}
The key ideas for the landscape product mirror those from the case $a=0$, where we used the  estimates of $\lambda$ and $u$ for the free Laplacian of maximal size. The difference now is that we must deal with the correlation between the Laplacian blocks  of strength $a>0$. Let \begin{align}\label{eq:SN}
    S_N=\frac{\log N}{\log  p^{-1} }
\end{align}
be the asymptotic maximal length of the Laplacian blocks as in \eqref{eq:l-asym-ber}. 
We have the following asymptotic behavior of the ground state eigenvalue $\lambda_N$ of $H_N$ and landscape maximum $\|u_N\|_\infty$ of $H_N$ in terms of $S_N$, just as in the free case. 
\begin{theorem}[Off-diagonal Bernoulli disorder asymptotics]\label{thm:Ber}
Let $a_j$, $\lambda_N$, $u_N$, and $S_N$ be given as above. Then
\begin{align} \label{eq:u-bound-ber}
  \lim_{N\to \infty}  \frac{\|u_N\|_{\infty}}{S_N^2 } = \frac{1}{8}, \ \ a.s.
\end{align}
and 
\begin{align}\label{eq:lam-bound-ber}
    \lim_{N\to \infty} \, \frac {\lambda_N S_N^2}{\pi^2}  = 1, \ \ a.s.
\end{align}
\end{theorem}
The remainder of this subsection will be devoted to the proof of Theorem~\ref{thm:Ber}, which then implies Theorem~\ref{thm:HNW1-intro}(i).
First we prove the landscape limit \eqref{eq:u-bound-ber}, and then the ground state eigenvalue limit \eqref{eq:lam-bound-ber}.

For any realization of $(a_j)_j\in \{a,1\}^{N-1}$, let $I=\intbr{l}{r}\subset \intbr{1}{N}$ be a longest interval where $a_j=1$ for $j\in \check I:=\intbr{l+1}{r}$.   Define its length $\ell_N=|I|=r-l+1$. By \eqref{eq:l-asym-ber} and Remark \ref{rem:l-asym}, we know that $\ell_N/S_N\to1$ a.s. 
Let 
\begin{align}\label{eqn:w-I}
  w(i)=-\Delta_{\ell_N}^{-1}\one \, (i), i\in I,  
\end{align}
the local landscape function for $-\Delta_{\ell_N}$ on $I$. Combing the domain monotonicity of the landscape function from Lemmas \ref{lem:domain-mono} and \ref{lem:free}, we obtain
\begin{align}
    \|u_N\|_{\infty}\ge 
   \max_{n\in I}   w(n) \ge \frac{\ell_N^2}{8},
\end{align}
which implies that a.s. 
\begin{align}
    \liminf_{N\to\infty} \frac{\|u_N\|_{\infty}}{S_N^2 }\ge \frac{1}{8}. 
\end{align}

On the other hand, let $\{I_k=\intbr{l_k}{r_k}\}$ be the collection of all the one-wells of $a_j$ as how we define $I$, i.e.,  $a_j=1$ for $j\in \check I_k:=\intbr{l_k+1}{r_k}$ and $a_{l_k}=a_{r_k+1}=a$. The restriction of $H_N$ on each $I_k$ will be a free (negative) Laplacian $-\Delta_{|I_k|}$ satisfying $\max_k |I_k| \le \ell_N$. Now let 
\begin{align}\label{eqn:wtu-Ik}
    \wt u(n)=\begin{cases}-\Delta_{I_k}^{-1}\one \, (i)+c, \ \ i\in I_k \\
    c, \ \  {\rm otherwise}
    \end{cases}, 
\end{align}
where 
\begin{align}\label{eqn:new-c}
    c=\frac{a\ell_N/2 +1}{2(1- a)}.
\end{align}
For $I_k=\intbr{l_k}{r_k}$, if $n\in \intbr{l_k+1}{r_k-1}$, then $a_n=a_{n+1}=1$, leading to $H_N \wt u (n)= -\Delta_{I_k}\wt u (n) =1$. Near the left boundary, by \eqref{eq:u-free}, $-\Delta_{I_k}^{-1}\one \, (l_k)=|I_k|/2$, and $-\Delta_{I_k}^{-1}\one \, (l_k+1)=|I_k|-1$. Then at $n=l_k$, 
\begin{align*}
    H_N \wt u (l_k) =-a\wt u(l_k-1)+2\wt u (l_k)-\wt u (l_k+1) & \ge  -ac+2(|I_k|/2+c)-(|I_k|-1+c)\\
    &=  (1-a)c+  1\ge 1.
\end{align*}
At $n=l_k-1$, 
\begin{align*}
    H_N \wt u (l_k-1) =-a\wt u(l_k-2)+2\wt u (l_k-1)-a\wt u (l_k) & \ge -ac+2c-a(\ell_N/2+c) =  2(1-a)c-a\ell_N/2= 1.
\end{align*}
Similarly, near the right boundary,   $-\Delta_{I_k}^{-1}\one \, (r_k)=|I_k|/2$, and $-\Delta_{I_k}^{-1}\one \, (r_k-1)=|I_k|-1$. Then at $n=r_k$, 
\begin{align*}
    H_N \wt u (r_k) =-\wt u(r_k-1)+2\wt u (r_k)-a\wt u (r_k+1) & \ge  -(|I_k|-1+c)+2(|I_k|/2+c)-ac\\
    &=  (1-a)c+  1\ge 1.
\end{align*} At $n=r_k+1$ 
\begin{align*}
    H_N \wt u (r_k+1) =-a\wt u(r_k)+2\wt u (r_k+1)-a\wt u (r_k+2) & \ge -a(\ell_N/2+c)+2c-ac\\
    &=  2(1-a)c-a\ell_N/2 = 1,
\end{align*}
where we used the choice of $c$ in \eqref{eqn:new-c}. For $n<l_k-1$ or $n>r_k+1$, 
$H_N \wt u (n) =-a\wt u (n-1)+2\wt u (n)-a\wt u (n+1) = (2-2a)c\ge1$. Therefore, $H_N\wt u(n)\ge 1$ for all $n$ and by Lemma \ref{lem:maxP}, then
\[\|u_N\|_{\infty} \le \|\wt u\|_{\infty}\le \frac{1}{8}\ell_N^2+ \ell_N+\frac{a\ell_N +1}{2-2a}.\]
This implies $\limsup_{N\to\infty} \frac{\|u_N\|_{\infty}}{S_N^2 }\le  \frac{1}{8}$ since $\ell_N/S_N\to1$ a.s. and completes the proof of \eqref{eq:u-bound-ber}. \\

Now we turn to the estimates for $\lambda_N$. Let $-\Delta_{\ell_N}$ be a largest block with all off-diagonal entries $a_j=1$ as in the decomposition in \eqref{eq:HM-ber}. 
By the min-max principle and explicit formula \eqref{eq:lam-free} for the ground state eigenvalue of $-\Delta_{\ell_N}$, we see that
\[  \lambda_N  \le  {\textrm{the smallest eigenvalue of} }\ -\Delta_{\ell_N}  
    = 4\sin^2\Big(\frac{\pi}{2(\ell_N+1)}\Big)  \le  \frac{\pi^2}{\ell_N^2},\]
which  together with $\ell_N/S_N\to1$ a.s. implies the upper bound for $\lambda_N$.

The key ingredient for the lower bound of $\lambda_N$ is the `heavy island' argument used in \cite{BW} for estimating the ground state energy for a Bernoulli-Anderson model with diagonal disorder. The argument was later generalized by \cite{CWZ} to the continuous case, with refined choice of parameters.   Let $H_N\psi=\lambda \psi$ be the ground state pair. 
An interval  $I=\intbr{i}{j}$ is called heavy island (w.r.t.  {$\psi$}) if (i): $a_k=1$ for $k\in I$, i.e., the restriction of $H_N$ on $I$ is a free (negative Laplacian); (ii): 
 $\sum_{k\in I}\psi^2(k) \ge \delta^2 \sqrt {(1-a) \ell_N^{3 }} $
where 
$\delta=\max\{\psi(i-1),\psi(j+1)\}$. It is defined so that boundary value of $\psi$ on $I $ is relatively small, compared to the total mass of $\psi$ on $I$. With the choice of boundary condition, one can show that there is at least one heavy island of size $\approx\ell_N$. Then we solve the eigenvalue equation for the free Laplacian on $I$ explicitly, which leads to the following \begin{lemma}\label{lem:lam-low-ber}
Let $a,\lambda_N$ and $\ell_N$ be given as above. Then there is a constant $c$ only depending on $a$ such that if $\ell_N>c$, then 
     \begin{align}
     \lambda_N 
     \ge \frac{\pi^2}{\ell_N^2}\Big(1-\frac{2}{(1-a)^{1/4}\ell_N^{1/4}}\Big).
 \end{align}
\end{lemma} 
The above lemma is close in spirit to \cite[Lemma 3]{BW} and \cite[Lemma 9]{CWZ}.  We omit the details here, and include a short proof of Lemma~\ref{lem:lam-low-ber} in Appendix~\ref{app:lam-low}. 
 Using this lemma and that $\ell_N/S_N\to1$ a.s., one then has a.s.
 \begin{align}
    \liminf_N  \lambda_N S_N^2 \ge \lim_N \frac{\pi^2S_N^2}{\ell_N^2}\Big(1-\frac{2}{(1-a)^{1/4}\ell_N^{1/4}}\Big)=\pi^2 ,
 \end{align}
 which completes the proof for \eqref{eq:lam-bound-ber}. \qedhere

%%%%%%%%%%%%%%%%%%%%%%%%%%%%%%%%%%%%%%%%%%%%%%%%%%%%%%%%%%%%%%%%%%%%%%%%%%%%%%%%%%%%%%%%%%%%%%%%%%%%%%%%%%%%%%%%%%
\subsection{Uniform case}\label{subsec:unif}
Let $H_N=H_N(a_2,\cdots,a_N)$ be as as in \eqref{eq:HW1}, 
where $\{a_j\}_{j=2}^N$ are now i.i.d. uniform random variables taking values in $[a,1]$  for some $0\le a<1$. Without loss of generality, we present the proof only for the case $a=0$. However, one can see in this section that all the proofs work for the case $0<a<1$ as well. The key idea is to consider diagonal blocks in $H_N$ with maximal size, which are ``$\eps_N$-close'' to a negative Laplacian, where the scale is $\eps_N=(\log N)^{-2}$. The asymptotic length of such a block will be 
\begin{align}\label{eq:TN}
    T_N=\frac{\log N}{  \log \eps_N^{-1}}=\frac{\log N}{2 \log\log N},
\end{align}
as given in \eqref{eq:l-asym-uni}.
Let $\lambda_N$ and $u_N$ be the ground state energy and the landscape function for $H_N$. 
We have the following asymptotic behavior of  $\lambda_N$ and $u_N$ in terms of $T_N$. 
\begin{theorem}[Off-diagonal uniform disorder asymptotics]\label{thm:unif}
Let $a_j$, $\lambda_N$, $u_N$, and $T_N$ be given as above. 
\begin{align} \label{eq:u-bound-unif}
  \lim_{N\to \infty}  \frac{\|u_N\|_{\infty}}{T_N^2 } = \frac{1}{8}, \ \ a.s.
\end{align}
and 
\begin{align}\label{eq:lam-bound-unif}
\frac{8}{\pi^2} \le \liminf_{N\to \infty} \, \frac  {\lambda_N T_N^2}{\pi^2}  \le  \limsup_{N\to \infty} \, \frac {\lambda_N T_N^2}{\pi^2}  \le 1, \ \ a.s.
\end{align}
\end{theorem}

\begin{remark}
    In the uniform case, we obtained the approximate order $\|u_N\|_{\infty}= \frac{T_N^2}{8}(1+o(1))$  for $u_N$.   
    We  conjecture that for the ground state eigenvalue, one has
 \begin{align} 
    \lim_{N\to \infty}  \frac{\lambda_N T_N^2}{\pi^2}= 1, \ \ \text{a.s.}, 
 \end{align}
 which would give the order $\lambda_N = \frac{\pi^2}{T_N^2}(1+o(1))$ and lead to Conjecture~\ref{conj:unif}. {As discussed just before the conjecture, for technical reasons we only obtain the upper bound, as we lack a precise enough Lifshitz tail result for the lower bound. Such asymptotic behavior for the ground state energy was obtained in \cite{sanchez2023principal} for the Schr\"odinger case (diagonal disorder case), using the very precise Lifshitz tails result of the associated parabolic Anderson model from \cite{BiskupKoenig}.}
\end{remark}

 Theorem \ref{thm:HNW1-intro}(ii) follows immediately from Theorem \ref{thm:unif}. The rest of this section will thus be devoted to the estimates of $\lambda_N$ and $u_N$ stated in Theorem \ref{thm:unif}.  
 %%%%%%%%%%%%%%%%%%%%%%%%%%%%%%%%%%%%%%%%%%%%%%%%%%%%%%%%%%%%%%%%%%%%%%%%%%%%%%%%%%%%%%%%%%%%

\subsubsection{Lower bound for the landscape function}
We start with the quantitative lower bound of the landscape $u_N=(H_N)^{-1}\one $. 
We will construct a subsolution to the equation $H_Nf=\one$ and then apply the maximum principle Lemma \ref{lem:maxP}. 

Let $\eps_N$ and $T_N$ be as in \eqref{eq:TN}. 
For any realization of $\{a_j\}$, let $I=\intbr{l}{r}\subset \intbr{1}{N}$ be a longest interval where $a_j\ge 1-\eps_N$ for $j\in \check I:=\intbr{l-1}{r}$, and let $\ell_N=|I|=r-l+1$. This describes $H_N$ in the form,
\begin{align}\label{eq:HM}
    {H_N=
\left(\begin{smallmatrix}
    &&&& \\
     &\ddots&&& \\
      &&  \fbox{$M_I$}  && \\
       &&&\ddots & \\
        &&&& 
  \end{smallmatrix}\right), \text{ where }   M_I=\left(\begin{smallmatrix}
	2 &  \ddots  &   &    \\
  \ddots &   \ddots   &  -a_j  &   \\
  & -a_j &  &  \ddots  \\
  &  &  \ddots &  2    
\end{smallmatrix}\right)_{\ell_N\times\ell_N}\!\!\!\!\!\!\!\!\!\!\!\!, \ \ a_j\ge 1-\eps_N, \ j\in \check I.} 
\end{align}
Similar to the Bernoulli case, by \eqref{eq:l-asym-ber} and Remark \ref{rem:l-asym}, $\ell_N/T_N\to1$ as $N\to\infty$  a.s. 

We now consider the local landscape function of $M_I$ on the sublattice $I$. 
Rewrite $M_I =-\Delta_{\ell_N}+E$,  where  
\begin{align}\label{eq:E}
    E=\begin{pmatrix}
	0 &  \ddots  &   &  0  \\
  \ddots &   \ddots   &  1-a_j  &   \\
  & 1-a_j &  &  \ddots  \\
 0 &  &  \ddots & 0    
\end{pmatrix}_{\ell_N\times\ell_N}.
\end{align}
The choice of $M_I $ in \eqref{eq:HM} implies  $\|E\|_\infty\le 2\eps_N$. The resolvent identity gives $-\Delta_{\ell_N}^{-1}=M_I^{-1}-\Delta_{\ell_N}^{-1}EM_I^{-1}$. Let $w=M_I^{-1}\one$ and $w_0=-\Delta_{\ell_N}^{-1}\one$ be the landscape function for $M_I$ and $-\Delta_{\ell_N}$, respectively.  Then 
 \[w_0=w-\Delta_{\ell_N}^{-1}Ew,\]
which implies 
\[\frac{\ell_N^2}{8}\le \|w_0\|_\infty\le \|w\|_\infty\Big(1+\|\Delta_{\ell_N}^{-1}\|_\infty\|E\|_\infty\Big)\le \|w\|_\infty\Big(1+\ell_N^2 \eps_N\Big),  \]
where we used $\ell_N^2/8\le \|w_0\|_\infty=\|\Delta_{\ell_N}^{-1}\|_\infty \le \ell_N^2/4$, both from \eqref{eq:u-free-max}.  Therefore, 
\[
\|w\|_\infty\ge \frac{1}{1+\ell_N^2 \eps_N}\frac{\ell_N^2}{8}. 
\]
Finally, by the domain monotonicity in Lemma \ref{lem:domain-mono}, one has $   \|u_N\|_{\infty}\ge \|w\|_{\infty}\ge\frac{1}{1+\ell_N^2 \eps_N}\frac{\ell_N^2}{8}.  $
The choice of $\eps_N,T_N$ in \eqref{eq:TN} implies $\ell_N^2\eps_N=1/(2\log \log N)^2\to 0$. Together with the asymptotic behavior $\ell_N/T_N\to1$ in \eqref{eq:l-asym-uni}, we have that a.s. 
\begin{align}
    \liminf_{N\to\infty} \frac{\|u_N\|_{\infty}}{T_N^2 }\ge \frac{1}{8}. 
\end{align}

\subsubsection{The upper bound for the landscape function}

%%%%%%%%%%%%%%%%%%%%%%%%%%%%%%%%%%%%%%%%%%%%%%%%%%%%%%%%%%%%%%%%%%%
For technical reasons, we will consider $\{a_j\}_{j\in\Z}$ and extend the matrix $H_N(a_2\cdots,a_N)$ to $H_\Lambda(\cdots,a_2,\cdots,a_N,\cdots)$ on a slightly large domain $\Lambda=\intbr{-N}{2N}$ as in \eqref{eq:H-ext}. 
By the domain monotonicity Lemma \ref{lem:domain-mono}, $u(n)=H_N^{-1}\one (n) \le H_{\Lambda}^{-1}\one (n) $. To bound $\|u\|_\infty$ from above, it is enough to bound  $\max_{\intbr{1}{N}} H_{\Lambda}^{-1}\one (n)$. In an abuse of notation we also denote the landscape function $H_{\Lambda}^{-1}\one $ on the larger domain by $u$. 

We  first derive  a rough upper bound for $ \|u\|_\infty=\max_{\intbr{-N}{2N}}u(n)$. Let $\wt \eps_N=(\log N)^{-1/2}$ and let $I_j=\intbr{l_j}{r_j}\subset \intbr{-N}{2N}$ be all intervals (largest connected components) where $a_n\ge 1-\wt \eps_N$ for all $n\in I_j$. Let $\ell_N$ be the length of a longest interval $I_j$. By Lemma \ref{lem:l-asym} and Remark \ref{rem:wt-eps}, $ \ell_N^{-1}\frac{\log N}{\log  \wt \eps^{-1}_N}=\ell_N^{-1}\frac{2\log N}{\log \log N}\to1$, a.s. as $N\to \infty$. 
Thus $\wt \eps_N\ell_N \to \infty$. Let $\wt u(n)= -\Delta_{I_j}^{-1}\one (n)+{\ell_N}/{ \wt \eps_N}$ if $n\in I_j$ and $\wt u(n)={\ell_N}/{ \wt \eps_N}$ otherwise. 
One can verify by direct computation that $(H_\Lambda \wt u)(n)\ge 1$ for all $n\in \Lambda$. 
Therefore, by the maximum principle Lemma \ref{lem:maxP} and \eqref{eq:u-free-max}, 
\[  \max_{\Lambda} u(n) \le \max_{\Lambda} \wt u(n) \le  \frac{\ell_N^2}{8}+\frac{2\ell_N}{ \wt \eps_N} 
    \le \ell_N^2\bigg(\frac{1}{8}+\frac{2}{\wt \eps_N\ell_N}\bigg)\to \frac{1}{8}\bigg(\frac{2\log N}{\log \log N}\bigg)^2=2T_N^2,\]
    where $T_N=\frac{\log N}{2\log \log N}$. Hence, a.s. 
    \begin{align}\label{eq:u-rough-upper}
        \limsup_{N\to \infty} \frac{\max_{\intbr{-N}{2N}}u(n)}{T_N^2}\le 2. 
    \end{align}

Next, we use the argument developed in \cite{sanchez2023principal} for the Schr\"odinger case (with diagonal disorder) to improve the rough upper bound \eqref{eq:u-rough-upper}. The key ingredient is fine estimates for the determinant of $H_N$. 

For any interval $I=\intbr{i}{j}\subset \Lambda=\intbr{-N}{2N}$, let $G_{\Lambda} (i,j)=H_{\Lambda}^{-1}(i,j)$, $G_I(i,j)= H_I^{-1}(i,j)$, where $ H_I=P_IH_{\Lambda}P_I$ is the restriction of $H_{\Lambda}$ on the sublattice $I$. 
Let  $ u_I=H_{I}^{-1} \one$ be the associated landscape functions.   The geometric resolvent identity  \eqref{eq:GRI} implies that 
for $n\in I\subset \intbr{-N}{2N}$, one has
\begin{align}
    u (n)=&u _I(n)+a_iG_I(n,i)u ({i-1})+a_{j+1}G_I(n,j)u ({j+1}) \\
    \le & u _I(n) +\| u  \|_\infty    \big (   G_I(n,i)+ G_I(n,j) \big),\label{eq:79}
\end{align}
where $\| u  \|_\infty=\max_{\intbr{-N}{2N}}u(n)$.

Now we state the estimate for the local Green's function, which leads to the optimal choice of the interval $I$.  
\begin{lemma}\label{lem:G-decay}
Suppose $I=\intbr{i}{j}\subset \intbr{-N}{2N}$. Then for any $y\in I$,  
\begin{align}\label{eq:41}
    G_I(i,y)\le \Big(1+\sum_{i+1\le k\le y}(k-1)(1-a_k^2)\Big)^{-1}
\end{align}
and 
\begin{align}\label{eq:42}
    G_{I}(y,j)\le  \Big(1+\sum_{y+1\le k\le j}(n-k+1)(1-a_k^2)\Big)^{-1} .
\end{align}
Consequently, if $I=\intbr{ x-Z^{-} }{x+Z^{+}}\subset \intbr{-N}{2N}$ for some $x\in \intbr{1}{N}$ and $N\ge Z^{\pm}\ge 0$, then 
\begin{align}\label{eq:GI1}
    G_I(x-Z^{-},x)\le \Big(\sum_{j=x-1}^{x-Z^{-}}(x-j)(1-a_j)\Big)^{-1}
\end{align}
and 
\begin{align}\label{eq:GI2}
    G_I(x,x+Z^{+})\le \Big(\sum_{j=x+1}^{x+Z^{+}}(j-x)(1-a_j)\Big)^{-1}.
\end{align}
\end{lemma}
\begin{proof}
Since $G_I= H_I ^{-1}$ is determined by the off-diagonal coefficients $a_j$ for $j\in I$, it is enough to consider 
 $y \in I=\intbr{1}{n}\subset \intbr{-N}{2N}$  
for   $1\le y\le n \le N$.  Let $  H_I= H_I(a_2,\cdots, a_y,\cdots,a_n)$ be as in \eqref{eq:H-ext}, and $  \wt H_I= H_I(a_2,\cdots, a_y,1,\cdots,1)$ be the matrix replacing $a_{y+1},\cdots,a_n$ by $1$. The monotonicity of off-diagonal coefficients in  Lemma \ref{lem:G-monot} implies $ G_I(i,j)=H^{-1}_I(i,j)\le  \wt  G_I(i,j)=  \wt H^{-1}_I(i,j).$
By Cramer's rule, 
\begin{align}\label{eq:Cramer}
    \wt  G_I(1,y)=\frac{(-1)^{1+y}\det   \wt H^\ast_I (1,y) }{\det \wt  H_I} ,
\end{align}
where $\wt  H^\ast_I (1,y)$ is the adjugate matrix at $(1,y)$ given by
\begin{align}
   \wt   H^\ast_I (1,y)=\begin{pmatrix}
  \begin{matrix}
  -a_2 & 2 & & \\
  0 & -a_3 & 2 &\\
 \ddots & \ddots &  \ddots & \ddots\\
  \ddots & \ddots & 0 & -a_y
  \end{matrix}
  & \rvline & \begin{matrix}
   && \\
  \vdots && \\
  0 && \\
   -1 &0& \cdots
  \end{matrix} \\
\hline
  \bigzero & \rvline &
  -\Delta_{\intbr{1}{n-y}}
\end{pmatrix}.
\end{align}
Clearly, $ \wt  H^\ast_I (1,y)$ has a upper triangular (block) form which implies 
\begin{align}\label{eq:wtH}
    |\det  \wt  H^\ast_I (1,y)| =a_2\cdots a_y\,  \big|\det \Delta_{\intbr{1}{n-y}}\big|\le n-y+1.
\end{align}
Note that for $y=1$, the desired estimate is trivially $  G_I(1,1)\le   G_I(1,1)=\frac{n}{n+1} \le 1$. 
It is enough to bound $\det \wt  H_I =\det H_I(a_2,\cdots, a_y,1,\cdots,1)$ from below. This relies on the following claim, which follows from the polynomial expansion of $\det H_I(a_2,\cdots, a_n)$ in terms of $1-a_j^2$.  
\begin{claim}\label{claim:Pn-poly}
Let 
\[P_n( a_2\cdots,  a_n)=\det H_I(a_2,\cdots, a_n),\ a_j\in[0,1],\ j=2,\cdots,n.\]
Then $P_n$ is a polynomial in the form 
\begin{align}
     P_n(a_2,\cdots,a_n)=c+\sum_{k=2}^nb_k(1-a_k^2)+\sum_{j\ge 2} d_{i_1,\cdots,i_j}(1-a_{i_1}^2)\cdots(1-a_{i_j}^2)
\end{align}
where  $c=n+1$ and  $b_k=(k-1)(n-k+1)$ and all the coefficients  $d_{i_1,\cdots,i_j}\ge 0$. 
\end{claim}
The proof of the claim is based on direct expansion of the determinant with respect to rows containing $a_j$, and is elementary. We include a brief argument in Appendix \ref{app:claim-Pn} for the readers' convenience. A direct consequence of this claim (with application to $\det  \wt  H_I$) is
\begin{align*}
  \det \wt  H_I=P_n( a_2,\cdots, a_y,1\cdots,1) 
  \ge &  n+1+\sum_{k=2}^y(k-1)(n-k+1)(1-a_k^2) \\ 
  \ge &   n+1+(n-y+1)\sum_{k=2}^y(k-1)(1-a_k).
\end{align*}
Together with \eqref{eq:Cramer} and \eqref{eq:wtH}, one has
\begin{align*}
G_{I}(1,y) \le \wt   G_{I}(1,y)\le    \frac{|\det  \wt  H^\ast_I (1,y)| }{\det \wt  H_I} 
 \le & \frac{n-y+1}{n+1+(n-y+1)\sum_{k=2}^y(k-1)(1-a_k)}\\
 \le & \frac{1}{1+\sum_{k=2}^y(k-1)(1-a_k)}.
\end{align*}
By the same argument, one also has
\begin{align*}
 G_{I}(y,n)\le    \frac{y}{P_n(1,\cdots,1,a_{y+1},\cdots,a_n )} \ 
 \le   \frac{1}{1+\sum_{k=y+1}^n(n-k+1)(1-a_k)},
\end{align*}
thus proving Lemma~\ref{lem:G-decay}.
\end{proof}

Now for any $\delta>0$ and $x\in \intbr{1}{N}$, let 
\begin{align}\label{eq:Z-}
  Z_{\delta}^{-}:=\min\Big\{n\in \N \big|\sum_{j=x-1}^{x-n}(x-j)(1-a_j)\ \ge \delta^{-1}\Big\} ,
\end{align} 
and 
\begin{align}\label{eq:Z+}
  Z_{\delta}^{+}:=\min\Big\{n \in \N \big|\sum_{j=x+1}^{x+Z^{+}}(j-x)(1-a_j)\ \ge \delta^{-1}\Big\}.
\end{align}

For any $\delta>0$, we see that $Z^{\pm}_\delta$ always exists (finite) as long as $\{a_j\}_{j\in \Z}$ are not identically zero. The following lemma in \cite{sanchez2023principal} shows $Z_{\delta}^{-}+ Z_{\delta}^{+}$ is bounded from above by $T_N+o(1)$, a.s. as $N\to \infty$. 
\begin{lemma}[Proposition 12, \cite{sanchez2023principal}]\label{lem:SdProp12}
Suppose  $\{a_j\}_{j\in \Z}$ are i.i.d. random samples satisfying  the $[0,1]$-uniform distribution.  For all $\delta>0$, 
\begin{align} \label{eq:ldelta-Tn}
    \limsup_{N\to \infty} \frac{\max_{x\in \intbr{1}{N}} \ell_\delta(x)}{T_N} \le 1, \ \ a.s.,
\end{align}
where $\ell_\delta(x)=Z_{\delta}^{-}+ Z_{\delta}^{+}$ and  $T_N=\frac{\log N}{2 \log \log N}$.
\end{lemma}
Proposition 12 of \cite{sanchez2023principal} was proved for more a general class of distributions.  We include a short proof for the uniform case for the reader's convenience in Appendix \ref{sec:Sdprop12}, following the main steps in \cite{sanchez2023principal}. Equation \eqref{eq:ldelta-Tn} also guarantees that $Z_{\delta}^{\pm}\le T_N+o(1) \le N$. Then for any $x\in \intbr{1}{N}$, $I_\delta=\intbr{x- Z_{\delta}^{-} }{x+ Z_{\delta}^{+}}\subset \intbr{-N}{3N}$. Combining \eqref{eq:GI1}, equation \eqref{eq:GI2} of Lemma \ref{lem:G-decay}, and the definition of $Z_{\delta}^{\pm}$, we have 
\begin{align}
    G_{I_\delta}(x-Z_{\delta}^{-},x), G_{I_\delta}(x,x+Z_{\delta}^{+})\le \delta.
\end{align}
Then by \eqref{eq:79}, for $x\in I_{\delta}\subset \intbr{-N}{2N}$,
\begin{align}
      u(x)\le &   u_{I_\delta}(x)+ \|   u \|_\infty \Big(  G_{I_\delta}(x-Z_{\delta}^{-},x)+ G_{I_\delta}(x,x+Z_{\delta}^{+})\Big)  
    \le \frac{1}{8}\ell_\delta^2(x)+2\delta\|   u \|_\infty, 
\end{align}
where in the last inequality we used $ u_{I_\delta}(x)=H_{I_\delta}^{-1} \one(x)\le -\Delta_{I_\delta}^{-1} \one(x)\le \frac{1}{8}\ell_\delta^2(x)$ from Lemma \ref{lem:G-monot} and Lemma \ref{lem:free}. 

Taking the maximum for $x\in \intbr{1}{N}$ and dividing by $T_N^2$ gives
\[  \frac{\max_{x\in \intbr{1}{N}} u(x) }{T_N^2}
    \le \frac{1}{8} \frac{\max_{x\in \intbr{1}{N}}\ell_\delta^2(x)}{T_N^2}+2\delta \frac{\|   u \|_\infty}{T_N^2} \]
Finally, by \eqref{eq:u-rough-upper} and \eqref{eq:ldelta-Tn}, we have for any $\delta>0$,
\begin{align}
    \limsup_{N\to \infty}  \frac{\max_{x\in \intbr{1}{N}} u(x) }{T_N^2}
    \le \frac{1}{8} +4\delta ,
\end{align}
which completes the proof of \eqref{eq:u-bound-unif} by taking $\delta\to0$. 

%%%%%%%%%%%%%%%%%%%%%%%%%%%%%%%%%%%%%%%%%%%%%%%%%%%%%%%%%%%%%%%%%%%%%%%%%%%%%%%%%%%%%%%%%%%%%%%%%%%%%%%%%%%%%%%%%%%%%%%%%%%%%%%%%%%%%%%%%%%

\subsubsection{Estimate for the ground state eigenvalue}

Notice that the upper bound $\limsup_N  \frac{\|u\|_{\infty}}{T_N^2} \le 1/8$  and the general bound $\lambda_N \|u\|_\infty\ge 1$ in \eqref{eq:general-lower} imply, 
\[\liminf_{N\to\infty}  {\lambda_N}{T_N^2} \ge \liminf_{N\to\infty} \, \Big(\lambda_N   \|u_N\|_\infty\Big)\, \liminf_{N\to\infty} \frac{ T_N^2} {\|u_N\|_\infty}\ge 8 ,\]
which is the desired lower bound in \eqref{eq:lam-bound-unif}. It is enough to bound $\lambda$ from above. 
Let $\eps_N,T_N,I, M_I$ be as in \eqref{eq:HM}, i.e., 
\begin{align}\label{eq:MlN}
   M_I=\begin{pmatrix}
	2 &  \ddots  &   &    \\
  \ddots &   \ddots   &  -a_j  &   \\
  & -a_j &  &  \ddots  \\
  &  &  \ddots &  2    
\end{pmatrix}_{\ell_N\times\ell_N}, \ \ a_j\ge 1-\eps_N, \ j\in I , 
\end{align}
is the maximal block in $H_N$ which is ``$\eps_N$-close'' to a free Laplacian $-\Delta_{\ell_N}$. 
By the min-max principle, $\lambda_N\le \wt \lambda $ where $ \wt \lambda$ is the smallest eigenvalue of $M_I$. 
On the other hand, as in \eqref{eq:E}, we write 
\[M_I-(-\Delta_{\ell_N})=E=\begin{pmatrix}
	0 &  \ddots  &   &  0  \\
  \ddots &   \ddots   &  1-a_j  &   \\
  & 1-a_j &  &  \ddots  \\
 0 &  &  \ddots & 0    
\end{pmatrix}_{\ell_N\times\ell_N}\]
Clearly, all eigenvalues of $E$ are contained in $[-2\eps_N, 2\eps_N]$. 
By the min-max principle again,
\begin{align*}
  \wt \lambda   \le \Big({\textrm{the smallest eigenvalue of} }\ -\Delta_{\ell_N} \Big)+2\eps_N  
     = 4\sin^2\Big(\frac{\pi}{2(\ell_N+1)}\Big) \, + 2\eps_N \le  \frac{\pi^2}{\ell_N^2}+2\eps_N  .
\end{align*}
Putting everything together with the choice of $\eps_N,T_N$ in \eqref{eq:l-asym-uni}, we obtain 
\begin{align*}
   \lambda_N T_N^{2}\le \pi^2\frac{T_N^2}{\ell_N^2}+2(\log N)^{-2} \big(\frac{\log N}{ 2\log \log N}\big)^2= \pi^2\frac{T_N^2}{\ell_N^2}+\frac{1}{ 2(\log \log N)^2},
\end{align*}
which implies that a.s. $\limsup_N \lambda_N T_N^{2}\le \pi^2$ .

%%===============================
\section{Band matrices with $W>1$}\label{sec:num}
%%===============================

In this section, 
we consider random band matrices $H_{N,W}$ in $d=1$ with a fixed band width $W\ge 1$. 
Let $H_{N,W}=(h_{ij})_{N\times N}$ be defined via,
\begin{align*}
    h_{ii}&=2W, \\
\numberthis\label{eqn:hwn-d}    h_{ij}&=-a_{ij}\in [-1,0],\quad \text{if }1\le |i-j|\le W,\\
    h_{ij}&=0,\quad{\rm otherwise},
\end{align*}
 as in \eqref{eq:HWN}, with an example given in \eqref{eq:H73}.
Let  
$ \lambda _{N,W} $
be the ground state energy, and let 
$u_{N,W}$
be the landscape function. We continue to study the approximation of $\lambda_{N,W}$ by $\|u_{N,W}\|_\infty$ for the case $W\ge 2$. We will display numerical  experiments supporting our Conjecture \ref{conj:HW-intro}, especially the approximation 
\begin{align} \label{eq:HW-appro}
    \lambda_{N,W} \|u_{N,W}\|_\infty \approx \frac{\pi^2}{8},
\end{align}
for random hopping terms $a_{ij}\in [0,1]$.

\subsection{The non-random graph Laplacian}

Let $-\Delta_{N,W}$ be the negative sub-graph Laplacian defined in \eqref{eqn:hwn-d}. 
We will discuss first this non-random case and prove Theorem \ref{thm:LapNW-intro}. The case $W=1$ which amounts to the standard 1D Laplacian was covered in Lemma~\ref{lem:free}, where both the ground state eigenvalue and landscape function were computed explicitly. 
For $W\ge 2$, we do not compute these explicitly, but instead determine the appropriate asymptotic behaviors.

\begin{theorem}[Banded free Laplacian asymptotics]\label{thm:LapNW-proof}
Let $\lambda_{N,W}=\lambda(-\Delta_{N,W})$  
be the ground state energy, and  $u_{N,W}= (-\Delta_{N,W})^{-1}\one  $
be the landscape function for $-\Delta_{N,W}$.  Let 
\begin{align*} 
    \sigma_{N,W}=\frac{1^2+\cdots+W^2}{N^2}=\frac{W(W+1)(2W+1)}{6N^2}.
\end{align*}
Then for any fixed $W$, 
\begin{align}\label{eq:LapNW-lam}
    \lim_{N\to\infty}\frac{\lambda_{N,W}}{\sigma_{N,W} }=\pi^2,
\end{align}
and 
\begin{align}\label{eq:LapNW-u}
    \lim_{N\to\infty}  \sigma_{N,W}\|u_{N,W}\|_{\infty} =\frac{1}{8}.
\end{align}
As a consequence,
\begin{align}
    \lim_{N\to\infty} \lambda_{N,W} \|u_{N,W}\|_{\infty}=\frac{\pi^2}{8}.
\end{align}
\end{theorem}
\begin{remark}
    As discussed in Example \ref{ex:band-graph} in Section~\ref{sec:kernel}, the graph $\Gamma$ induced by $-\Delta_{N,W}$ is roughly isometric to $\Z$. It is well known that the following  Hardy's inequality holds on $\Z$ with a uniform constant:
    \[ \sum_{n=1}^N\frac{f^2(n)}{n^2}\le 4 \sum_{n=2}^N \big(f(n)-f(n-1)\big)^2, \ f:\Z\to \R,\]
which easily implies the Hardy inequality in the form of \eqref{eq:hardy} on $\Gamma$ with a constant depending on $W$ (and independent of $N$). 
  Then by Theorem  \ref{thm:isometric} and Corollary \ref{cor:hardy}, equations \eqref{eq:hardy-u} and \eqref{eq:hardy-lam} hold for $-\Delta_{N,W}$, i.e., $\lambda \propto_W N^{-2}$ and $\|u\|_\infty \propto_W N^2$. Here, Theorem \ref{thm:LapNW-proof} gives the precise asymptotic behavior for these values in terms of $W$. 
\end{remark}
\begin{proof}
We first rewrite $-\Delta_{N,W}$ as  
\[-\Delta_{N,W}=A^1+\cdots+A^W,\]
where 
\begin{align}\label{eq:106}
    (A^j\psi)(n)=\begin{cases}2\psi(n)-\psi(n+j)-\psi(n-j), & j< n\le N-j \\
    2\psi(n)-\psi(n+j), & 1\le n\le  j \\ 
    2\psi(n)-\psi(n-j) , & N-j< n\le N
    \end{cases},
\end{align}
so that $A^j$ represents all the bonds that cover a distance $j$.
For example, $A^1=-\Delta_N$ is the standard nearest-neighbor Laplacian. Let $\lambda^j$ be the smallest eigenvalue of $A^j$.  As in Lemma \ref{lem:free} that $ \lambda^1$ has the explicit expression as
\begin{align}
    \lambda^1=4 \sin^2 \Big(\frac{\pi}{2(N+1)}\Big)\approx_N \frac{\pi^2}{N^2},
\end{align}
where we use the notation $X\approx_NY$ to mean that $\lim_{N\to \infty}X/Y=1$.
The normalized (in $\ell^2$ norm) ground state can also be computed explicitly,  see e.g. \cite{chung2000discrete}, 
\begin{align}\label{eq:108}
    \psi^1(n)=\sqrt{\frac{2}{N+1}} \sin \Big(\frac{n\pi}{N+1}\Big), \ n=1,\cdots,N. 
\end{align}
For $j\ge 2$,  the matrix $A^j$ is a direct sum, after suitable permutation, of $j$  nearest-neighbor Laplacians of size $N/j$. Therefore, 
\[\lambda^j=\textrm {the smallest eigenvalue of } (-\Delta_{N/j})\approx_N \frac{j^2\pi^2}{N^2}.\]
By the min-max principle, 
\[\lambda_{N,W} \ge \lambda^1+\cdots+\lambda^W\approx_N (1^2+\cdots+W^2)\frac{\pi^2}{N^2}:=\pi^2\sigma_{N,W}\]
On the other hand, plugging \eqref{eq:108} into \eqref{eq:106}, one has
\begin{align}
     (A^j\psi^1)(n)=\begin{cases}4 \sin^2 \Big(\frac{j\pi}{2(N+1)}\Big)\ \psi^1(n), \ \ j< n\le N-j \\
    \Theta_W\Big(\frac{1}{N^{3/2}}\Big),\ \  {\rm otherwise}
    \end{cases},
\end{align}
{where  we   use the asymptotic notation $X=\Theta_W(Y)$ if $c_W\le X/Y \le C_W$ for two positive constants $c_W,C_W$ only depending on $W$ (independent of $N$).}
Therefore,
\begin{align}
    \ipc{\psi^1}{A^j\psi^1}&=4 \sin^2 \Big(\frac{j\pi}{2(N+1)}\Big)\Big(1-\Theta_W\Big(\frac{1}{N^{3}}\Big)\Big)+\Theta_W\Big(\frac{1}{N^{3}}\Big) \\
  &\approx_N \frac{j^2\pi^2}{N^2}  \Big(1-\Theta_W\Big(\frac{1}{N}\Big)\Big)
\end{align}
By the min-max principle again,
\begin{align*}
    \lambda_{N,W} \le \ipc{\psi^1}{-\Delta_{N,W} \psi^1 }=&\sum_{j=1}^W \ipc{\psi^1}{A^j\psi^1 }\\
    \approx_N&\sum_{j=1}^W \frac{j^2\pi^2}{N^2}  \Big(1-\Theta_W\Big(\frac{1}{N}\Big)\Big)=\pi^2\sigma_{N,W}  \Big(1-\Theta_W\Big(\frac{1}{N}\Big)\Big),
\end{align*}
which leads to the $\limsup$ of $\lambda_{N,W} /\sigma_{N,W}$ and therefore completes the proof of \eqref{eq:LapNW-lam}. 

Next we estimate the landscape function $u$. Let 
\begin{align*}
    u^1(n)=\frac{1}{2}n(N+1-n), 1\le n\le N
\end{align*}
be the landscape function of $A^1$. Evaluating $u^1$ by \eqref{eq:106}, one has that for all $1\le n\le N$, 
\begin{align}\label{eq:114}
     (A^ju^1)(n)&=\begin{cases}  j^2, & j< n\le N-j \\
   j^2+\frac{1}{2}(n-j)(N+1-n+j), & 1\le n\le  j \\ 
    j^2+\frac{1}{2}(n+j)(N+1-n-j) , & N-j< n\le N
    \end{cases} \\
    &\le  j^2.
\end{align}
Therefore, $ (-\Delta_{N,W} u^1)(n)=\sum_{j=1}^W \,  (A^j u^1)(n)\le 1^2+\cdots+W^2,$ which implies 
\[ \frac{-\Delta_{N,W} u^1(n)}{1^2+\cdots+W^2}  \le 1= -\Delta_{N,W} u(n),\]
for all $1\le n \le N$. Together with the maximum principle Lemma \ref{lem:maxP}, one has 
\begin{align}
    \|u_{N,W}\|_{\infty} \ge  \frac{1}{1^2+\cdots+W^2} \|u^1\|_{\infty}(n) \approx_N \frac{1}{1^2+\cdots+W^2} \frac{N^2}{8} =\frac{1}{8 \sigma_{N,W}}.
\end{align}
On the other hand, \eqref{eq:114}   implies that \[ (A^ju^1)(n)\ge j^2 -W(N+W), \ \ \forall \ 1 \le n \le N\]
Let $\wt u= u^1+W(N+W)$. Then $(A^ju^1)(n)\ge j^2 $, for all $1 \le n \le N$,
which implies 
\begin{align}
    (-\Delta_{N,W} \wt u)(n)=\sum_{j=1}^W \,  (A^j \wt u)(n) \ge  1^2+\cdots+W^2.
\end{align}
Therefore, by the maximum principle Lemma \ref{lem:maxP}, 
\begin{align*}
    \|u_{N,W}\|_{\infty}  \le  \frac{1}{1^2+\cdots+W^2} \|\wt u\|_{\infty}\approx_N & \frac{1}{1^2+\cdots+W^2} \Big(\frac{N^2}{8}+\Theta_W(N)\Big) \\
    \approx_N & 
   \frac{N^2}{1^2+\cdots+W^2} \Big(\frac{1}{8}+\Theta_W(\frac{1}{N})\Big) \approx_N\frac{1}{8 \sigma_{N,W}},
\end{align*}
which completes the proof for \eqref{eq:LapNW-u}. 
\end{proof}
\begin{remark}
One can also prove the landscape function estimate using random walk exit times since $u(x)=\frac{1}{\#\mathcal{B}_W}\mathbb{E}_x[T_{\intbr{1}{N}}]$, where $T_{\intbr{1}{N}}$ is the exit time of a discrete-time SRW.
\end{remark}

A numerical illustration is shown in Figures~\ref{Free_RB} and \ref{Free_W} below.

\begin{figure}[!ht]
	\centering
	\includegraphics[width=0.48\linewidth]{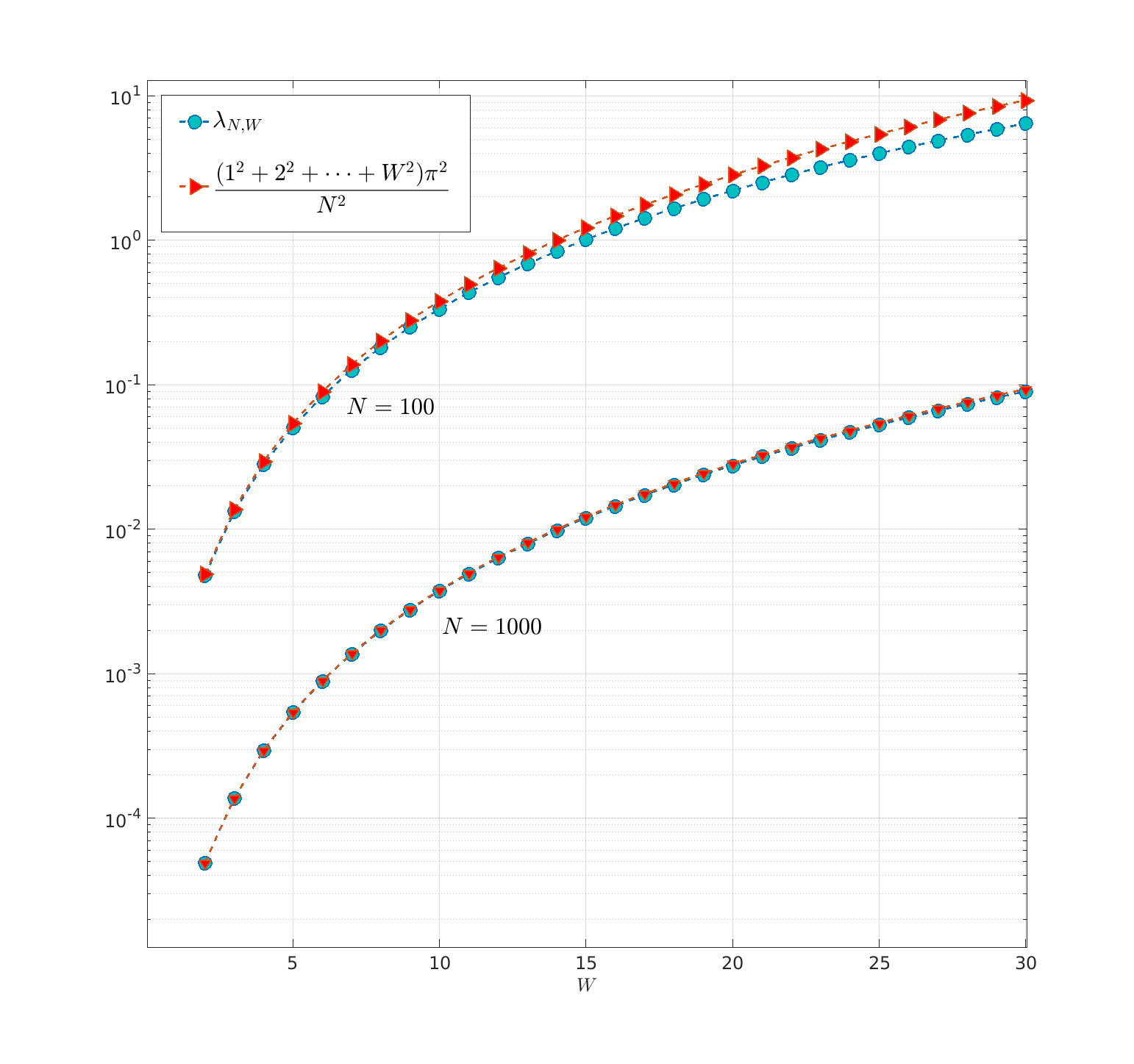}
        \includegraphics[width=0.48\linewidth]{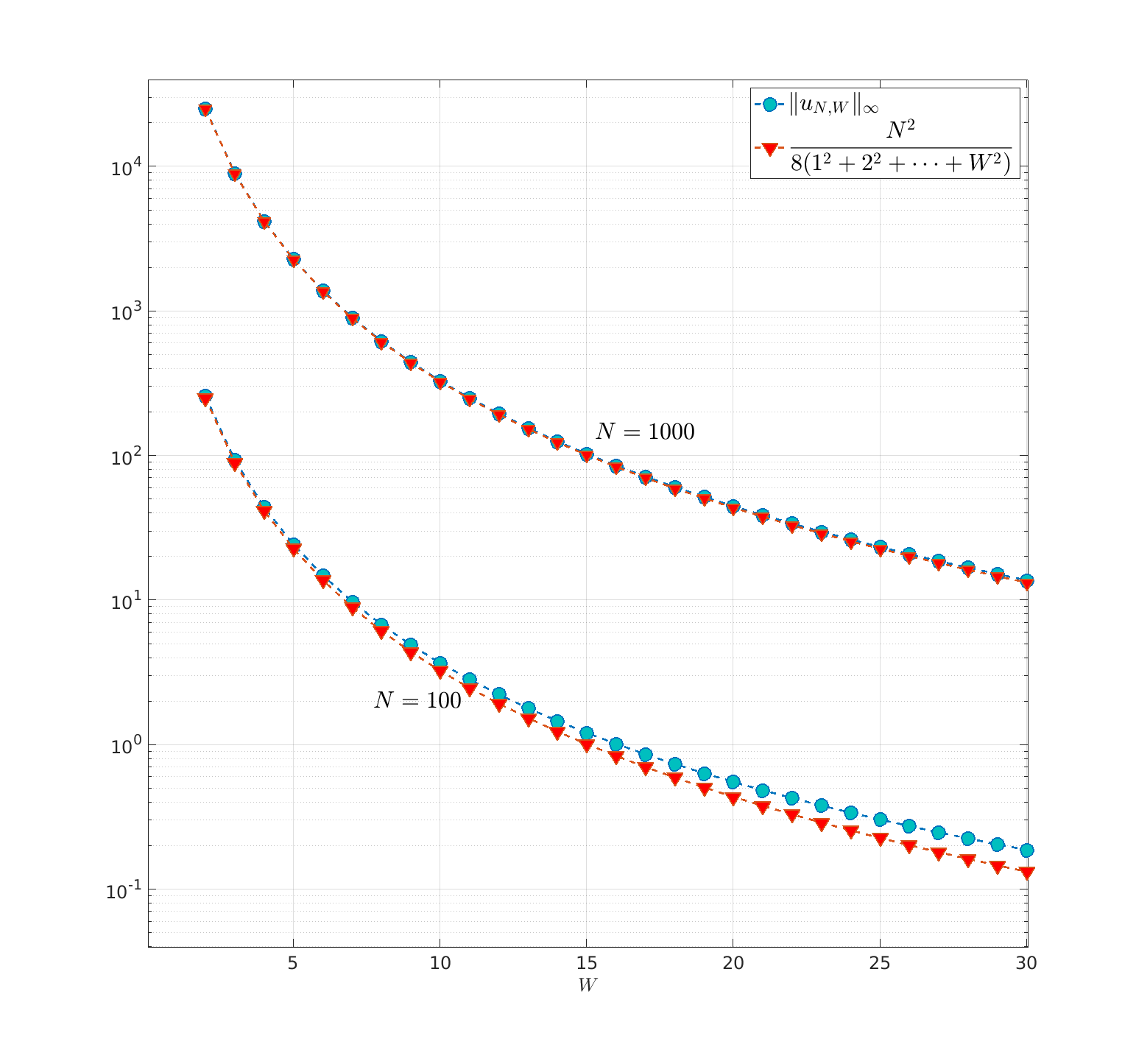}
	\caption{Left: Comparison of $\lambda_{N,W}$ of $-\Delta_{N,W}$ and $\frac{(1^2+2^2+\cdots+W^2)\pi^2}{N^2}$ with various $W$ and $N$. Right:  Comparison of $\Vert u_{N,W}\Vert_{\infty}$ of $-\Delta_{N,W}$ and $\frac{N^2}{8(1^2+2^2+\cdots+W^2)}$ with various $W$ and $N$. }
	\label{Free_RB}
\end{figure}

\begin{figure}[!ht]
	\centering
	\includegraphics[width=0.48\linewidth]{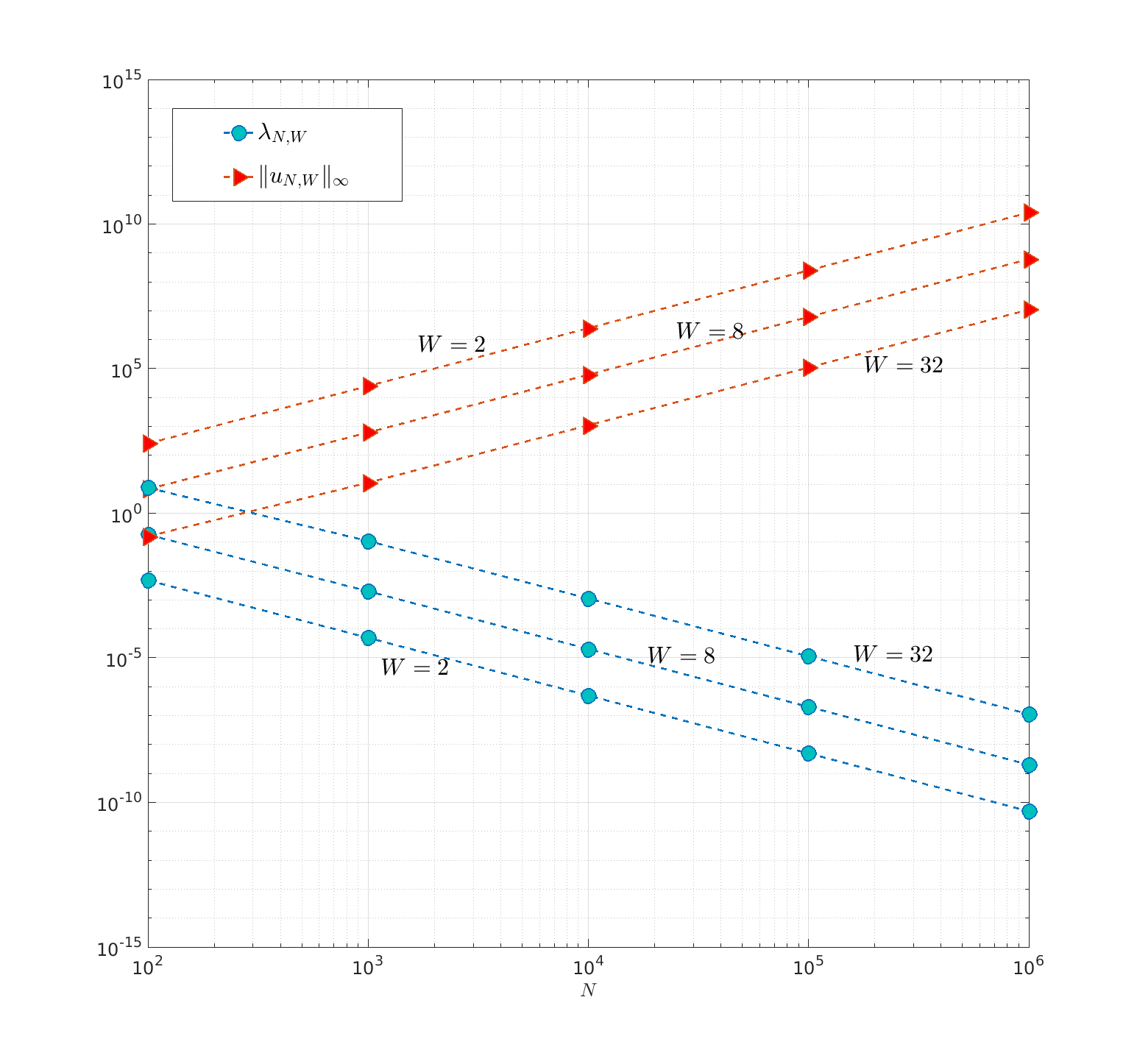}
        \includegraphics[width=0.48\linewidth]{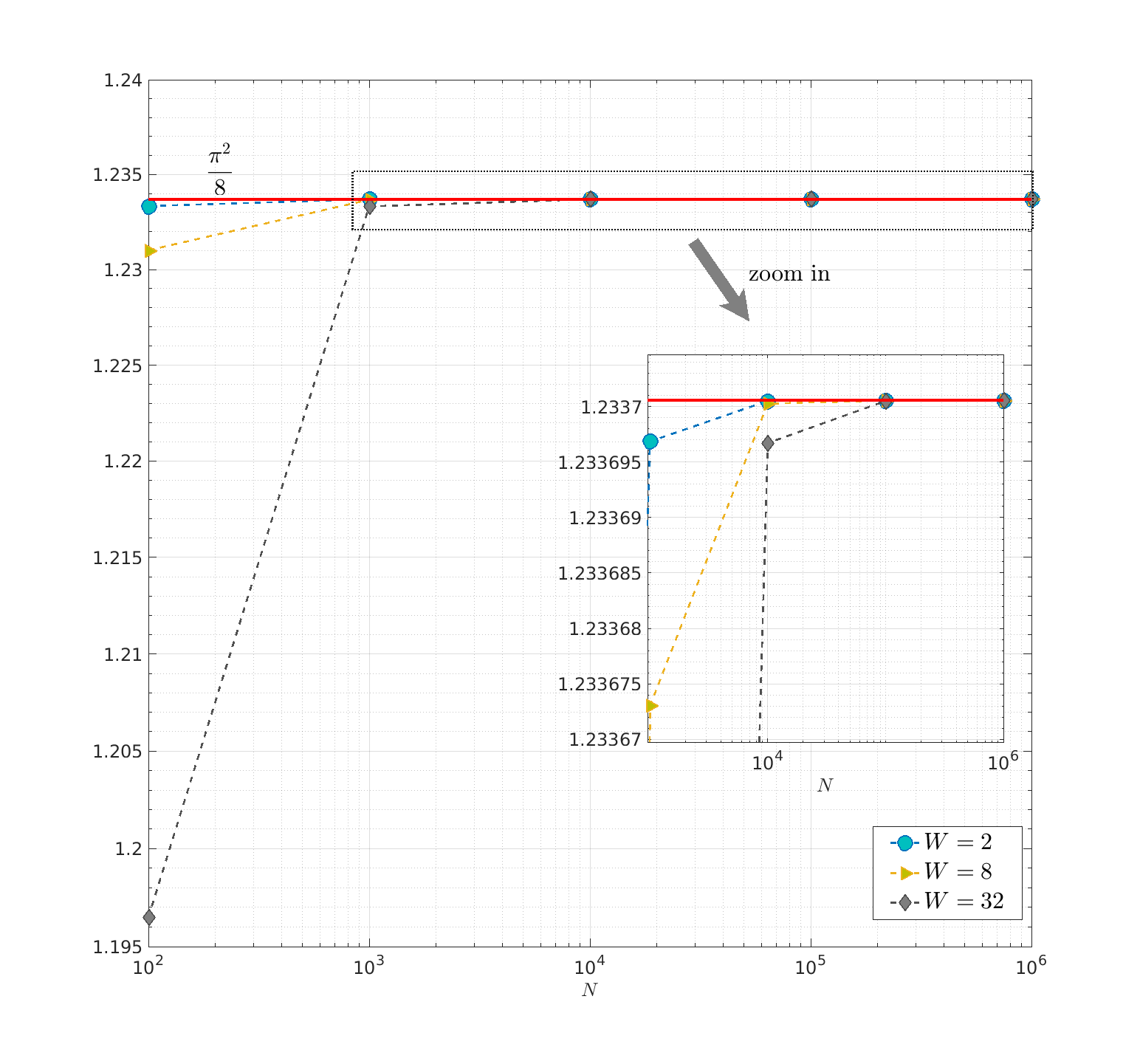}
	\caption{Left: $\lambda_{N,W}$ and $\Vert u_{N,W}\Vert_{\infty}$ of $-\Delta_{N,W}$ with $W=2,8,32$. Right: the associated $\lambda_{N,W}\Vert u_{N,W}\Vert_{\infty}$ with $W=2,8,32$. }
	\label{Free_W}
\end{figure}

\subsection{Random cases}

\begin{figure}[!ht]
	\centering
	\subfigure[$N=10^3$; Mean=0.0116; Std=0.0463 ]{\label{W2_Ba}
	 \includegraphics[width=0.4\linewidth]{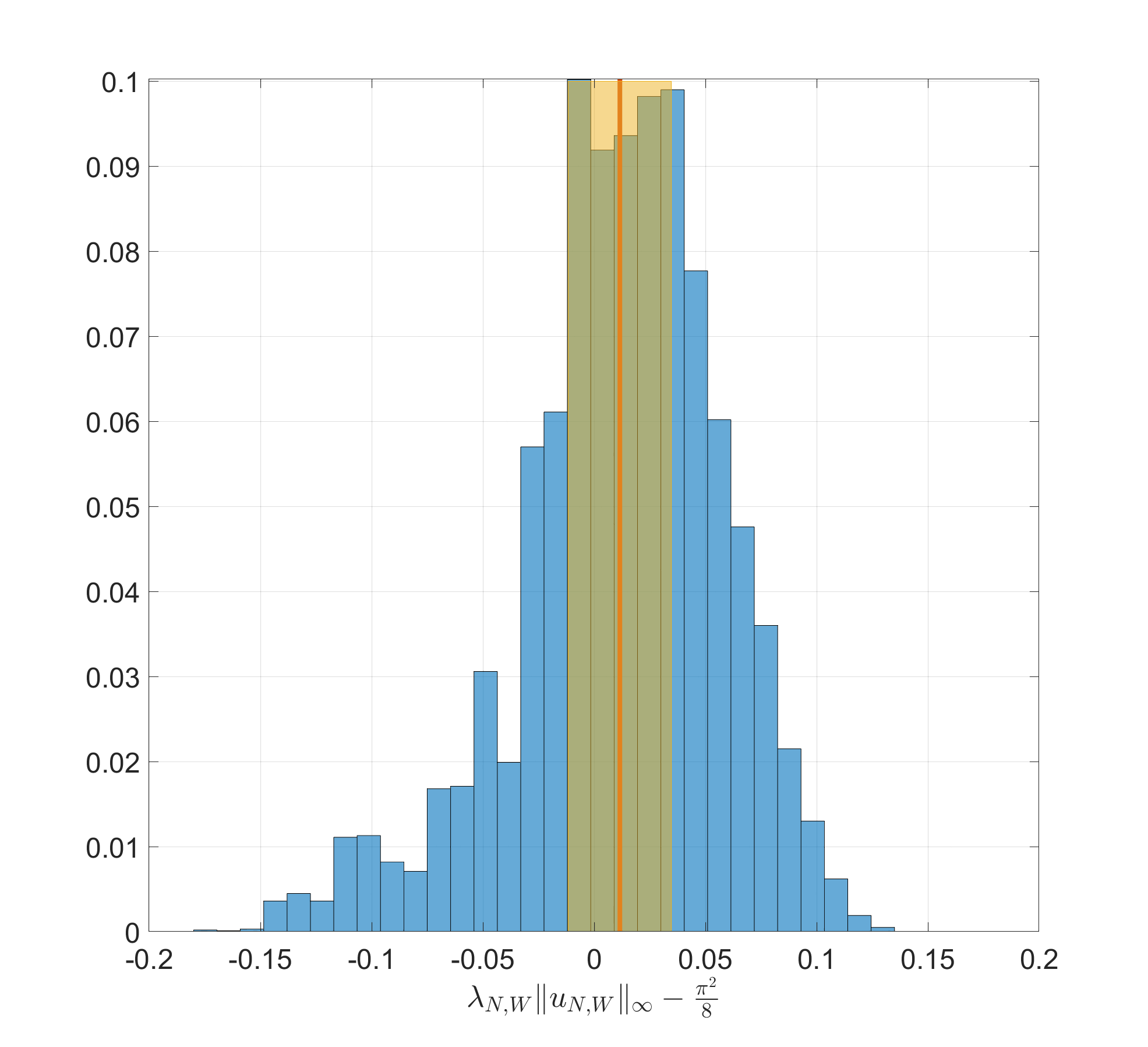}
	}
	\quad
	\subfigure[$N=10^4$; Mean=0.0151; Std=0.0300 ]{\label{W2_Bb}
			\includegraphics[width=0.4\linewidth]{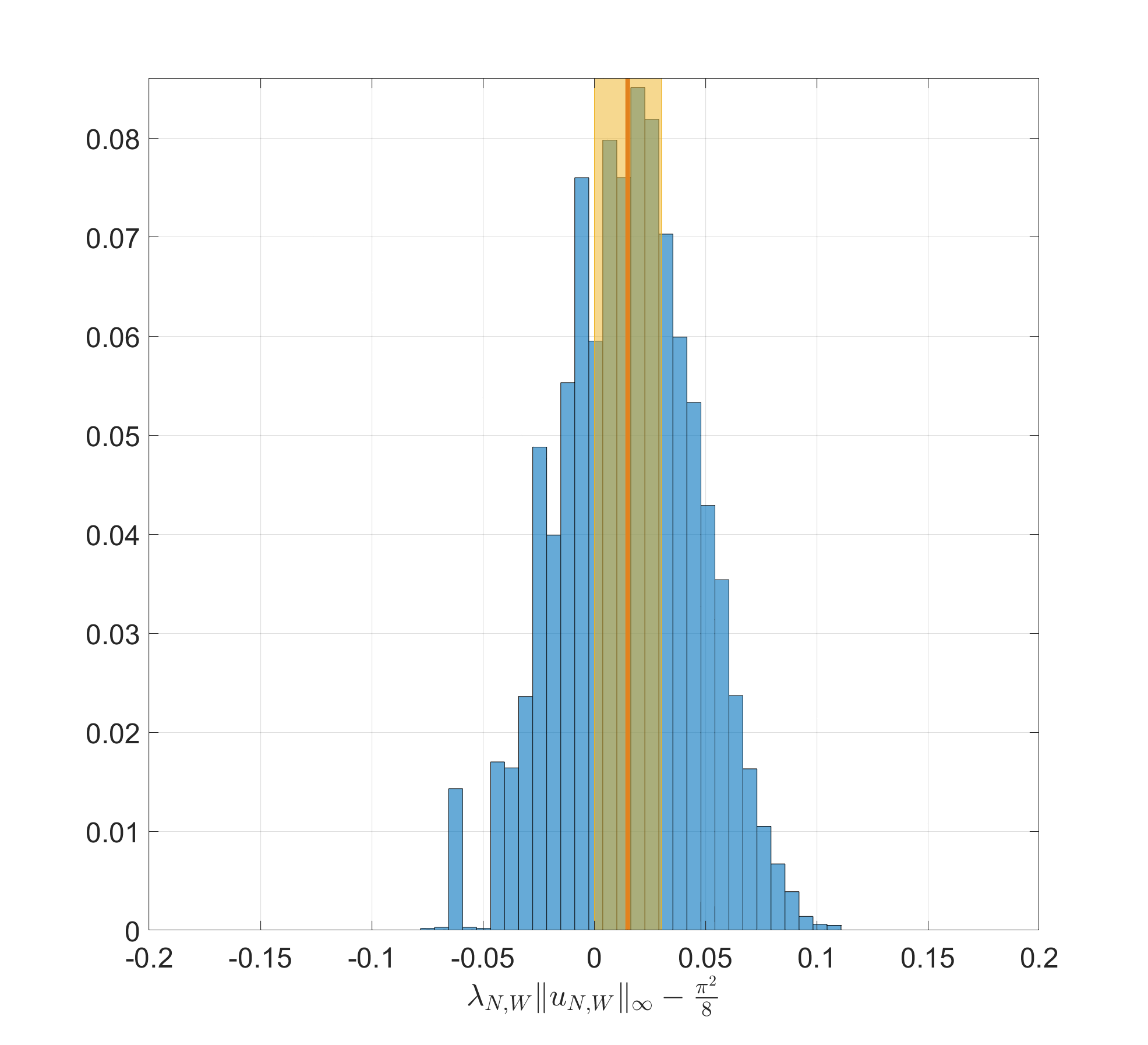}
	}
	\quad
	\subfigure[$N=10^5$; Mean=0.0070; Std=0.0286 ]{\label{W2_Bc}
		\includegraphics[width=0.4\linewidth]{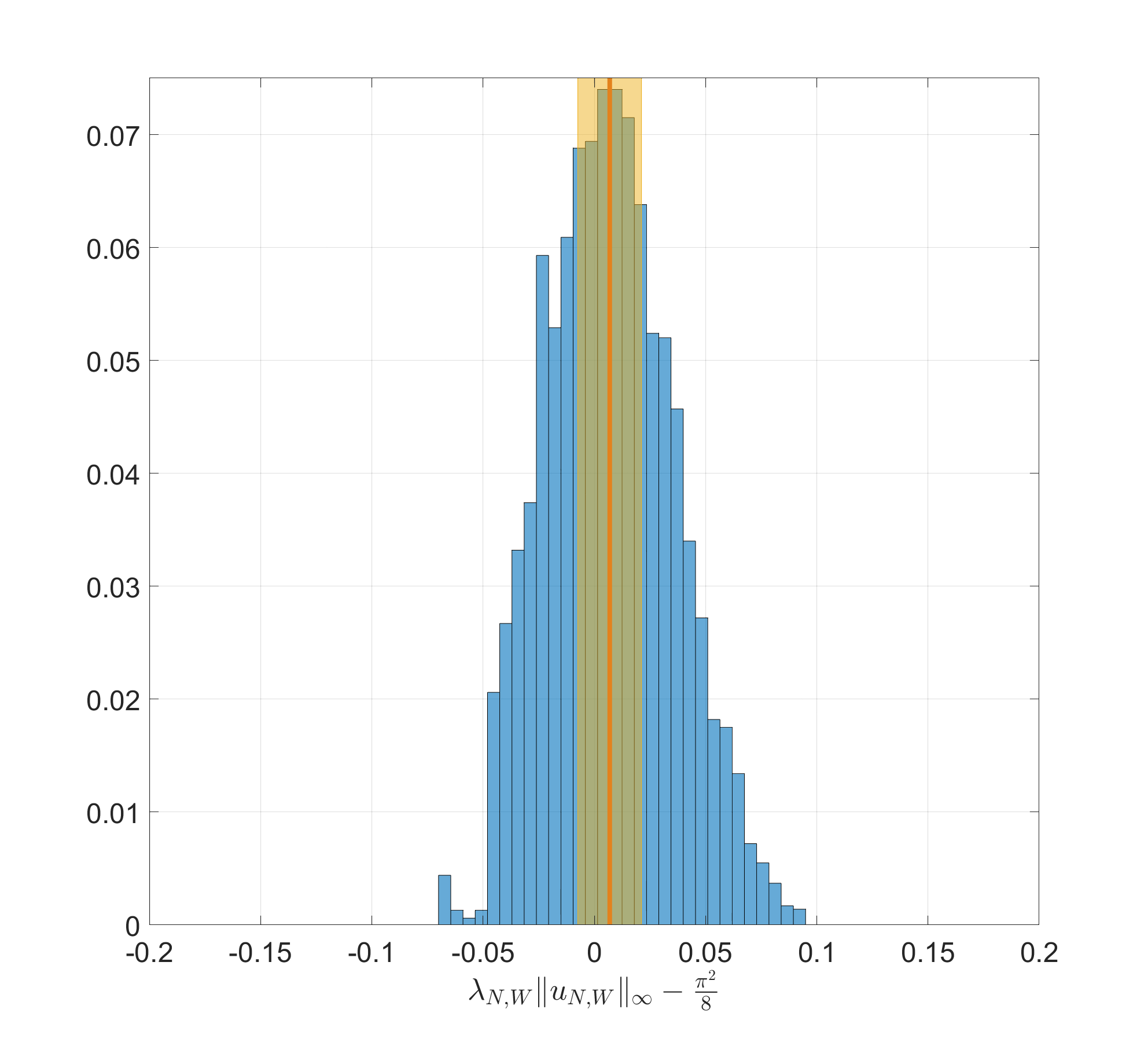}
	}
	\quad
	\subfigure[$N=10^6$;  Mean=0.0046; Std=0.0244 ]{\label{W2_Bd}
		\includegraphics[width=0.4\linewidth]{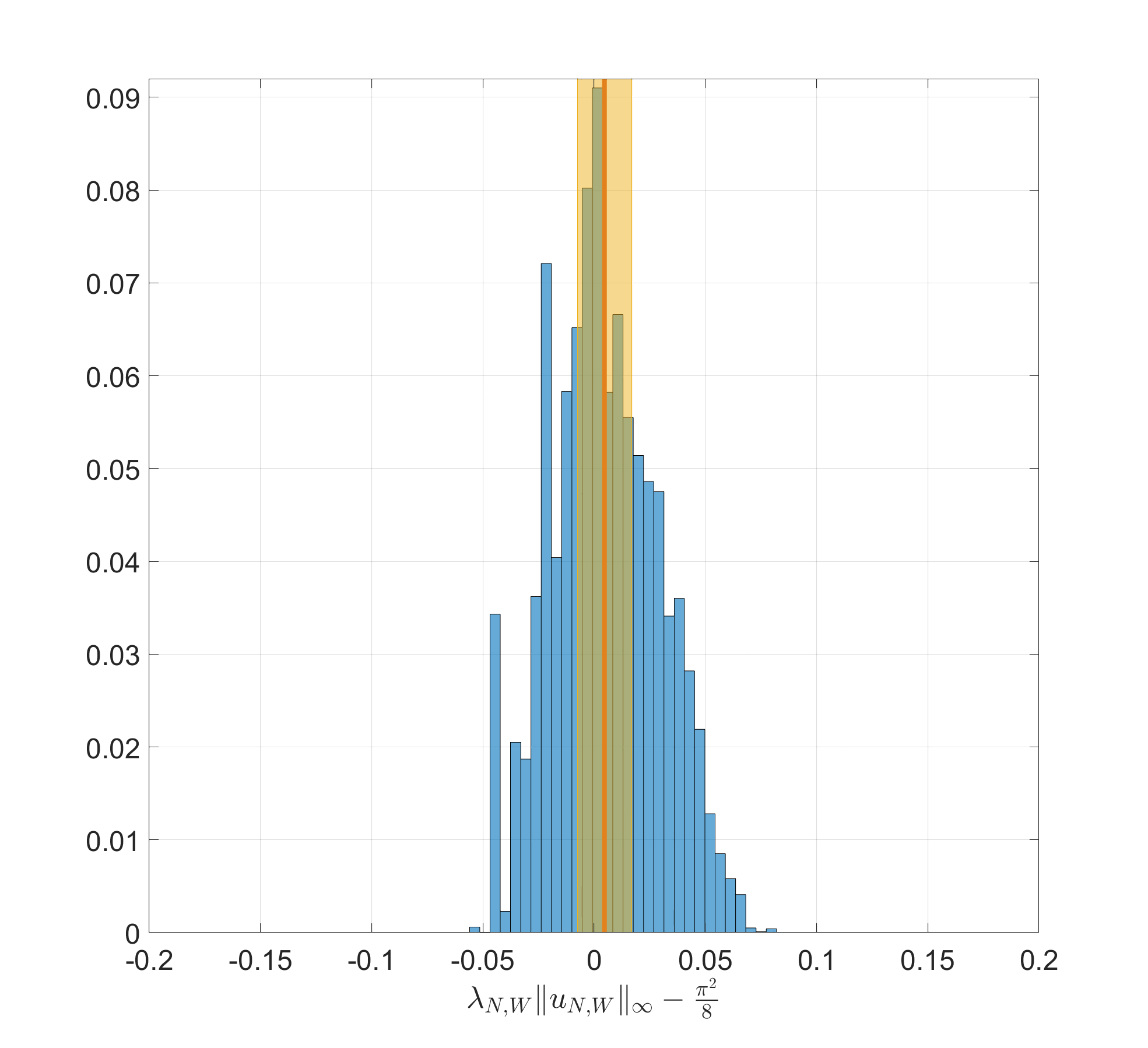}
	}
	\caption{Histograms of the distribution of $(\lambda_{N,W}\Vert u_{N,W}\Vert_{\infty}-\frac{\pi^2}{8})$ from $10^4$ random realizations with $W=2$. The numerical results here are quite similar to the simulations in \cite{sanchez2023principal}, which considers the Anderson model. }	
	 \label{W2_B}
\end{figure}

In this part, we will extend the discussion to the random cases where $W>1$ numerically. We first consider one example, where the bandwidth $W=2$. The entries $a_{ij}$ are from  Bernoulli distribution of 1 and 0, with probabilities 30\% and 70\%. Figure \ref{W2_B} displays the histgrams of the distribution of $(\lambda_{N,W}\Vert u_{N,W}\Vert_{\infty}-\frac{\pi^2}{8})$ over $10^4$ random realizations for various $N$.

\begin{figure}[!ht]
	\centering
	\subfigure[]{\label{MoreEigsB}
	 \includegraphics[width=0.45\linewidth]{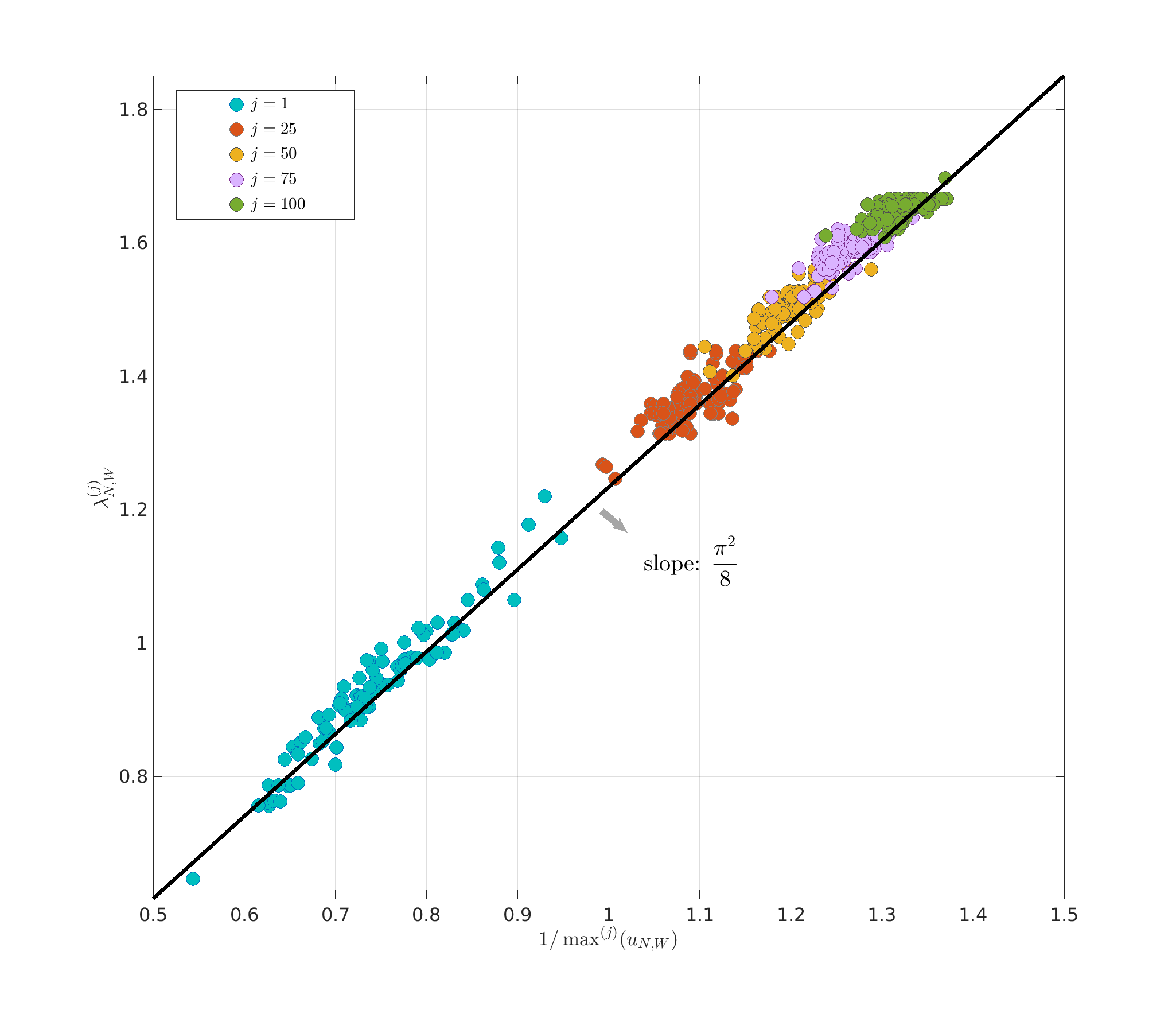} 
	}
	\quad
	\subfigure[]{\label{MoreEigsU}
			\includegraphics[width=0.45\linewidth]{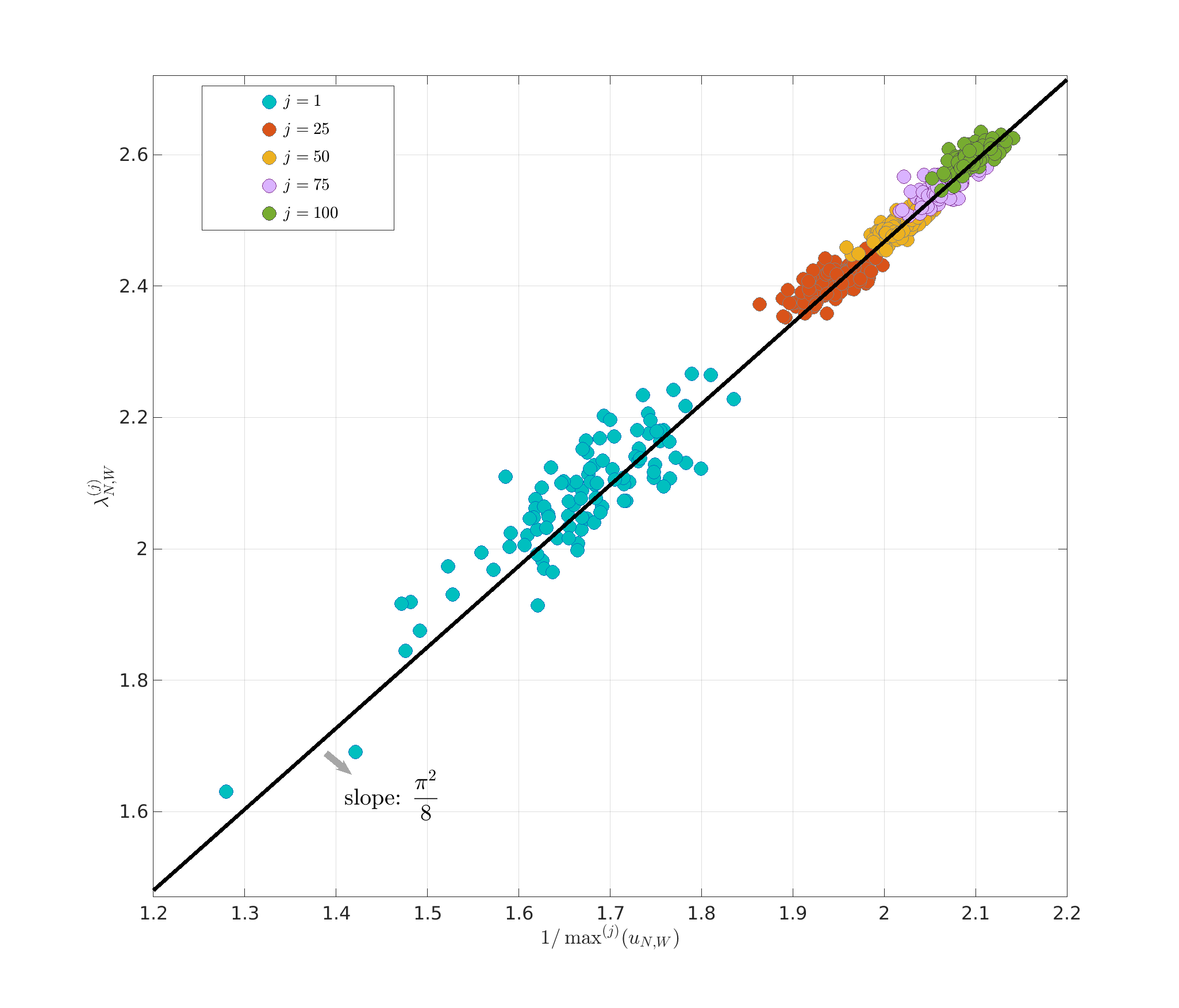}
	}
       \caption{The 1st, 25th, 50th, 75th and 100th eigenvalues versus the reciprocal of corresponding local maxima values $1/\max^{(j)}(u_{N,W})$ on the x-axis, for 100 independent realizations. Left: $\{a_{ij}\}$ takes Bernoulli distribution of 1 and 0, with probabilities 30\% and 70\%, $N=10^4$ and $W=2$.  
       Right: $\{a_{ij}\}$ takes uniformly random numbers from [0,1], $N=10^4$ and $W=3$. }
\end{figure}

Afterwards, we consider $\lambda_{N,W}^{(j)},j=1,2,\cdots$, and their relation to $\max^{(j)}(u_{N,W})$, which are the local minima of the landscape function. 
We first  extend Figure \ref{W2_Bb} to  more low-lying energies, and show the relation of $\lambda_{N,W}^{(j)}$ and $1/\max^{(j)}(u_{N,W})$ ($j=1,25,50,75,100$) in Figure \ref{MoreEigsB}. Each eigenvalue includes 100 random realizations, and the slope of the reference line is $\dfrac{\pi^2}{8}$. Likewise, we consider the uniform case for more low-lying energies in Figure~\ref{MoreEigsU}, where $N=10000, W=3$. 

%%%%%%%%%%%%%%%%%%%%%%%%%%%%%%%%%%%%%%%%%%%%%%%%%%%%%%%%%%%%%%%%%%%%%%%%%%%%%%%%%%%%%%%%%%%%
\appendix

\section{Proof of Lemma \ref{lem:lam-low-ber}}\label{app:lam-low}

In this section, we prove Lemma~\ref{lem:lam-low-ber}, which provides a lower bound on the ground state eigenvalue $\lambda_N$ for the 1D nearest neighbor hopping model $H_N$ on $\intbr{1}{N}$ with $\{a,1\}$-Bernoulli off-diagonal disorder.

We first partition $\intbr{1}{N}$ into islands $\isl$ and walls $\wall$. We define $I=\intbr{i}{j}$ to be an island if $a_{i}=\cdots= a_j=1$ and $a_{i-1}\neq 1$ (always assuming $a_1=1$ for convenience) and $a_{j+1}\neq 1$, and we define walls $\mathcal{W}=\{W\}$ as all the connected components created by removing all islands from $\intbr{1}{N}$. An example of such a matrix with islands and walls is, on $\intbr{1}{8}$, 
\begin{align}  \label{eqn:island-wall}
   &\left(\begin{smallmatrix}
	2 & -1 &    &   &   &   & &   \\
	-1 &   2 &-a &  & &   &  &  \\
 	& -a  & 2 & -a   &    &   & &  \\
  &  &  -a & 2  & -1 &  &  & \\
   &  &   &  -1& 2 & -a    &  &  \\
   &  &   &  & -a  & 2  &   -1 & \\
   &  &  & &    &  -1  & 2 & -1 \\
   &  &  & &    &    & -1 & 2
\end{smallmatrix}\right),
\end{align}
where $a_3,a_4,a_6\neq 1$, and $\mathcal{I}=\big\{\intbr{1}{2},\{5\}, \intbr{7}{8}\big\}$, $\mathcal{W}=\big\{ \intbr{3}{4}, \{6\}\big\}$. Notice that the islands defined here are different from those used in Theorem \ref{thm:Ber} and Lemma \ref{lem:lam-low-ber}, see e.g. \eqref{eqn:w-I}, \eqref{eqn:wtu-Ik}. We actually make the island one point less (from the left) which is the shrunken island $\check I$.  We omit the checks and still denote the shrunken islands by $I$. The longest length of the islands $\ell_N$ will only differ by one and will not affect the asymptotics as $\ell_N\to \infty$. The main reason to take such shrunken islands is to keep those single site walls and to separate the two islands   $\{5\}$ and $\intbr{7}{8}$.

\begin{lemma}
{Let $\psi$ be the ground state for $H_N$, and let $\ell_N$ be the length of the longest sequence of consecutive $a_j=1$.} Then 
\begin{align}
    \sum_{j\in\wall}\psi^2(j)\lesssim \frac{1}{(1-a)\ell_N^2}.
\end{align}

\end{lemma}
\begin{proof}
By the upper bound of $\lambda_N$ and the explicit form of $H_N$, 
\begin{align*}
     \frac{\pi^2}{\ell_N^2}\ge  \lambda = \ipc{\psi}{H_N \psi}&=2\ipc{\psi}{\psi}-  \sum_{j=2}^{N} a_j\psi(j)\psi(j-1)-\sum_{j=1}^{N-1} a_{j+1}\psi(j)\psi(j+1)\\
     &=2\ipc{\psi}{\psi}-  2\sum_{j=1}^{N-1} a_{j+1}\psi(j)\psi(j+1).
\end{align*}
By Cauchy–Schwarz and the non-negativity of $a_j$, we obtain
\begin{align*}
    2\sum_{j=1}^{N-1} a_{j+1}\psi(j)\psi(j+1)\le &  \sum_{j=1}^{N-1} a_{j+1} \psi(j)^2+\sum_{j=1}^{N-1} a_{j+1} \psi(j+1)^2\\
  = & \sum_{j=1}^{N-1} a_{j+1} \psi(j)^2+\sum_{j=2}^{N} a_{j} \psi(j)^2 \\
  =& \sum_{\substack{1\le j\le N-1 \\ a_{j+1}=1}}   \psi(j)^2+a\sum_{\substack{1\le j\le N-1 \\ a_{j+1}=a}}    \psi(j)^2+\sum_{\substack{2\le j\le N  \\ a_{j}=1}}   \psi(j)^2+a\sum_{\substack{2\le j\le N  \\ a_{j}=s}}  \psi(j)^2\\
  \le & \sum_{\substack{1\le j\le N  \\ a_{j+1}=1}}   \psi(j)^2+a\sum_{\substack{1\le j\le N-1 \\ a_{j+1}=a}}    \psi(j)^2+\sum_{\substack{1\le j\le N  \\ a_{j}=1}}   \psi(j)^2+a\sum_{\substack{2\le j\le N  \\ a_{j}=a}}  \psi(j)^2.
\end{align*}
 On the other hand,
 \begin{align*}
     \ipc{\psi}{\psi}-\sum_{\substack{1\le j\le N  \\ a_{j+1}=1}}   \psi(j)^2=\sum_{\substack{1\le j\le N  \\ a_{j+1}=a}}   \psi(j)^2\ge \sum_{\substack{1\le j\le N-1  \\ a_{j+1}=a}}\psi(j)^2,
 \end{align*}
 and 
 \begin{align*}
     \ipc{\psi}{\psi}-\sum_{\substack{1\le j\le N  \\ a_{j}=1}}   \psi(j)^2=\sum_{\substack{1\le j\le N  \\ a_{j}=a}}   \psi(j)^2\ge \sum_{\substack{2\le j\le N  \\ a_{j}=a}}   \psi(j)^2. 
 \end{align*}
Putting all together, we have 
\begin{align*}
  2\ipc{\psi}{\psi}-  2\sum_{j=1}^{N-1} a_{j+1}\psi(j)\psi(j+1)\ge & (1-a)\sum_{\substack{1\le j\le N-1  \\ a_{j+1}=a}}\psi(j)^2+  (1-a)\sum_{\substack{2\le j\le N  \\ a_{j}=a}}   \psi(j)^2 \\
 \ge &  (1-a)\sum_{\substack{2\le j\le N  \\ a_{j}=a}}   \psi(j)^2=(1-  a)\sum_{j\in\wall}\psi^2(j),
\end{align*}
    which implies that 
    \[\sum_{j\in\wall}\psi^2(j)\le \frac{\pi^2}{(1-a)\ell_N^2}. \]
\end{proof}
We further split islands into heavy and light ones according to the boundary values of $\psi$.
\begin{definition}
An island $I_H=\intbr{i}{j}$ is called \emph{heavy} if 

\begin{align}
   \sum_{k=i}^j\psi^2(k) \ge \delta^2 (1-a)^{1/2}\ell_N^{3/2},
\end{align}
where 
$\delta=\max\{\psi(i-1),\psi(j+1)\}$.
Otherwise, an island is called \emph{light}. 
Indices belonging to heavy islands are denoted by $\islh$, and those belong to light island by $\isll$.
\end{definition}

\begin{lemma}
There is at least one heavy island.
\end{lemma}
\begin{proof}
We have the partition,
\begin{align*}
    1=\sum_{j\in \intbr{1}{N}}\psi^2(j)=\sum_{j\in\wall}\psi^2(j)+\sum_{j\in\isll}\psi^2(j)+\sum_{j\in\islh}\psi^2(j).
\end{align*}
Notice that if $I=\intbr{i}{j}$ is an island, then $i-1,j+1\in \wall$. In particular, for all light islands $\isll$,
\begin{align*}
     \sum_{k\in \isll}\psi^2(k) \le (1-a)^{1/2}\ell_N^{3/2} \sum_{k\in\wall}\delta^2\lesssim (1-a)^{1/2}\ell_N^{3/2} \frac{1}{(1-a)\ell_N^2} \le \frac{1}{(1-a)^{1/2}\ell_N^{1/2}}.
\end{align*}
Therefore,
\begin{align*}
    \sum_{j\in\islh}\psi^2(j)&\ge 1-\bigg(\sum_{j\in\wall}+\sum_{j\in\isll}\bigg)\psi^2(j)\\
   &\ge  1- \frac{1}{(1-a)\ell_N^2}-\frac{1}{(1-a)^{1/2}\ell_N^{1/2}}>0,
\end{align*}
since  $\ell_N\to \infty$. 
\end{proof}

Finally, on a heavy island $I=\intbr{i}{j}$, where $\ell=|I|\le \ell_N$, we study the ground state eigenvalue equation $H_N  \psi(k)=\lambda \psi(k)$ for $k\in I$. This becomes,
\begin{align}\label{eq:ev-eqn}
    -\psi(k-1)+2\psi(k)-\psi(k+1)=\lambda\psi(k),\quad k=i,\ldots,j,
\end{align}
with boundary values $\psi(i-1)=\delta_1$ and $\psi(j+1)=\delta_2$, which satisfy 
\begin{align}
    \delta_i^2\le \frac{1}{(1-a)^{1/2}\ell_N^{3/2}}m^2,\ \ 
\end{align}
where $m^2=\sum_I\psi^2(k)$.

The equation \eqref{eq:ev-eqn} has solution of the form 
\begin{align*}
    \psi(k)=A\sin(Bk+C),
\end{align*}
which implies 
\begin{align*}
    \lambda=2-2\cos B=B^2+\mathcal{O}(B^4).
\end{align*}
{Note that the ground state $\psi$ must be pointwise positive, $\psi(k)>0$.} Without loss of generality, assume that $A>0$ and $i=1,j=\ell$. 
 The boundary conditions become,
 \begin{align*}
     \delta_1\psi(0)=A\sin C, \ \ \delta_2=\psi(\ell+1)= A\sin \big(B(\ell+1)+C\big) .
 \end{align*}
 Using the upper bound for $\lambda$, one has
 \begin{align*}
     2-2\cos B=\lambda \le \pi^2/\ell_N^2 \ll 1,
 \end{align*}
 which implies $B\in (0, \pi/2)$. All the phases
 \begin{align*}
     C,C+B,C+2B,\cdots, C+(\ell+1)B
 \end{align*}
 must be in the same zone $(2\Z\pi, (2\Z+1)\pi)$; without loss of generality, we assume it is $(0,\pi)$. 
 
The ground state $\psi(k)$ can not be monotone in $k$, otherwise,
 \begin{align*}
   m^2=  \sum_{I}\psi(k) \le \ell \, \max(\delta_1^2,\delta_2^2)\le \ell_N\, \frac{m^2}{(1-a)^{1/2}\ell_N^{3/2}}=\frac{m^2}{(1-a)^{1/2}\ell_N^{1/2}},
 \end{align*}
 which is a contraction since $\ell_N\to \infty$. 
 
 On the other hand, 
 \begin{align*}
     \delta_1^2=A^2\sin^2C \le \frac{1}{(1-a)^{1/2}\ell_N^{3/2}}m^2
&{= \frac{1}{(1-a)^{1/2}\ell_N^{3/2}}\sum_{I}A^2\sin^2(Bk+C)}\\
&\le \frac{1}{(1-a)^{1/2}\ell_N^{3/2}}\, A^2\, \ell_N,
 \end{align*}
 which implies $
     \sin^2C\le \frac{1}{(1-a)^{1/2}\ell_N^{1/2}} \to 0$. 
 Therefore, $C\searrow 0$. The same computation on $\delta_2$ implies $ \sin^2(B(\ell+1)+C)\le \frac{1}{(1-a)^{1/2}\ell_N^{1/2}} \to 0$, leading to 
  $B(\ell+1)+C	\nearrow \pi$. Putting everything together, we obtain,
 \begin{align}
     B(\ell+1)\ge \pi- \frac{1}{(1-a)^{1/4}\ell_N^{1/4}}-C\ge \pi- \frac{2}{(1-a)^{1/4}\ell_N^{1/4}}.
 \end{align}
 Therefore,
 \begin{align}
     \lambda=2-2\cos B\ge B^2(1-B^2)
     \ge \frac{\pi^2}{\ell_N^2}\bigg(1-\frac{2}{(1-a)^{1/4}\ell_N^{1/4}}\bigg),
 \end{align}
 which completes the proof of Lemma \ref{lem:lam-low-ber}.

\section{Proof of Claim \ref{claim:Pn-poly}}\label{app:claim-Pn}
In this section, we prove Claim~\ref{claim:Pn-poly} concerning
$P_n(a_2,\cdots,a_n)=\det H(a_2,\cdots,a_n)$, where 
\[ H(a_2,\cdots,a_n)=\begin{pmatrix}
	2 & -a_2 &0  & \cdots &  0 \\
	-a_2 &   2 &\ddots   &  \vdots \\
	0 & \ddots& \ddots&\ddots  & 0 \\
	\vdots  &\ddots & \ddots&  2   &-a_n \\
	0  &\cdots & 0&-a_n & 2
\end{pmatrix}, \ a_j\in[0,1], j=2\cdots,n.\]
We start with direct computation for $n=2,3$:
\[P_2(a_2)=4-a_2^2=3+(1-a_2^2), \ \ P_3(a_2,a_3)=4+2(1-a_2^2)+2(1-a_3^2).\]
For $n\ge 4$, 
\begin{multline}
    P_n(a_2,\cdots,a_n)
    =2P_{n-1}(a_3,\cdots,a_n)-a_2^2P_{n-2}(a_4,\cdots,a_n) \\
    =2P_{n-1}(a_3,\cdots,a_n)-P_{n-2}(a_4,\cdots,a_n)+(1-a_2^2)P_{n-2}(a_4,\cdots,a_n). \label{eq:96}
\end{multline}
Inductively, one can easily verify that
\begin{align}
P_n(a_2,\cdots,a_n)=c_n+\sum_{k=2}^nb^n_k(1-a_k^2)+\sum_{j\ge 2} d^n_{i_1,\cdots,i_j}(1-a_{i_1}^2)\cdots(1-a_{i_j}^2),
\end{align}
which is a multilinear polynomial in $z_k=1-a_k^2$, $k=2,\cdots,n$. 
For simplicity, we refer to $c_n$ as the constant term, $b^n_k(1-a_k^2)$ as the linear term, and refer the rest as the higher order terms, e.g., $d^n_{3,5}(1-a_3^2)(1-a_5^2)$ is considered as a quadratic term. It remains to compute the constant and linear coefficients $c_n,b^n_k$, and show that all higher order coefficients $d^n_{i_1,\cdots,i_j}\ge 0$.  By \eqref{eq:96}, the constant term $c_n$ satisfies the recurrent relation $c_n=2c_{n-1}-c_{n-2}$ with initial values $c_2=3,c_3=4$. Then $c_n=n+1,n=2,\ldots$ by induction. In fact, by the polynomial expansion \eqref{eq:ldelta-Tn}, $c_n=P_n(1,\cdots,1)=\det(-\Delta_{[1,n]})=n+1$. We may assume $c_0=\det I=1$ and $c_1=\det (2)=2$ for consistency.  The first two terms in \eqref{eq:96} do not contain $a_2$. Hence only the last term in in \eqref{eq:96} will contribute (non-trivial) terms containing $1-a_2^2$. In particular, the coefficient of the linear term $b_2^n$ comes from the constant term in the expansion of  $P_{n-2}(a_4,\cdots,a_n)$, i.e., $b_2^n=c_{n-2}=n-1$. Similar expansion can be done with respect to the last row to obtain $b_n^n=n-1$. 
For $3\le i\le  n-1$, we can expand  $ P_n(a_2,\cdots,a_n)$ with respect to the $i$th row to obtain
\begin{align}\label{eq:98}
    P_n(a_2,\cdots,a_n)
     :=& f(\check a_i)+ (1-a_i^2)P_{i-2}(a_2,\cdots,a_{i-2})P_{n-i}(a_{i+2}\cdots,a_n),
\end{align}
where $f(\check a_i)$ is a polynomial which does not contain $a_i$. Only the last term will contribute non-trivial terms containing $1-a_i^2$, and the coefficient of the linear term comes from the product of the constant terms of $P_{i-3}(a_2,\cdots,a_{i-2})$ and $P_{n-i}(a_{i+1}\cdots,a_n)$, i.e., $b_i^n=c_{i-2}c_{n-i}=(i-1)(n-i+1)$, for $3\le i\le n-1$. Notice this expression of $b_i^n$ is also true for $i=2$ and $i=n$.  

The coefficients for the higher order terms can be analyzed in a similar manner, via the further expansion of either \eqref{eq:96} or \eqref{eq:98}. For example,  continuing from the expansion of $1-a_i^2$ in \eqref{eq:98}, if we want to obtain the quadratic term $(1-a_i^2)(1-a_j^2)$ for some $2 \le j\le i-2$, it is enough to expand  $P_{i-2}(a_2,\cdots,a_{i-2})$ and $P_{n-i}(a_{i+2}\cdots,a_n)$ as in \eqref{eq:ldelta-Tn}. And the coefficient of $(1-a_i^2)(1-a_j^2)$ would come from the product of the linear coefficient in $P_{i-2}(a_2,\cdots,a_{i-2})$ for $1-a_j^2$ and the constant term in $P_{n-i}(a_{i+2}\cdots,a_n)$. Hence, this quadratic coefficient $d^n_{i,j}$ is nonnegative. (The quadratic coefficient $d^n_{i,j}$ is actually strictly positive if $|i-j|\ge 2$. It is zero if and only if $|i-j|=1$.) Inductively, one can show that all the higher order coefficients $d^n_{i_1,\cdots,i_j}$ in \eqref{eq:ldelta-Tn} are nonnegative. This completes the proof of Claim \ref{claim:Pn-poly}.

%%%%%%%%%%%%%%%%%%%%%%%%%%%%%%%%%%%%%%%%%%%%%%%%%%%%%%%%%%%%%%%%%%%%%%%%%%%%%%%%%%%%%%%%
\section{Proof of Lemma \ref{lem:SdProp12}}\label{sec:Sdprop12}
In this section we include a short proof for the uniform case in Lemma~\ref{lem:SdProp12} for completeness, following the main steps in \cite{sanchez2023principal}. Suppose  $\{a_j\}_{j\in \Z}$ are i.i.d. random variables with the $[0,1]$-uniform distribution. Let $Z_{\delta}^{\pm}$  be given as in \eqref{eq:Z-},\eqref{eq:Z+}, and  
 $\ell_\delta=Z_{\delta}^{+}+Z_{\delta}^{-}$. Then, for any $\delta>0,n\in\Z^+$, 
\begin{align*}
    \Pr{\ell_\delta>n}=& \Pr{Z_{\delta}^{+}>n-1}+\sum_{j=1}^{n-1}\Pr{Z_{\delta}^{+}=j}\Pr{Z_{\delta}^{-}>n-j} \\
    \le & \Pr{Z_{\delta}^{+}>n-1}+\Pr{Z_{\delta}^{-}>n-1}+\sum_{j=2}^{n-1}\Pr{Z_{\delta}^{+}>j-1}\Pr{Z_{\delta}^{-}>n-j}.
\end{align*}
For any $t>0$ and $m\ge1$,
\begin{align*}
    \Pr{Z_{\delta}^{+}>m}&\le  \Pr{\sum_{j=x+1}^{x+m}(j-x)(1-a_j)\ \le  \delta^{-1}}  \\
    &\le  e^{t\delta^{-1}} \prod_{j=1}^m   \Ev{e^{-tj(1-a_j)}} 
    \le    e^{t\delta^{-1}} \prod_{j=1}^m \frac{1}{tj}  
   =  e^{t\delta^{-1}} \frac{1}{t^m m!},
\end{align*}
where we used the Chernoff bound (exponential Markov inequality) and the fact that $\{1-a_j\}_{j\in\Z}$ are also i.i.d. uniformly distributed on $[0,1]$. 
Now taking $t=m\delta$, one has $\Pr{Z_{\delta}^{+}>m}\le  e^{m} \frac{1}{m^m \delta^m m!} \le (e^2/\delta)^m m^{-2m}$, where we used Stirling's approximation $m!>m^m/e^m$. 
By the same argument, $\Pr{Z_{\delta}^{-}>m} \le C_\delta^m m^{-2m}$, where we denote $C_\delta=e^2/\delta$ for convenience. 
Therefore, 
\begin{align*}
    \Pr{\ell_\delta>n}\le & 2C_\delta^{n-1}  (n-1)^{-2(n-1)}+\sum_{j=2}^{n-1}C_\delta^{j-1} (j-1)^{-2(j-1)}C_\delta^{n-j} (n-j)^{-2(n-j)} \\
    =& 2C_\delta^{n-1}  (n-1)^{-2(n-1)}+C_\delta^{n-1}\sum_{j=2}^{n-1} (j-1)^{-2(j-1)} (n-j)^{-2(n-j)}.
\end{align*}
Meanwhile, $f(j)=(j-1)^{-2(j-1)} (n-j)^{-2(n-j)},\, j\in [2,n-1]$ attains its maximum at $j=(n+1)/2$, which implies 
\begin{align*}
    \Pr{\ell_\delta>n}\le & 2C_\delta^{n-1}  (n-1)^{-2(n-1)}+C_\delta^{n-1}(n-2) (\frac{n-1}{2})^{-2(n-1)}\\ 
    \le & 2n(4C_\delta)^{n} (n-1)^{-2(n-1)}\\
    =&2\exp\{\log n+n\log(4C_\delta)-2(n-1)\log(n-1)\}
    \le 2\exp\{
    -2n\log n+C'n\} ,
\end{align*}
where $C'=\log (4C\delta)+2\log 2+3$ provided $n$ is large enough. 
Now take $n=(1+\eps)T_N=(1+\eps)\frac{\log N}{2 \log\log N} $ for some $\eps>0$. 
Then 
$ \log n\ge \log T_N=\log \log N-\log (2\log\log N)$
which implies that 
\begin{align*}
  -2n\log n +C'n \le & -2(1+\eps)\frac{\log N}{2 \log\log N}(\log \log N-\log 2\log\log N)+C'(1+\eps)\frac{\log N}{2 \log\log N}\\
  \le & -(1+\eps/2)\log N.
\end{align*}
In the last inequality, we picked $N$ large and $\eps$ small so that $(1+\eps)\frac{\log (2\log\log N)}{ \log\log N}+C'\frac{1+\eps}{2 \log\log N}\le \eps/2 $. Notice that the above estimate is independent of of $x$.  
Therefore,
\begin{align*}
\Pr{ {\max_{x\in \intbr{1}{N}}  \ell_\delta(x)}\ge (1+\eps){T_N}} \le & \sum_{x\in \intbr{1}{N}}\Pr{  \ell_\delta(x)\ge (1+\eps){T_N}}\\
\le &  N\cdot 2\exp\{-(1+\eps/2)\log N\}=2N^{-\eps/2}.
\end{align*}
Then for $N_k=\lfloor e^K \rfloor$,  
\[\sum_{k}\Pr{ {\max_{x\in \intbr{1}{N_k}}  \ell_\delta(x)}\ge (1+\eps){T_{N_k}}}<\infty.\] 
Finally, as done in Proposition 3 of \cite{sanchez2023principal}, combing  the monotonicity of $n\mapsto \max_{x\in \intbr{1}{N}} $, with the Borel–Cantelli Lemma, and sending $\eps\to 0$ gives that a.s. 
\[\limsup_{N\to \infty}\frac{\max_{x\in \intbr{1}{N}}}{T_N}\le 1.  \]

\bibliographystyle{abbrv}
\bibliography{landscape.bib}

 {
  \bigskip
  \vskip 0.08in \noindent --------------------------------------

\footnotesize
\medskip

L.~Shou, \textsc{School of Mathematics, University of Minnesota, 206 Church St SE, Minneapolis, MN 55455 USA}\par\nopagebreak

\textit{Present address}: Condensed Matter Theory Center and Joint Quantum Institute, Department of Physics, University of Maryland, College Park, MD 20742, USA\par\nopagebreak
    \textit{E-mail address}:  \href{mailto:lshou@umd.edu}{lshou@umd.edu}

\vskip 0.4cm

  W. ~Wang, \textsc{LSEC, Institute of Computational Mathematics and Scientific/Engineering Computing, Academy of Mathematics and Systems Science, Chinese Academy of Sciences, Beijing 100190, China}\par\nopagebreak
  \textit{E-mail address}: \href{mailto:ww@lsec.cc.ac.cn}{ww@lsec.cc.ac.cn}
  
\vskip 0.4cm

S.~Zhang, \textsc{Department of Mathematical Sciences, University of Massachusetts Lowell, 
Southwick Hall, 
11 University Ave.
Lowell, MA 01854
 }\par\nopagebreak
  \textit{E-mail address}: \href{mailto:shiwen\_zhang@uml.edu}{shiwen\_zhang@uml.edu}
  }

\end{document}